%% file: gif.tex
\documentstyle[12pt,a4,epsf]{book}

\catcode`\@=11
\def\marginnote#1{}
\newcount\hour
\newcount\minute
\newtoks\amorpm
\hour=\time\divide\hour by60
\minute=\time{\multiply\hour by60 \global\advance\minute by-\hour}
\edef\standardtime{{\ifnum\hour<12 \global\amorpm={am}
        \else\global\amorpm={pm}\advance\hour by-12 \fi
        \ifnum\hour=0 \hour=12 \fi
        \number\hour:\ifnum\minute<10 0\fi\number\minute\the\amorpm}}
\edef\militarytime{\number\hour:\ifnum\minute<10 0\fi\number\minute}
\def\draftlabel#1{{\@bsphack\if@filesw {\let\thepage\relax
   \xdef\@gtempa{\write\@auxout{\string
      \newlabel{#1}{{\@currentlabel}{\thepage}}}}}\@gtempa
   \if@nobreak \ifvmode\nobreak\fi\fi\fi\@esphack}
        \gdef\@eqnlabel{#1}}
\def\@eqnlabel{}
\def\@vacuum{}
\def\draftmarginnote#1{\marginpar{\raggedright\scriptsize\tt#1}}
\def\draft{\oddsidemargin -.5truein
        \def\@oddfoot{\sl preliminary draft \hfil
        \rm\thepage\hfil\sl\today\quad\militarytime}
        \let\@evenfoot\@oddfoot \overfullrule 3pt
        \let\label=\draftlabel
        \let\marginnote=\draftmarginnote
   \def\@eqnnum{(\theequation)\rlap{\kern\marginparsep\tt\@eqnlabel}
\global\let\@eqnlabel\@vacuum}  }

\def\nodraft{\oddsidemargin -.5truein
        \def\@oddfoot{\hfil\rm\thepage\hfil}
        \let\@evenfoot\@oddfoot \overfullrule 3pt
   \def\@eqnnum{(\theequation)\rlap{\kern\marginparsep\tt \@eqnlabel }
\global\let\@eqnlabel\@vacuum}  }

\def\preprint{\twocolumn\sloppy\flushbottom\parindent 1em
        \leftmargini 2em\leftmarginv .5em\leftmarginvi .5em
        \oddsidemargin -.5in    \evensidemargin -.5in
        \columnsep 15mm \footheight 0pt
        \textwidth 250mmin      \topmargin  -.4in
        \headheight 12pt \topskip .4in
        \textheight 175mm
        \footskip 0pt
        \def\@oddhead{\thepage\hfil\addtocounter{page}{1}\thepage}
        \let\@evenhead\@oddhead \def\@oddfoot{} \def\@evenfoot{} }

\def\titlepage{\@restonecolfalse\if@twocolumn\@restonecoltrue\onecolumn
     \else \newpage \fi \thispagestyle{empty}\c@page\z@
        \def\thefootnote{\fnsymbol{footnote}} }

\def\endtitlepage{\if@restonecol\twocolumn \else  \fi
        \def\thefootnote{\arabic{footnote}}
        \setcounter{footnote}{0}}  

\catcode`@=12
\relax
\def\bea{\begin{array}}
\def\bem{\begin{displaymath}}
\def\beq{\begin{equation}}

\def\eea{\end{array}}
\def\eem{\end{displaymath}}
\def\eeq{\end{equation}}
\def\no{\noindent}
\def\half{\frac{1}{2}}

\def\ov{\overline}

\def\Tr{\mathop{\rm Tr}}

\def\crbig{\\\noalign{\vspace {3mm}}}

\def\vac{|0\rangle}
\def\bra#1{\left\langle #1\right|}
\def\ket#1{\left| #1\right\rangle}
\def\sldc{${\rm Sl}(2,{\bf C})$\ }
\def\tr{{\rm tr}\, }
\def\ha{{1\over 2}}
\def\wt{\widetilde}
\def\ra{\rangle}
\def\la{\langle}
\def\a{\alpha}
\def\b{\beta}
\def\g{\gamma}
\def\G{\Gamma}
\def\e{\epsilon}
\def\f{\phi}
\def\fb{{\ov \phi}}
\def\vf{\varphi}
\def\m{\mu}
\def\n{\nu}
\def\r{\rho}

\def\s{\sigma}
\def\sb{\ov\sigma}
\def\l{\lambda}
\def\L{\Lambda}
\def\p{\psi}
\def\pb{\ov\psi}
\def\cb{\ov\chi}
\def\d{\partial}
\def\dag{\dagger}

\def\da{{\dot\alpha}}
\def\db{{\dot\beta}}
\def\dg{{\dot\gamma}}
\def\dd{{\dot\delta}}
\def\t{\theta}
\def\tb{{\ov \theta}}
\def\lb{{\ov \lambda}}
\def\eb{{\ov \epsilon}}
\def\zb{{\ov z}}
\def\ib{{\ov i}}
\def\jb{{\ov j}}
\def\kb{{\ov k}}
\def\mb{{\ov m}}

\def\D{\Delta}
\def\DD{\Delta^\dag}
\def\Db{\ov D}
\def\M{{\cal M}}
\def\rd{\sqrt{2}}
\def\Tr{{\rm Tr\, }}
\def\F{{\cal F}}
\def\Dint{\int\, {\rm d}^2\theta {\rm d}^2\overline\theta\ }
\def\Fint{\int\, {\rm d}^2\theta\ }
\def\Fbarint{\int\, {\rm d}^2\overline\theta\ }
\def\xint{\int\, {\rm d}^4 x}

\relax

\nodraft
\begin{document}
\voffset=-1.6cm
\parskip .33cm
\title{\bf Introduction to Supersymmetry 
\vskip 1.cm
\large{\bf Adel Bilal}\break
\vskip .5cm
\centerline{\small Institute of Physics, University of Neuch\^atel}
\centerline{\small rue Breguet 1, 2000 Neuch\^atel, Switzerland }
\centerline{\small \tt adel.bilal@unine.ch}
}
\date{}
\maketitle

\vskip 1.cm
\hoffset 1.cm
\centerline{Abstract}
\noindent
These are expanded notes of lectures given at the summer school 
``Gif 2000" in Paris. They constitute the first part of an 
``Introduction to supersymmetry and supergravity" with the second 
part on supergravity by J.-P. Derendinger to appear soon. 

\noindent
The present introduction is elementary and pragmatic. I discuss: 
spinors and the Poincar\'e group, the susy algebra and susy multiplets, 
superfields and susy lagrangians, susy gauge theories, spontaneously 
broken susy, the non-linear sigma model, N=2 susy gauge theories, and 
finally Seiberg-Witten duality.

\pagenumbering{roman}
\tableofcontents 


\chapter{Introduction}
\pagenumbering{arabic}\setcounter{page}{1}

\input{ch0.tex}

\chapter{Spinors and the Poincar\'e group}\label{chap1}

\input{ch1.tex}

\chapter{The susy algebra and its representations}\label{chap2}

\input{ch2.tex}

\chapter{Superspace and superfields}\label{chap3}

\input{ch3.tex}

\chapter{Supersymmetric gauge theories}\label{chap4}

\input{ch4.tex}

\chapter{Spontaneously broken supersymmetry}\label{chap7}

\input{ch7.tex}

\chapter{The non-linear sigma model}\label{chap5}

\input{ch5.tex}

\chapter{$N=2$ susy gauge theory}\label{chap6}

\input{ch6.tex}







\chapter{Seiberg-Witten duality in $N=2$ gauge theory}\label{chap11}

\input{ch11.tex}

{\large Acknowledgements}

These lectures have grown out of previous ones on the subject.
Chapters 1 to 5 as well as 8 and 9 are elaborations of lectures 
given on several occasions, in particular at the Ecole 
Normale Sup\'erieure in Paris in 1995-96, and I wish to thank all the 
members of the audience for critical remarks and suggestions. 

I am particularly grateful to J.-P. Derendinger for providing access to 
his unpublished lecture notes on supersymmetry: the present 
chapters 6 and 7 on the non-linear sigma model and susy breaking are heavily 
inspired from his notes.









\end{document}

%% file: ch0.tex
%

Supersymmetry not only has played a most important role in the development of
theoretical physics over the last three decades, but also has strongly
influenced experimental particle physics. 

Supersymmetry first appeared in the early 
seventies  in the context of string theory where it was a symmetry of the 
{\it two-dimensional} world sheet theory. At this time it was more considered 
as a purely theoretical tool. Shortly after it was realised that supersymmetry
could be a symmetry of four-dimensional quantum field theories and as such 
could well be directly relevant to elementary particle physics. String theories 
with supersymmetry on the world-sheet, if suitably modified, were shown to 
actually exhibit supersymmetry in space-time, much as the four-dimensional 
quantum field theories: this was the birth of superstrings. Since then, 
countless supersymmetrc theories have been developed with minimal or 
extended global supersymmetry or with a local version of supersymmetry which is 
supergravity.

There are several reasons why an elementary particle physicist wants to 
consider supersymmetric theories. The main reason is that radiative 
corrections tend to be less important in supersymmetric theories, due 
to cancellations between fermion loops and boson loops. As a result 
certain quantities that are small or vanish classically (i.e. at tree 
level) will remain so once radiative corrections (loops) are taken into 
account. Famous examples include the vanishing or extreme smallness of 
the cosmological constant, the hierarchy problem (why is there such a 
big gap between the Planck scale / GUT scale and the scale of electroweak 
symmetry breaking) or the issue of renormalisation of quantum gravity.
While supersymmetry could solve most if not all of these questions, it
cannot be the full answer, since we know that supersymmetry cannot be 
exactly realised in nature: it must be broken at experimentally accessible 
energies since otherwise one certainly would have detected many of the 
additional particles it predicts.

Supersymmetric models often are easier to solve than non-supersymmetric 
ones since they are more 
constrained by the higher degree of symmetry. Thus they may serve as 
toy models where certain analytic results can be obtained and may 
serve as a qualitative guide to the behaviour of more realistic theories. 
For example the study of supersymmetric versions of QCD have given quite 
some insights in the strong coupling dynamics responsible for phenomena 
like quark confinement. In this type of studies the basic property is a 
duality (a mapping) between a weakly and a strongly coupled theory. It 
seems that dualities are difficult to realise in non-supersymmetric 
theories but are rather easily present in supersymmetric ones. The study of 
dualities in superstring theories has been particular fruitful over the 
last five years or so.

Supersymmetry has also appeared outside the realm of elementary particle 
physics and has found applications in condensed matter systems, in 
particular in the study of disordered systems.

In these lectures, I will try to give an elementary and pragmatic introduction 
to supersymmetry. In the first four chapters, I introduce the 
supersymmetry algebra and its basic representations, i.e. the supermultiplets 
and then present supersymmetric field theories with emphasis on 
supersymmetric gauge theories. The presentation is pragmatic in the sense 
that I try to introduce only as much mathematical structure 
as is necessary to arrive at the supersymmetric field theories.
No emphasis is put on uniqueness theorems or the like. On the 
other hand, I very quickly introduce superspace and superfields as a 
useful tool because it allows to easily and efficiently 
construct supersymmetric 
Lagrangians. The discussion remains classical and due to lack of time the 
issue of renormalisation is not discussed here. Then 
follows a brief discussion of spontaneous breaking of supersymmetry.
The supersymmetric non-linear sigma model is discussed in some detail as 
it is relevant to the coupling of supergravity to matter multiplets. 
Finally I focus on $N=2$ extended supersymmetric gauge theories followed by a
rather detailed introduction to the determination of their low-energy 
effective action, taking advantage of duality and the rigid 
mathematical structure of $N=2$ supersymmetry. 

There are many textbooks and review articles on supersymmetry (see e.g.
\cite{FF} to \cite{PS}) that
complement the present introduction and also contain many references to 
the original literature which are not given here.

%% file: ch1.tex
%

We begin with a review of the Lorentz and Poincar\'e groups and 
spinors in four-dimensional
Minkowski space. The signature is taken to be +,-,-,- so that $p^2=+m^2$
and $\m,\n,\ldots$ always are space-time indices, while $i,j,\ldots$ are 
only space indices. Then the metric $g_{\m\n}$ is diagonal with 
$g_{00}=1$, $g_{11}=g_{22}=g_{33}=-1$. 

\section{The Lorentz and Poincar\'e groups}\label{1sec11}

\underline{The Lorentz group} has six generators, 
three rotations $J_i$ and three boosts 
$K_i$, $i=1,2,3$ with commutation relations
\beq\label{1lor}
[J_i,J_j]=i \e_{ijk}J_k\ , \quad [K_i,K_j]=-i \e_{ijk}J_k\ , \quad
[J_i,K_j]=i \e_{ijk} K_j\ .
\eeq
To identify the mathematical structure and to construct representations of 
this algebra one introduces the  linear combinations 
\beq\label{1complcomb}
J_j^\pm=\ha (J_j\pm i K_j)
\eeq 
in terms of which the algebra separates into 
two commuting ${\rm SU}(2)$ algebras:
\beq\label{1su2}
[J_i^\pm, J_j^\pm]=i \e_{ijk}J_k^\pm\ , \quad [J_i^\pm, J_j^\mp]=0\ .
\eeq
These generators are not hermitian however, and we see that the Lorentz 
group is a complexified version of ${\rm SU}(2)\times {\rm SU}(2)$: this 
group is \sldc. (More precisely, \sldc is the universal cover of the 
Lorentz group, just as ${\rm SU}(2)$ is the universal cover of 
${\rm SO}(3)$.) To see that this group is really \sldc is easy: introduce 
the four $2\times 2$ matrices $\s_\m$ where $\s_0$ is the identity matrix 
and $\s_i$, $i=1,2,3$ are the three Pauli matrices.
(Note that we always write the Pauli matrices with a lower index $i$, while $\s^0=\s_0$ and $\s^i=-\s_i$.) Then for every 
four-vector $x^\m$ the $2\times 2$ matrix $x^\m \s_\m$ is hermitian 
and has determinant equal to $x^\m x_\m$ which is a Lorentz invariant. 
Hence a Lorentz transformation  preserves the determinant and the 
hermiticity of this matrix, and thus must act as 
$x^\m \s_\m \to A x^\m \s_\m A^\dag$ with $|\det A |=1$. We see that 
up to an irrelevant phase, A is a complex $2\times 2$ matrix of unit 
determinant, i.e. an element of \sldc. This establishes the mapping 
between an element of the Lorentz group and the group \sldc.

\underline{The Poincar\'e group} contains, 
in addition to the Lorentz transformations, 
also the translations. More precisely it is a semi-direct product of the 
Lorentz-group and the group of translations in space-time. The generators 
of the translations are usually denoted $P_\m$. In addition to the commutators 
of the Lorentz generators $J_i$ (rotations) and $K_i$ (boosts) one has the 
following commutation relations involving the $P_\m$:
\beq\label{1poincare}
\begin{array}{rcl}
[P_\m,P_\n]&=&0\ , \crbig
[J_i,P_j]&=&i \e_{ijk} P_k \ , \ [J_i,P_0]=0\ , \
[K_i,P_j]=-i P_0\ ,\ [K_i,P_0]=-i P_j \ ,
\end{array}
\eeq
which state that translations commute among themselves, that the $P_i$ are a 
vector and $P_0$ a scalar under space rotations and how $P_i$ and $P_0$ mix 
under a boost. The Lorentz and Poincar\'e algebras are often written in a more 
covariant looking, but less intuitive form. One defines the Lorentz generators 
$M_{\m\n}=-M_{\n\m}$ as $M_{0i}=K_i$ and $M_{ij}=\e_{ijk} J_k$. Then the full 
Poincar\'e algebra reads
\beq\label{1poincarebis}
\begin{array}{rcl}
[P_\m,P_\n]&=&0\ , \crbig
[M_{\m\n},M_{\r\s}]&=&i g_{\n\r} M_{\m\s} -i g_{\m\r} M_{\n\s}
-i g_{\n\s} M_{\m\r} + i g_{\m\s} M_{\n\r}\ , \crbig
[M_{\m\n},P_\r]&=& -i g_{\r\m} P_\n + i g_{\r\n} P_\m \ . \cr
\end{array}
\eeq

\section{Spinors}\label{1sec12}

\no\underline{Two-component spinors}

There are various equivalent ways to introduce spinors. Here we define 
spinors as the objects carrying the basic representation of \sldc. Since 
elements of \sldc are complex $2\times 2$ matrices, a spinor is a two 
complex component object $\p=\pmatrix{\p_1\cr \p_2\cr}$ transforming 
under an element $\M=\pmatrix{\a&\b\cr \g&\delta\cr}\in $ \sldc as
\beq\label{1spintrans}
\p_\a \to \p'_\a=\M_\a^{\ \b}\ \p_\b \ ,
\eeq
with $\a,\b=1,2$ labeling the components. Now, unlike for ${\rm SU}(2)$, 
for \sldc a representation and its complex conjugate are not equivalent. 
$\M$ and $\M^*$ give inequivalent representations. A two-component object 
$\pb$ transforming as
\beq\label{1dotspintrans}
\pb_\da \to \pb'_\da={\M^*}_\da^{\ \db}\ \pb_\db \ 
\eeq
is called a dotted spinor, while the above $\p$ is called an undotted one. 
Comparing the complex conjugate of (\ref{1spintrans}) with (\ref{1dotspintrans})
we see that we can identify $\pb_\da$ with $(\p_\a)^*$.

The representation carried by the $\p_\a$ is called $(\ha,0)$ 
(matrices $\M$) and the one carried by the $\pb_\da$ is called 
$(0,\ha)$ (matrices $\M^*$). They are both irreducible. Now, any 
\sldc matrix can be written as
\beq\label{1mmetoile}
\begin{array}{rcl}
\M &=& \exp(a_j \s_j +i b_j \s_j) \crbig
\M^*&=& \exp(a_j \s_j^* -i b_j \s_j^*) \ . \cr
\end{array}
\eeq
This explicitly displays the generators as the spin $\ha$ representation 
of the complexified ${\rm SU}(2)$, in accordance with (\ref{1complcomb}).

It proves very useful to now introduce some notations and conventions.
We intoduce the antisymmetric two-index tensors $\e^{\a\b}$ and  $\e_{\a\b}$
\beq\label{1epstensor}
\e^{\a\b}=\e^{\da\db}=\pmatrix{0&1\cr -1&0\cr}\ , \quad
\e_{\a\b}=\e_{\da\db}=\pmatrix{0&-1\cr 1&0\cr}
\eeq
which are used to raise and lower indices as follows:
\beq\label{1raiseandlower}
\p^\a=\e^{\a\b}\p_\b\ , \quad \p_\a=\e_{\a\b}\p^\b \ , \quad
\pb^\da=\e^{\da\db}\pb_\db\ , \quad \pb_\da=\e_{\da\db}\pb^\db \ .
\eeq
One can then easily show that the transformation under an element 
$\M$ of \sldc is ${\p'}^\a = \p^\b (\M^{-1})_\b^{\ \a}$ and 
${\pb'}^\da = \pb^\db ({\M^*}^{-1})_\db^{\ \da}$.

The four $\s_\m$ matrices introduced above naturally have a dotted and 
an undotted index. Recalling that our signature is +,-,-,- we have
\beq\label{1sigmu}
(\s^\m)_{\a\da}= ({\bf 1}, -\s_i)_{\a\da} \ .
\eeq
Raising the indices using the $\e$ tensors 
yields
\beq\label{1sigbarmu}
(\sb^\m)^{\da\a} = \e^{\da\db}\e^{\a\b} (\s^\m)_{\b\db}= 
({\bf 1}, +\s_i)^{\da\a} \ .
\eeq

Whenever we consider expressions involving more 
than one spinor we have to remember that spinors anticommute. Hence 
(with two-component spinors)
$\p_1 \chi_2=- \chi_2 \p_1$, as well as $\p_1 \cb_{\dot 2}=- \chi_{\dot 2} \p_1$
etc. The scalar products $\p\chi$ and $\pb\cb$ are defined as
\beq\label{1scalar}
\begin{array}{rcl}
\p\chi\equiv \p^\a \chi_\a &=&\e^{\a\b} \p_\b\chi_\a 
=-\e^{\a\b}\p_\a\chi_\b = -\p_\a\chi^\a = \chi^\a\p_\a = \chi\p\crbig
\pb\cb\equiv\pb_\da\cb^\da &=& \ldots = \cb_\da \pb^\da\crbig
(\p\chi)^\dag &=& \cb_\da\pb^\da = \cb\pb=\pb\cb \ . \cr
\end{array}
\eeq
Note that by convention undotted indices are always contracted from upper 
left to lower right, while dotted indices are always contracted from lower 
left to upper right. Note however that this rule does not apply when rising 
or lowering spinor indices with the $\e$-tensor. With this rule we also have
\beq\label{2psisigmuchi}
\p\s^\m\cb = \p^\a \s^\m_{\a\db}\cb^\db \quad , \quad 
\pb\sb^\m \chi = \pb_\da\sb^{\m\da\b}\chi_\b\ .
\eeq
One can then prove a certain amount of useful identities which we summarise 
here:
\beq\label{1spinident}
\begin{array}{rcl}
\chi\s^\m\pb = -\pb\sb^\m\chi \quad &,& \quad
\chi \s^\m \sb^\n \p = \p \s^\n \sb^\m \chi \crbig
(\chi\s^\m \pb)^\dag = \p \s^\m \cb \quad &,& \quad
(\chi\s^\m \sb^\n\p)^\dag=\pb\sb^\n\s^\m\cb \crbig
\p\chi=\chi\p \quad , \quad \pb\cb &=& \cb\pb \quad , \quad
(\p\chi)^\dag = \pb\cb \ . \cr
\end{array}
\eeq

\no\underline{Dirac spinors}

One introduces the Dirac matrices in the Weyl representation as
\beq
\g^\m=\pmatrix{0&\s^\m\cr \sb^\m& 0 \cr}\ , 
\quad \g_5=i \g^0\g^1\g^2\g^3=\pmatrix{{\bf 1}&0\cr 0& -{\bf 1}\cr}
\eeq
A four-component Dirac spinor is made from a two-component undotted and a 
two-component dotted spinor as $\pmatrix{\p_\a\cr \cb^\da\cr}$. Clearly it
transforms as the reducible $(\ha,0)\oplus (0,\ha)$ representation of the 
Lorentz group. Then $\pmatrix{\p_\a\cr 0\cr}$ and $\pmatrix{0\cr \cb^\da\cr}$
are chiral Dirac (or Weyl) spinors. A Majorana spinor is a Dirac spinor with $\chi\equiv\p$, i.e. it is of the form $\pmatrix{\p_\a\cr \pb^\da\cr}$.
The Lorentz generators are
\beq\label{1gammalorentz}
\Sigma^{\m\n}={i\over 2} \g^{\m\n} \ , \quad 
\g^{\m\n}={1\over 2}(\g^\m\g^\n-\g^\n\g^\m)
= {1\over 2} 
\pmatrix{ \s^\m \sb^\n-\s^\n\sb^\m & 0 \cr 0 &  \sb^\m \s^\n-\sb^\n\s^\m \cr } 
\ .
\eeq
We see that indeed the undotted and dotted spinors transform separately, 
the generators being $i\s^{\m\n}$ for $\p_\a$ and $i\sb^{\m\n}$ for $\pb^\da$ 
with
\beq\label{1weylgen}
\begin{array}{rcl}
(\s^{\m\n})_\a^{\ \b} &=& {1\over 4} \left( \s^\m_{\a\dg} \sb^{\n\dg\b} 
- ( \m \leftrightarrow \n) \right) \crbig
(\sb^{\m\n})^\da_{\ \db} &=& {1\over 4} \left( \sb^{\m\da\g} \s^\n_{\g\db} 
- ( \m \leftrightarrow \n) \right)\ .\cr
\end{array}
\eeq
Note that e.g. $\s^{12}=\sb^{12} =-{i\over 2} \s_3\equiv -{i\over 2} \s_z$ 
so that the rotation generator $M_{12}=M^{12}$ is $\ha \s_z$ as expected.

\no\underline{Casimirs: mass and helicity}

A useful quantity is the Pauli-Lubanski vector
\beq\label{1pauliluba}
W^\m= \ha \e^{\m\n\r\s} P_\n M_{\r\s} \ ,
\eeq 
which can be easily shown to commute 
with the $P_\m$ and behaves as a four-vector under commutation with the 
Lorentz generators. It follows that $W^2\equiv W^\m W_\m$ as well as 
$P^2\equiv P_\m P^\m$ commute with all the generators, i.e. they are 
two (and the only two) Casimirs of the Poincar\'e group. For a massive 
particle one can go to the rest frame where $P_\m=(m,0,0,0)$ and then 
$P^2=m^2$ and $W^2=-m^2 s(s+1)$ where $s$ is the spin of the particle. 
The different states of this irreducible representation are distinguished 
by the value 
of $p_i$ and of $M_{12}\equiv J_3 = S_3$ in the rest frame
because in the rest frame only the 
spin and not the orbital part contributes to the angular momentum.  In the 
above representation of dotted or undotted spinors one has of course $s=\ha$. 
For a massless particle $P^2=0$ and also $W^2=0$. We may take 
$P_\m=(E,0,0,E)$ so that $W^\m= M_{12} P^\m$ with $M_{12}=\pm s$ 
being the  helicity. For such massless 
particles $s$ is fixed and the different states of this irreducible 
representation 
are distinguished by the sign of the helicity and by the values of $p_i$.

%% file: ch2.tex
%

\section{The supersymmetry algebra}

We want to enlarge the Poincar\'e algebra by generators that transform 
either as undotted spinors $Q^I_\a$ or as dotted spinors $\ov Q^I_\da$ under 
the Lorentz group and that commute with the translations. The extra index 
$I=1, \ldots N$ labels the different spinorial generators in case there are 
more than one pair. This means that according to (\ref{1weylgen})
\beq\label{2algmixed}
\begin{array}{rcl}
[P_\m,Q^I_\a]&=&0\ , \crbig
[P_\m,\ov Q^I_\da]&=&0 \ ,\crbig
[M_{\m\n},Q^I_\a]&=&i (\s_{\m\n})_\a^{\ \b} Q^I_\b , \crbig
[M_{\m\n},\ov Q^{I\da}]&=& i (\sb_{\m\n})^\da_{\ \db} \ov Q^{I\db} \ . \cr
\end{array}
\eeq
In particular, $M_{12}\equiv J_3$ and thus $[J_3,Q^I_1] =\ha Q^I_1$ and
$[J_3,Q^I_2] =-\ha Q^I_2$. Since 
$\ov Q^{I1}=-(Q^I_2)^\dag$ and $\ov Q^{I2}=(Q_1^I)^\dag$ one similarly has
$[J_3,(Q^I_2)^\dag] =\ha (Q^I_2)^\dag$ and
$[J_3,(Q^I_1)^\dag] =-\ha (Q^I_1)^\dag$. We conclude that $Q^I_1$ and 
$(Q^I_2)^\dag$ rise  the 
$z$-component of the spin (helicity) by half a unit, while $Q^I_2$ and
$(Q^I_1)^\dag$ lower it by half a unit.

Since the $Q^I_\a$ transform in the $(\ha,0)$ representation and the 
$\ov Q^I_\da$ in the 
$(0,\ha)$, the anticommutator of $Q^I_\a$ and $\ov Q^J_\db$ must transform 
as $(\ha,\ha)$, i.e. as a four vector. The obvious candidate is $P_\m$
so that we arrive at
\beq\label{2algqqbar}
\{Q^I_\a, \ov Q^J_\db \}=2 \s^\m_{\a\db} P_\m \delta^{IJ} \ .
\eeq
The $\delta^{IJ}$ can always be achieved by diagonalising an a priori 
arbitrary symmetric matrix and by rescaling the $Q$ and $\ov Q$. 
Furthermore, since $\ov Q$ is the adjoint of $Q$, positivity of the 
Hilbert space excludes zero eigenvalues of this matrix. Finally
\beq\label{2algqqandqbarqbar}
\{Q^I_\a, Q^J_\b\}= \e_{\a\b} Z^{IJ}\ , \quad 
\{\ov Q^I_\da, \ov Q^J_\db\} = \e_{\da\db} (Z^{IJ})^*\ .
\eeq
The $Z^{IJ}=-Z^{JI}$ are central charges which means they commute with 
all generators of the full algebra. The simplest algebra has $N=1$, i.e. 
there are no indices $I,J$ and there is no possibility of central charges. 
This is the unextended susy algebra. If $N>1$ one talks about extended 
supersymmetry. In the simplest extended case, $N=2$, there is just one 
central charge $Z\equiv Z^{12}$. From the algebraic point of view there 
is no limit on $N$, but we will see that with increasing $N$ the theories 
also must contain particles of increasing spin and there seem to be no 
consistent quantum field theories with spins larger than one (without gravity) 
or larger than two (with gravity) leading to $N\le 4$, resp. $N\le 8$.

\section{Some basic properties}

Using the above susy algebra it is easy to establish some basic properties 
of supersymmetric theories. Since the full susy algebra contains the 
Poincar\'e algebra as a subalgebra, any representation of the full susy 
algebra also gives a representation of the Poincar\'e algebra, although 
in general a reducible one. Since each irreducible representation (of the 
type considered above) of the Poincar\'e algebra corresponds to a particle, 
an irreducible representation of the susy algebra in general corresponds 
to several particles. The corresponding states are related to each other 
by the $Q^I_\a$ and $\ov Q^J_\db$ and thus have spins differing by units 
of one half. They form a supermultiplet. By abuse of language we will call 
an irreducioble representation of the susy algebra simply a supermultiplet. 
Clearly, using the spin-statistics 
theorem,  the $Q$ and $\ov Q$ change bosons into fermions and vice versa.
One then has:

\no\underline{All particles belonging to an irreducible representation of 
susy, i.e. within one}\break\hfill
\underline{supermultiplet, have the same mass}. This is obvious 
since $P^2$ commutes with all generators of the susy algebra, i.e. it is 
still a Casimir operator.

\no\underline{In a supersymmetric theory the energy $P_0$ is always positive}.
To see this, let $\ket \Phi$ be any state. Then by the positivity of the Hilbert space we have
\beq\label{2posenergy}
\begin{array}{rcl}
0&\le& ||Q^I_\a \ket \Phi ||^2 + || (Q^I_\a)^\dag \ket \Phi ||^2 
=\bra\Phi \left( (Q^I_\a)^\dag Q^I_\a 
+ Q^I_\a  (Q^I_\a)^\dag \ket \Phi \right) \crbig
&=& \bra\Phi \{Q^I_\a, \ov Q^I_\da\} \ket\Phi 
= 2 \s^\m_{\a\da} \bra\Phi P_\m \ket\Phi \cr
\end{array}
\eeq
since $ (Q^I_\a)^\dag\equiv \ov Q^I_\da$. Summing this over 
$\a\equiv \da = 1,2$ and using $\tr \s^\m=2 \delta^{\m 0}$ 
yields $0\le 4 \bra\Phi P_0 \ket\Phi$, which was to be shown.

\no\underline{A supermultiplet always contains an equal number of 
bosonic and fermionic}\break\hfill
\underline{degrees of freedom}. By degrees of freedom 
one means physical (positive norm) states. Hence a photon has two 
degrees of freedom corresponding to the two helicities $+1$ and $-1$ 
(the two polarizations). Let the fermion number be $N_F$ equal one 
on a fermionic state and 0 on a bosonic one. Equivalently $(-)^{N_F}$
is $+1$ on bosons and $-1$ on fermions. We want to show that 
\beq\label{2ftrace}
{\rm{Tr}}\ (-)^{N_F}=0
\eeq 
if the trace is taken over any finite-dimensional 
representation. Note that $ (-)^{N_F}$ anticommutes with $Q$.
Using the cyclicity of the trace, one has
\beq\label{2fermiontrace}
\begin{array}{rcl}
0&=&{\rm{Tr}} \left( - Q_\a (-)^{N_F} \ov Q_\db 
+ (-)^{N_F} \ov Q_\db Q_\a \right)
={\rm{Tr}} \left( (-)^{N_F} \{Q_\a, \ov Q_\db\} \right) \crbig
&=&2 \s^\m_{\a\db} {\rm{Tr}}  \left( (-)^{N_F} P_\m \right) \ .\cr
\end{array}
\eeq
Choosing any non-vanishing momentum $P_\m$ gives the desired result.

\section{Massless supermultiplets}

We will first assume that all central charges $Z^{IJ}$ vanish. Below we will 
see that for massless representations this is necessarily a consequence of the 
positivity of the Hilbert space. Then all $Q^I_\a$ anticommute among 
themselves, 
and so do the $\ov Q^J_\db$. Since $P^2=0$ we choose a reference frame where 
$P_\m=E(1,0,0,1)$ so that $\s^\m P_\m=\pmatrix{0&0\cr 0& 2E\cr}$ and thus
\beq\label{2masslessalg}
\{ Q^I_\a, \ov Q^J_\db\} =\pmatrix{0&0\cr 0& 4E\cr}_{\a\db} \delta^{IJ} \ .
\eeq
In particuler, $\{Q^I_1, \ov Q^J_{\dot 1}\}=0,\ \forall I,J$. On a positive 
definite Hilbert space we must then set $Q^I_1=\ov Q^J_{\dot 1} =0,\ \forall I,J$. 
The argument is similar to the one above:
\beq\label{2qunzero}
0=\bra\Phi \{Q^I_1,\ov Q^I_{\dot 1}\}\ket\Phi
=||Q^I_1\ket\Phi ||^2 + ||\ov Q^I_{\dot 1} \ket\Phi ||^2 \Rightarrow 
Q^I_1=\ov Q^J_{\dot 1} =0
\eeq
Thus we are left with only the $Q^I_2$ and $\ov Q^J_{\dot 2}$, i.e. $N$ of the 
initial $2N$ fermionic generators. If we define
\beq\label{2oscil}
a_I={1\over\sqrt{4E}} Q^I_2 \ , \quad a_I^\dag = {1\over\sqrt{4E}} \ov Q^I_{\dot 2}
\eeq
the $a_I$ and $a_I^\dag$ are anticommuting annihilation and creation operators:
\beq\label{2oscilalg}
\{a_I,a_J^\dag \}= \delta_{IJ}\ , \quad \{a_I,a_J \}=\{a_I^\dag ,a_J^\dag \}=0 \ .
\eeq

One then chooses a ``vacuum state", i.e. a state annihilated by all the $a_I$. 
Such a state will carry some irreducible representation of the Poincar\'e algebra, 
i.e. in addition to its zero mass it is characterised by some helicity $\l_0$. We 
denote this state as $\ket {\l_0}$. From the commutators of $Q^I_2$ and 
$\ov Q^J_{\dot 2}$ with the helicity operator which in the present frame is 
$J_3=M_{12}$ one sees that $Q^I_2$ lowers the helicity by one half and 
$\ov Q^J_{\dot 2}$ rises it by one half. (For simplicity, we suppose here 
that the state $\ket {\l_0}$ transforms as a singlet under the ${\rm SU}(N)$ that 
acts on the indices $I,J$. One could easily drop this restriction.)
The supermultiplet then is of the form
\beq\label{2supermult}
\begin{array}{rcl}
&&\ket {\l_0}\crbig
a_I^\dag \ket {\l_0} &=& {\ket {\l_0+\ha}}_I\crbig
a_I^\dag a_J^\dag \ket {\l_0} &=& {\ket {\l_0+1}}_{IJ}\crbig
&&\ldots\cr
a_1^\dag a_2^\dag \ldots a_N^\dag \ket {\l_0} &=&  {\ket {\l_0+{N\over 2}}} \ .\cr
\end{array}
\eeq
Due to the antisymmetry in $I,J,\ldots$ there are $\pmatrix{N\cr k\cr}$ states 
with helicity $\l=\l_0+{k\over 2}$, $k=0,1,\ldots N$. Summing the binomial 
coefficients gives a total of $2^N$ states with $2^{N-1}$ having integer helicity 
(bosons) and $2^{N-1}$ having half-integer helicity (fermions). In general, in 
such a supermultiplet, except if $\l_0=-{N\over 4}$, the helicities will not be 
distributed symmetrically about zero. Such supermultiplets cannot be invariant 
under CPT, since CPT flips the sign of the helicity. To satisfy CPT one then 
need to double these multiplets by adding their CPT conjugate with opposite 
helicities and opposite quantum numbers.

For unextended susy, $N=1$, each massless supermultiplet only contains two states 
$\ket \l_0$ and $\ket {\l_0+\ha}$. We denote these multiplets by $(\l_0, \l_0+\ha)$. 
They can never be CPT self-conjugate and one needs to double them. Thus
one arrives at the following massless $N=1$ multiplets:

\no\underline{The chiral multiplet} consists of $(0,\ha)$ and its CPT 
conjugate $(-\ha,0)$, corresponding to a Weyl fermion and a complex scalar.

\no\underline{The vector multiplet} consists of $(\ha,1)$ plus $(-1,-\ha)$, 
corresponding to a gauge boson (massless vector) and a Weyl fermion, both 
necessarily in the adjoint representation of the gauge group.

\no\underline{The gravitino multiplet} contains $(1,{3\over 2})$ and 
$(-{3\over 2}, -1)$, i.e. a gravitino and a gauge boson.

\no\underline{The graviton multiplet} contains $({3\over 2},2)$ and 
$(-2,-{3\over 2})$, corresponding to the graviton and the gravitino.

Since we so not want helicities larger than two, we must stop here. Also the 
gravitino should be present only in a theory with gravity, so if $N=1$ it must 
only occur once and then in the gravity multiplet. Hence the gravitino multiplet 
cannot appear in unextended susy. However it does appear in extended susy when 
decomposing larger multiplets into $N=1$ multiplets.

For $N=2$ a supermultiplet contains $(\l_0, \l_0+\ha, \l_0+\ha, \l_0+1)$. 
Restricting ourselves to the cases where the helicity does not exceed one, 
we have two possibilities.

\no\underline{The $N=2$ vector multiplet} contains $(0,\ha,\ha,1)$ and its CPT 
conjugate \break\hfill
$(-1,-\ha,-\ha,0)$, corresponding to a vector (gauge boson), 
two Weyl fermions and a complex scalar, again all in the adjoint 
representation of the gauge group. In terms of $N=1$ representations this is 
a vector and a chiral $N=1$ multiplet.

\no\underline{The hypermultiplet}: If $\l_0=-\ha$ we get $(-\ha,0,0 \ha)$. 
This may or not be  CPT self-conjugate. If it is, it is called a 
half-hypermultiplet. If it is not we have to add its CPT conjugate to 
get a (full) hypermultiplet $2\times (-\ha,0,0 \ha)$. 

For $N=4$, restricting again to helicities not exceeding one, there is a single 
\underline{$N=4$ multiplet} which always is CPT self-conjugate. It is 
$(-1, 4\times (-\ha), 6 \times 0, 4\times \ha, 1)$, containing a vector 4 Weyl 
fermions and 3 complex scalars. In terms of $N=2$ multiplets it is just the sum 
of the $N=2$ vector multiplet and a hypermultiplet, however now all 
transforming in the adjoint of the gauge group.

\section{Massive supermultiplets}

We now consider the case $P^2>0$ and a priori arbitrary central charges
$Z^{IJ}$. 
Going to the rest frame $P_\m=(m,0,0,0)$, the susy algebra becomes
\beq\label{2susyalgmassive}
\begin{array}{rcl}
\{ Q^I_\a, ( Q^J_\b)^\dag \} &=& 2 m \delta_{\a\b} \delta^{IJ} \crbig
\{ Q^I_\a,  Q^J_\b\} &=& \e_{\a\b} Z^{IJ}\crbig
\{ (Q^I_\a)^\dag,  (Q^J_\b)^\dag\} &=& \e_{\a\b} (Z^{IJ})^* \ .\cr
\end{array}
\eeq
By an appropriate ${\rm U}(N)$ rotation among the $Q^I$ the antisymmetric 
matrix of central charges can be brought into standard form:
\beq\label{2zstandard}
Z^{IJ}=\pmatrix{ 0&q_1&0&0& \cr -q_1&0&0&0& \ldots\cr
0&0&0&q_2& \cr 0&0&-q_2&0& \cr & &\vdots & & \cr} 
\eeq
with all $q_n \ge 0$, $n=1, \ldots {N\over 2}$. 
We assume that $N$ is even, since otherwise there is an extra zero 
eigenvalue of the $Z$-matrix which can be handled trivially.

It follows that if we let
\beq\label{2massiveoscil}
\begin{array}{rcl}
a_\a^1 &=& {1\over \sqrt{2}} \left( Q_\a^1 + \e_{\a\b} (Q^2_\b)^\dag \right)
\crbig
b_\a^1 &=& {1\over \sqrt{2}} \left( Q_\a^1 - \e_{\a\b} (Q^2_\b)^\dag \right)
\crbig
a_\a^2 &=& {1\over \sqrt{2}} \left( Q_\a^3 + \e_{\a\b} (Q^4_\b)^\dag \right)
\crbig
b_\a^2 &=& {1\over \sqrt{2}} \left( Q_\a^3 - \e_{\a\b} (Q^4_\b)^\dag \right)
\crbig
&\vdots&\cr
\end{array}
\eeq
then the $a_\a^r$ and $b_\a^r$, $r=1,\ldots {N\over 2}$ and their 
hermitian conjugates satisfy the following algebra of harmonic oscillators
\beq\label{2massiveoscilalg}
\begin{array}{rcl}
\{ a^r_\a, (a^s_\b)^\dag\} &=& (2m-q_r) \delta_{rs} \delta_{\a\b} \crbig
\{ b^r_\a, (b^s_\b)^\dag\} &=& (2m+q_r) \delta_{rs} \delta_{\a\b} \crbig
\{ a^r_\a, (b^s_\b)^\dag\} &=& \{ a^r_\a, a^s_\b \} = \ldots = 0 \cr
\end{array}
\eeq
Clearly, positivity of the Hilbert space requires that 
\beq\label{2bpsbound}
2m \ge |q_n| 
\eeq
for all $n$. If some of the $q_n$ saturate the bound, i.e. are equal to $m$, 
then the corresponding operators must be set to zero, as we did in the 
massless case  with the $Q^I_1$. Clearly, in the massless case the bound 
becomes $0\ge |q_n|$ and thus $q_n=0$ always. There cannot be central 
charges in the massless case and the bound is always saturated, thus only 
exactly have of the fermionic generators survive. 

In the more general 
massive case, if all $|q_n|$ are strictly less than $2m$ we have a total 
of $2N$ harmonic oscillators.  Then starting from a state of minimal 
``helicity" (i.e.  $z$ component of the angular momentum)  
$\l_0$ annihilated by all $a^n_\a$ and $b^n_\b$, application 
of the creation operators yields a total of $2^{2N}$ states with 
helicities ranging from  $\l_0$ to  $\l_0+N$. 

For $N=1$ this yields four states, again labeled by their helicities 
(or rather the $z$ component of the angular momentum), as $(-\ha,0,0,\ha)$ 
(which is the same as the CPT extended massless multiplet) or 
$(-1,-\ha,-\ha,0)$ to which we must add the CPT conjugate $(1,\ha,\ha,0)$. 
The latter  are the same states as a massless vector plus a massless chiral 
multiplet and can be obtained from them via a Higgs mechanism. In terms of 
massive representations this is a vector (3 dofs) a Dirac fermion (4 dofs) 
and a single real scalar (1 dof).  

For $N=2$ we already have 16 states with 
helicities ranging at least from $-1$ to $1$. Such a massive $N=2$ multiplet 
can be viewed as the union of a massless $N=2$ vector and hypermultiplet. 
A generic massive $N=4$ multiplet contains $2^8=256$ states including  at least a helicity $\pm 2$. Thus such a theory must include a massive spin two 
particle which is not believed to be possible in quantum field theory.

If $k<{N\over 2}$ of the $q_n$ are equal to $2m$ then we only have $2N -2k$ 
oscillators, and the supermultiplets will only contain $2^{2(N-k)}$ states. 
They are called short multiplets or  BPS multiplets. If all $q_n$ equal $2m$, 
i.e. $k={N\over 2}$ we get the shortest multiplets with only $2^N$ states, 
exactly as in the massless case. These BPS multiplets are also called 
ultrashort, and are completely analogous to the massless multiplets.

%% file: ch3.tex
%

Since we want to construct supersymmetric quantum field theories, we have to 
find representations of the susy algebra on fields. A convenient and compact 
way to do this is to introduce superspace and superfields, i.e. fields defined 
on superspace. This is particularly simple for unextended susy, so we will 
restrict here to $N=1$ superspace and superfields. Then we have two plus two 
susy generators $Q_\a$ and $\ov Q_\da$, as well as four generators $P_\m$ of 
space-time translations. The idea then is to enlarge space-time labelled by 
the coordinates $x^\m$ by adding two plus two anticommuting Grassmannian 
coordinates $\t_\a$ and $\tb_\da$. Thus coordinates on superspace are 
$(x^\m,\t_\a,\tb_\da)$. Rather than elaborating on the meaning of such a 
space we will simply use it as a very efficient recepee to perform 
calculations in supersymmetric theories.

\section{Superspace}

As already said, we restrict here to $N=1$. The ``odd" superspace coordinates 
$\t_\a$ and $\tb_\da$ just behave as constant ($x^\m$ independent) spinors. 
Recall that as all spinors they anticommute among themselves, i.e. 
$\t^1\t^2=-\t^2\t^1$, and idem for the $\tb^\da$. Spinor indices in 
bilinears are contracted acording to the usual rule, i.e. 
$\t\t=\t^\a\t_\a=-2\t^1\t^2=+2\t_2\t_1=-2\t_1\t_2$, and 
$\tb\tb=\tb_\da\tb^\da=2\tb_1\tb_2=\ldots$. One can then easily prove the 
following useful identities:
\beq\label{3thetaident}
\begin{array}{rcl}
\t^\a\t^\b=-\ha \e^{\a\b} \t\t \quad & , & \quad
\tb^\da\tb^\db =\ha \e^{\da\db} \tb\tb \ , \crbig
\t_\a\t_\b=\ha \e_{\a\b} \t\t \quad & , & \quad
\tb_\da\tb_\db =-\ha \e_{\da\db} \tb\tb \ , \crbig
\t\s^\m\tb\, \t\s^\n\tb = \ha \t\t\, \tb\tb g^{\m\n} \quad & , & \quad
\t\p\ \t\chi = -\ha \t\t \ \p\chi \ .
\end{array}
\eeq

Derivatives in $\t$ and $\tb$ are defined in an obvious way as 
${\d\over\t^\a} \t^\b = \delta_\a^\b$ and 
${\d\over\tb^\da} \tb^\db = \delta_\da^\db$. Since the $\t$'s anticommute, 
any product involving more than two $\t$'s or more than two $\tb$'s vanishes.
Hence an arbitrary (scalar) function on superspace, i.e. a superfield, 
can always be expanded as
\beq\label{3gensuperfield}
\begin{array}{rcl}
F(x,\t,\tb)&=&f(x)+\t\p(x) +\tb\cb(x) +\t\t\, m(x) +\tb\tb\,  n(x)\crbig
&+&\t\s^\m\tb\,  v_\m(x) +\t\t\,  \tb\lb(x) +\tb\tb\, \t\r(x) 
+\t\t\, \tb\tb\,  d(x) \ .
\crbig
\end{array}
\eeq
If $F$ carries extra vector indices then so do the fomponent fields 
$f,\p,\ldots$.

Integration on superspace is defined for a single Grassmannian variable, 
say $\t^1$ as $\int {\rm d}\t^1 (a+\t^1 b) = b$ so that 
$\int  {\rm d}\t^1  {\rm d}\t^2 \t^2 \t^1 =1$. Then since $\t\t=2\t^2\t^1$
and $\tb\tb=2\t^1\t^2$ 
we define ${\rm d}^2\t=\ha  {\rm d}\t^1  {\rm d}\t^2 $ and  
${\rm d}^2\tb = \ha {\rm d}\tb^2  {\rm d}\tb^1 = [{\rm d}^2\t]^\dag$ so that
\beq
\Fint \t\t = \Fbarint \tb\tb =1 \ .
\eeq
It is easy to check that
\beq\label{3fint}
\Fint = {1\over 4}\e^{\a\b} {\d\over\d \t^\a} {\d\over\d \t^\b} \quad , \quad
\Fbarint = -{1\over 4}\e^{\da\db} {\d\over\d \tb^\da} {\d\over\d \tb^\db} \ .
\eeq
Clearly one also has
\beq\label{3dint}
\Dint \t\t\tb\tb =1 \ .
\eeq
With these definitions it is easy to see that one has the hermiticity property
\beq\label{3herm}
\left( {\d\over\t^\a} \right)^\dag = + {\d\over\tb^\da} 
\eeq
with $\a\equiv \da$. Note the plus sign rather than a minus sign as one 
would expect from $(\d_\m)^\dag=-\d_\m$.

We now want to realise the susy generators $Q_\a$ and their hermitian 
conjugates $\ov Q_\da=(Q_\a)^\dag$ as differential operators on superspace. 
We want that $i\e^\a Q_\a$ generates a translation in $\t^\a$ by a constant 
infinitesimal spinor $\e^\a$ plus some translation in $x^\m$. The latter 
space-time translation is determined by the susy algebra since the 
commutator of two such susy transformations is a translation in space-time. 
Thus we want
\beq\label{3susytranseps}
(1+i\e Q) F(x,\t,\t)= F(x+\delta x, \t+\e, \tb)\ .
\eeq
Hence $i Q_\a={\d\over \d\t^\a}+\ldots$ where $+\ldots$ must be of the form $c (\s^\m\tb)_\a P_\m = -i c (\s^\m\tb)_\a \d_\m$ with some constant $c$ to be determined. We arrive at the ansatz
\beq\label{3qgen}
Q_\a= - i \left( {\d\over \d\t^\a} -i c (\s^\m\tb)_\a \d_\m \right) \ .
\eeq
Then the hermitian conjugate is
\beq\label{3qbargen}
\ov Q_\da=  i \left( {\d\over \d\tb^\da} -i c^* (\t\s^\m)_\da \d_\m \right) \ ,
\eeq
and they satisfy the susy algebra, in particular
\beq\label{3susyanticom}
\{ Q_\a,  Q_\db\}= 2 \s^\m_{\a\db} P_\m = - 2i \s^\m_{\a\db} \d_\m
\eeq
if ${\rm Re}\ c=1$. We choose $c=1$ so that
\beq\label{3susygendiff}
\begin{array}{rcl}
Q_\a &=& - i  {\d\over \d\t^\a} - \s^\m_{\a\db} \tb^\db \d_\m \crbig
\ov Q_\da&=& i {\d\over \d\tb^\da} + \t^\b \s^\m_{\b\da} \d_\m \ . \cr
\end{array}
\eeq
We can now give the action on the superfield $F$ and determine $\delta x$:
\beq\label{3susytransepsepsbar}
(1+i\e Q+i\eb \ov Q) F(x^\m,\t^\a,\tb^\db)
= F(x^\m -i\e\s^\m\tb+i\t\s^\m\eb, \t^\a+\e^\a, \tb^\db+\eb^\db)
\eeq
and the susy variation of a superfield is of course defined as
\beq\label{3susyvarF}
\delta_{\e,\eb} F= (i\e Q+i\eb \ov Q) F \ .
\eeq

Since a general superfield contains too many component fields to 
correspond to an irreducible representation of $N=1$ susy, it will 
be very useful to impose susy invariant condition to lower the number of
components. To do this, we first find covariant derivatives $D_\a$ and 
$\Db_\da$ that anticommute with the susy generators $Q$ and $\ov Q$. 
Then $\delta_{\e,\eb} (D_\a F) = D_\a (\delta_{\e,\eb} F)$ and idem for 
$\Db_\da$. It follows that $D_\a F=0$ or $\Db_\da F=0$ are susy 
invariant constraints one may impose to reduce the number of components
in a superfield. One finds
\beq\label{3covdiff}
\begin{array}{rcl}
D_\a &=&  {\d\over \d\t^\a} +i \s^\m_{\a\db} \tb^\db \d_\m \crbig
\Db_\da&=&=  {\d\over \d\tb^\da} + i \t^\b \s^\m_{\b\da} \d_\m  \cr
\end{array}
\eeq
where $\Db_\da=(D_\a)^\dag$ and
\beq\label{3DQanticom}
\begin{array}{rcl}
\{D_\a, \Db_\db\}= 2 i \s^\m_{\a\db}\d_\m \quad & , & 
\{D_\a,D_\b\}=\{\Db_\da, \Db_\db\}=0 \crbig
\{D_\a, Q_\b\} = \{\Db_\da,Q_\b\} 
&=& \{D_\a,\ov Q_\db\} = \{\Db_\da, \ov Q_\db \} =0 \ .\crbig 
\end{array}
\eeq

\section{Chiral superfields}

A chiral superfield $\f$ is defined by the condition
\beq\label{3chiralsup}
\Db_\da \f = 0
\eeq
and an anti-chiral one $\fb$ by
\beq\label{3antichiralsup}
D_\a \fb = 0\ .
\eeq
This is easily solved by observing that
\beq\label{3ycoord}
\begin{array}{rcl}
D_\a\tb=\Db_\da\t &=& D_\a \ov y^\m=\Db_\da y^\m=0 \ , \crbig
y^\m=x^\m+i\t\s^\m\tb \quad & , & \quad \ov y^\m = x^\m-i\t\s^\m\tb \ . \cr
\end{array}
\eeq
Hence $\f$ depends only on $\t$ and $y^\m$ (i.e. all $\tb$ dependence is 
through $y^\m$) and $\fb$ only on $\tb$ and $\ov y^\m$. Concentrating on 
$\f$ we have the component expansion
\beq\label{3chiralcompy}
\f(y,\t)=z(y)+\rd\t\p(y)-\t\t f(y)
\eeq
or Taylor expanding  in terms of $x,\ \t$ and $\tb$:
\beq\label{3chiralcompx}
\begin{array}{rcl}
\f(y,\t)&=&z(x) + \rd\t\p(x) + i\t\s^\m\tb \d_\m z(x) - \t\t f(x) \crbig
&-&{i\over\rd} \t\t\d_\m\p(x) \s^\m\tb - {1\over 4}\t\t\tb\tb \d^2 z(x) \ . \cr
\end{array}
\eeq
Physically, such a chiral superfield describes one complex scalar $z$ and 
one Weyl fermion $\p$. The field $f$ will turn out to be an auxiliary field. 
For $\fb$ we similarly have
\beq\label{3antichiralcompx}
\begin{array}{rcl}
\fb(y,\tb)&=&\zb(\ov y) +\rd \tb\pb(\ov y) -\tb\tb \ov f(\ov y) \crbig
&=&\zb(x) + \rd\tb\pb(x) - i\t\s^\m\tb \d_\m \zb(x) 
- \tb\tb \ov f(x) \crbig
&+&{i\over\rd} \tb\tb \t\s^\m\d_\m\pb(x) - {1\over 4}\t\t\tb\tb \d^2 \zb(x) \ . \cr
\end{array}
\eeq

Finally, let us find the explicit susy variations of the component fields 
as it results from (\ref{3susyvarF}): First, for chiral superfields it is 
useful to change variables from $x^\m,\t,\tb$ to $y^\m,\t,\tb$. Then
\beq\label{3susygeny}
Q_\a=-i{\d\over \d\t^\a} \quad , \quad
\ov Q_\da= i {\d\over \d \tb^\da} + 2 \t^\b \s^\m_{\b\da} {\d\over \d y^\m}
\eeq
so that
\beq\label{3susycompfields}
\begin{array}{rcl}
\delta\f(y,\t) &\equiv& \left( i \e Q + i \eb \ov Q \right) \f(y,\t) =
\left( \e^\a {\d\over \d\t^\a}  + 2i \t \s^\m\eb {\d\over \d y^\m} \right)
\f(y,\t) \crbig
&=& \rd \e\p -2 \e\t f + 2i\t\s^\m\eb (\d_\m z+\rd \t\d_\m\p ) \crbig
&=& \rd \e\p +\rd \t \left( -\rd\e f + \rd i \s^\m \eb \d_\m z \right) 
- \t\t \left( - i \rd \eb \sb^\m \d_\m \p \right) \ .
\end{array}
\eeq
Thus we read the susy transformations of the component fields:
\beq\label{3susycompfieldstwo}
\begin{array}{rcl}
\delta z &=& \rd \e\p \crbig
\delta \p &=& \rd i \d_\m z \s^\m \eb  -\rd f \e \crbig
\delta f &=&  \rd i \d_\m \p  \s^\m \eb \ . \crbig
\end{array}
\eeq
The factors of $\rd$ do appear because of our normalisations of the fields 
and the definition of $\delta\f$. If desired, they could be absorbed by a 
rescaling of $\e$ and $\eb$.

\section{Susy invariant actions}

To construct  susy invariant actions we now only need to make a few 
observations. First, products of superfields are of course superfields. 
Also, products of (anti) chiral superfields are still (anti) chiral superfields.
Typically, one will have a superpotential $W(\f)$ which is again chiral. 
This $W$ may depend on several different $\f_i$. Using the $y$ and $\t$ 
variables one easily Taylor expands
\beq\label{3superpot}
\begin{array}{rcl}
W(\f)&=& W(z(y)) +\rd {\d W\over \d z_i} \, \t\p_i(y) \crbig
&-& \t\t \left( {\d W\over \d z_i} \, f_i(y)
+\ha {\d^2 W\over \d z_i  \d z_j} \, \p_i(y) \p_j(y) \right) \  \crbig
\end{array}
\eeq
where it is understood that $\d W/\d z$ and $\d^2 W /\d z\d z$ are 
evaluated at $z(y)$.
The second and important observation is that any Lagrangian of the form
\beq\label{3susylagr}
\Dint F(x,\t,\tb) + \Fint W(\f) + \Fbarint [ W(\f)]^\dag
\eeq
is automatically susy invariant, i.e. it transforms at most by a total 
derivative in space-time. The proof is very simple. The susy variation 
of any superfield is given by (\ref{3susyvarF}) and, since the $\e$ and 
$\eb$ are constant spinors and the $Q$ and $\ov Q$ are differential 
operators in superspace, it is again a total derivative in all of 
superspace:
\beq\label{3susyvartwo}
\delta F = {\d\over \d\t^\a} (-\e^\a F) +  {\d\over \d\tb^\da} (-\eb^\da F)
+{\d\over \d x^\m} [ -i (\e\s^\m\tb-\t\s^\m\eb) F] \ .
\eeq
Integration $\Dint$ only leaves the last term which is a total space-time 
derivative as claimed. If now $F$ is a chiral superfield like $\f$ or 
$W(\f)$ one changes variables to $\t$ and $y$ and one has
\beq\label{3susyvarthree}
\delta \f = {\d\over \d\t^\a} (-\e^\a \f(y,\t)) 
+{\d\over \d y^\m} [ -i (\e\s^\m\tb-\t\s^\m\eb) \f(y,\t)] \ .
\eeq
Integrating $\Fint$ again only leaves the last term which becomes 
${\d\over \d x^\m} [ \ldots]$ and is a total derivative in space-time. 
The analogous result holds for an anti chiral superfield 
$\ov W(\fb) = [W(\f)]^\dag$ and integration $\Fbarint$. This proves 
the supersymmetry of the action resulting from the space-time integral 
of the Lagrangian (\ref{3susylagr}).

The terms $\Fint W(\f) + h.c.$ in the Lagrangian have the form of a 
potential. The kinetic terms must be provided by the term $\Dint F$. 
The simplest choice is $F=\f^\dag \f$. This is neither chiral nor anti 
chiral but real. To compute $\f^\dag \f$ one must first expand the 
$y^\m$ in terms of $x^\m$. We only need the terms $\sim \t\t\tb\tb$, 
called the $D$-term:
\beq\label{3kinDterm}
\begin{array}{rcl}
\f^\dag \f\Big\vert_{\t\t\tb\tb}&=& 
-{1\over 4}  z^\dag \d^2 z -{1\over 4} \d^2 z^\dag z 
+\ha \d_\m z^\dag \d^\m z 
+ f^\dag f 
+{i\over 2} \d_\m\p\s^\m \pb - {i\over 2} \p \s^\m\d_\m \pb  \crbig
&=&  \d_\m z^\dag \d^\m z  
+{i\over 2} (\d_\m\p\s^\m \pb -  \p \s^\m\d_\m \pb)
+ f^\dag f 
+ {\rm total \ derivative}\ . \crbig
\end{array}
\eeq
Then
\beq\label{3susyaction}
S= \int\, {\rm d}^4 x {\rm d}^2\t {\rm d}^2\tb \ \  \fb_i^\dag \f_i
+  \int\, {\rm d}^4 x {\rm d}^2\t \ W(\f_i) + h.c.
\eeq
yields
\beq\label{3susyactiontwo}
S= \int\, {\rm d}^4 x \Big[  
|\d_\m z_i|^2 -i \p_i\s^\m\d_\m \pb_i + f_i^\dag f_i 
- {\d W\over \d z_i} f_i + h.c. 
- \ha {\d^2 W\over \d z_i \d z_j} \p_i \p_j + h.c. \Big] \ . 
\eeq
More generally, one can replace $\f_i^\dag \f_i$ by a (real) K\"ahler 
potential $K(\f_i^\dag, \f_j)$. This leads to the non-linear $\s$-model 
discussed later. In any case, the $f_i$ have no kinetic term and hence 
are auxiliary fields. They should be eliminated by substituting their 
algebraic equations of motion
\beq\label{3feoms}
f^\dag_i=\left( {\d W\over \d z_i}\right) 
\eeq
into the action, leading to
\beq\label{3susyactionthree}
S= \int\, {\rm d}^4 x \Big[  
|\d_\m z_i|^2 -i \p_i\s^\m\d_\m \pb_i 
- \left|  {\d W\over \d z_i}\right|^2  
-  \ha {\d^2 W\over \d z_i \d z_j} \p_i \p_j 
- \ha \left( {\d^2 W\over \d z_i \d z_j}\right)^\dag \pb_i \pb_j 
\Big] \ . 
\eeq
We see that the scalar potential $V$ is determined in terms of the 
superpotential $W$ as
\beq\label{3scalarpot}
V=\sum_i \left|  {\d W\over \d z_i}\right|^2 \ .
\eeq

To illustrate this model, consider the simplest case of a single chiral 
superfield $\f$ and a cubic superpotential 
$W(\f)= {m\over 2} \f^2 +{g\over 3} \f^3$. Then 
${\d W\over \d z}=m\f+g\f^2$ and the action becomes
\beq\label{3susyactionWZ}
\begin{array}{rcl}
S_{\rm WZ}= \int\, {\rm d}^4 x &\Big[  &
|\d_\m z|^2 -i \p\s^\m\d_\m \pb
- m^2 |z|^2 -{m\over 2} (\p\p+\pb\pb)
  \crbig
&-&  m g (z^\dag z^2 + (z^\dag)^2 z) - g^2 |z|^4  + g(z\p\p+z^\dag\pb\pb)
\Big] \ . \crbig
\end{array}
\eeq
Note that the Yukawa interactions appear with a coupling constant $g$ that 
is related by susy to the bosonic coupling constants $m g$ and $g^2$.

\section{Vector superfields}

The $N=1$ supermultiplet of next higher spin is the  vector multiplet. 
The corresponding superfield $V(x,\t,\tb)$ is real and has the expansion
\beq\label{3vectorsuper}
\begin{array}{rcl}
V(x,\t,\tb)&=& C+i\t\chi-i\tb\cb +\t\s^\m\tb v_\m\crbig
&+& {i\over 2} \t\t (M+iN) -{i\over 2} \tb\tb (M-iN)\crbig
&+& i \t\t\, \tb\left( \lb +{i\over 2}\sb^\m\d_\m\chi\right)
- i \tb\tb\, \t\left( \l -{i\over 2}\s^\m\d_\m\cb\right) \crbig
&+&\ha \t\t\tb\tb \left( D-\ha \d^2 C\right) \crbig
\end{array}
\eeq
where all component fields only depend on $x^\m$. There are 8 bosonic 
components ($C,D,M,N,v_\m$) and 8 fermionic components ($\chi,\l$). 
These are too many components to describe a single supermultiplet. We want to reduce their number by making use of the supersymmetric generalisation of a gauge transformation. Note that the 
transformation
\beq\label{3vecttrans}
V\to V+\f+\f^\dag \ ,
\eeq
with $\f$ a chiral superfield, implies the component transformation
\beq\label{3vmutrans}
v_\m\to v_\m + \d_\m (2{\rm Im} z)
\eeq
which is an abelian gauge transformation. We conclude that (\ref{3vecttrans}) is 
its desired supersymmetric generalisation. If this transformation 
(\ref{3vecttrans})
is a symmetry (actually a gauge 
symmetry, as we just saw) of the theory then, by an appropriate choice of $\f$, one can 
transform away the components $\chi,C,M,N$ and one component of $v_\m$. 
This choice is called the Wess-Zumino gauge, and it reduces the vector 
superfield to
\beq\label{3vectWZ}
V_{\rm WZ}=\t\s^\m\tb v_\m(x) + i\t\t\, \tb\lb(x)
-i \tb\tb\, \t\l(x) +\ha \t\t\tb\tb D(x) \ .
\eeq
Since each term contains at least one $\t$, the only non-vanishing power 
of $V_{\rm WZ}$ is
\beq\label{3vectWZsquare}
V_{\rm WZ}^2=\t\s^\m\tb\ \t\s^\n\tb\ v_\m v_\n = \ha \t\t\tb\tb\ v_\m v^\m
\eeq
and $V_{\rm WZ}^n=0$, $n\ge 3$.

To construct kinetic terms for the vector field $v_\m$ one must act on 
$V$ with the covariant derivatives $D$ and $\Db$. Define
\beq\label{3walpha}
W_\a=-{1\over 4} \Db\Db D_\a V \quad , \quad 
\ov W_\da = -{1\over 4} DD \Db_\da V \ .
\eeq
(This is appropriate for abelian gauge theories and will be slightly 
generalized in the non-abelian case.) Since $D^3=\Db^3=0$, $W_\a$ is 
chiral and $\ov W_\da$ antichiral. Furthermore it is clear that they 
behave as anticommuting Lorentz spinors. Note that they are invariant 
under the transformation (\ref{3vecttrans}) since
\beq\label{3Winvar}
\begin{array}{rcl}
W_\a &\to& W_\a-{1\over 4} \Db\Db D_\a (\f+\f^\dag)
= W_\a +{1\over 4} \Db^\db \Db_\db D_\a \f\crbig
&=& W_\a  +{1\over 4} \Db^\db \{ \Db_\db , D_\a \}\f
= W_\a+{i\over 2} \s^\m_{\a\db} \d_\m \Db^\db \f = W_\a
\end{array}
\eeq
since $\Db\f=D\f^\dag=0$. It is then easiest to use the WZ-gauge to 
compute $W_\a$. To facilitate things further, change variables to 
$y^\m,\t^\a,\tb^\da$ so that
\beq\label{3ddby}
D_\a={\d\over \d\t^\a} +2i \s^\m_{\a\db}\tb^\db {\d\over \d y^\m} \quad , \quad
\Db_\da={\d\over \d\tb^\da}
\eeq
and write
\beq\label{3vectWZtwo}
V_{\rm WZ}=\t\s^\m\tb v_\m(y) + i\t\t\, \tb\lb(y)
-i \tb\tb\, \t\l(y) +\ha \t\t\tb\tb \left( D(y)-i\d_\m v^\m(y)\right) \ .
\eeq
Then, using $\s^\n\sb^\m-g^{\n\m}=2\s^{\n\m}$, it is straightforward to find
(all arguments are $y^\m$)
\beq\label{3DvectWZ}
\begin{array}{rcl}
D_\a V_{\rm WZ} &=& (\s^\m\tb)_\a v_\m +2 i \t_\a \, \tb\lb 
-i \tb\tb\, \l_\a +\t_\a \, \tb\tb D\crbig
&+& 2i (\s^{\m\n} \t)_\a \tb\tb \d_\m v_\n 
+ \t\t\tb\tb (\s^\m \d_\m \lb)_\a  \crbig
\end{array}
\eeq
and then, using $\Db\Db \tb\tb=-4$,
\beq\label{3Walphacomp}
W_\a=-i \l_\a(y) + \t_\a D(y) + i (\s^{\m\n}\t)_\a f_{\m\n}(y) 
+ \t\t (\s^\m\d_\m \lb(y))_\a
\eeq
with
\beq\label{3abelianfmn}
f_{\m\n}=\d_\m v_\n - \d_\n v_\m
\eeq
being the abelian field strength associated with $v_\m$.

Since $W_\a$ is a chiral superfield, $\Fint W^\a W_\a$ will be 
a susy invariant Lagrangian. To obtain its 
component expansion we need the $\t\t$-term ($F$-term) of $W^\a W_\a$:
\beq\label{3WWFterm}
W^\a W_\a \Big\vert_{\t\t}= -2i\l\s^\m\d_\m\lb + D^2 
- \ha (\s^{\m\n})^{\a\b} (\s^{\r\s})_{\a\b} f_{\m\n} f_{\r\s} \ ,
\eeq
where we used $(\s^{\m\n})_\a^{\ \b}=\tr\,  \s^{\m\n}=0$. Furthermore,
\beq\label{3sigsigtrace}
(\s^{\m\n})^{\a\b} (\s^{\r\s})_{\a\b}
=\ha\left( g^{\m\r}g^{\n\s}-g^{\m\s}g^{\n\r} \right) 
-{i\over 2} \e^{\m\n\r\s}
\eeq
(with $\e^{0123}=+1$) so that
\beq\label{3gaugekinetic}
\Fint W^\a W_\a = -\ha f_{\m\n} f^{\m\n} 
-2i \l\s^\m\d_\m\lb + D^2 
+{i\over 4} \e^{\m\n\r\s} f_{\m\n} f_{\r\s}
\ .
\eeq
Note that the first three terms are real while the last one is purely imaginary.

%% file: ch4.tex
%

We first discuss pure $N=1$ gauge theory which only involves the vector
multiplet 
and will be described in terms of the vector superfield of the previous 
section. We will need a slight generalization of the definition of  
$W_\a$ to the non-abelian case.  All members of the vector multiplet (the 
gauge boson $v_\m$ and the gaugino $\l$) necessarily are in the same 
representation of the gauge group, i.e. in the adjoint representation. 
Lateron we will couple chiral multiplets to this vector multiplet. The 
chiral fields can be in any representation of the gauge group, e.g. in 
the fundamental one.

\section{Pure $N=1$ gauge theory}

We start with the vector multiplet (\ref{3vectorsuper}) with every component 
now in the adjoint representation of the gauge group $G$, i.e. 
$V\equiv V_a T^a,\ a=1, \ldots {\rm dim}G$ where the $T_a$ are the 
generators in the adjoint. The basic object then is $e^V$ rather than $V$ 
itself. The non-abelian generalisation of the transformation (\ref{3vecttrans}) 
is now
\beq\label{4vecttrans}
e^V \to e^{i\L^\dag} e^V e^{-i\L} \ \Leftrightarrow
e^{-V} \to e^{i\L} e^{-V} e^{-i\L^\dag}
\eeq
with $\L$ a chiral superfield. To
first order in $\L$ this reproduces (\ref{3vecttrans}) with $\f=-i\L$. 
We will construct an action such that this non-linear transformation is a 
(local) symmetry. This transformation can again be used to set $\chi,C,M,N$ 
and one component of $v_\m$ to zero, resulting in the same component 
expansion (\ref{3vectWZ}) of $V$ in the Wess-Zumino gauge. From now on we 
adopt this WZ gauge. Then  $V^n=0, n\ge 3$. The same remains true if some 
$D_\a$ or $\Db_\da$ are inserted in the product, e.g. $V (D_\a V) V=0$. 
One simply has
\beq\label{4evexp}
e^V=1+V+\ha V^2 \ .
\eeq
The superfields $W_\a$ are now defined as
\beq\label{4Wnonab}
W_\a=-{1\over 4} \Db \Db\left( e^{-V} D_\a e^V\right) \quad , \quad
\ov W_\da=+{1\over 4} D D \left( e^V \Db_\da e^{-V}\right) \ ,
\eeq
which to first order in V reduces to the abelian definition (\ref{3walpha}).
Under the transformation (\ref{4vecttrans}) one then has
\beq\label{4wtrans}
\begin{array}{rcl}
W_\a &\to& -{1\over 4} \Db \Db\left(  e^{i\L} e^{-V} e^{-i\L^\dag}
 D_\a  \left( e^{i\L^\dag} e^V e^{-i\L}\right) \right)
 \crbig
&=&-{1\over 4} \Db \Db\left(  e^{i\L} e^{-V}  
 \left( (D_\a e^V) e^{-i\L} + e^V D_\a  e^{-i\L} \right) \right) \ .
\crbig
\end{array}
\eeq
The second term is $-{1\over 4} \Db \Db \left( e^{i\L} D_\a  e^{-i\L} \right)$
and vanishes for the same reason as  ${1\over 4} \Db \Db D_\a\f$ in 
(\ref{3Winvar}). Thus
\beq\label{4wtranstwo}
W_\a \to -{1\over 4}e^{i\L}  \Db \Db\left(e^{-V}D_\a e^V \right) e^{-i\L} 
= e^{i\L} W_\a e^{-i\L} 
\eeq
i.e. $W_\a$ transforms covariantly under (\ref{4vecttrans}). Similarly, one has \beq\label{4wtransthree}
\ov W_\da\to e^{i\L^\dag} W_\a e^{-i\L^\dag}  \ .
\eeq

Next, we want to obtain the component expansion of $W_\a$ in WZ gauge. 
Inserting the expansion (\ref{4evexp}) into the definition (\ref{4Wnonab}) 
gives
\beq\label{4Wexp}
W_\a=-{1\over 4} \Db\Db D_\a V +{1\over 8} \Db\Db [V, D_\a V] \ .
\eeq
The first term is the same as in the abelian case and has been computed in 
(\ref{3Walphacomp}), while for the new term we have (all arguments are $y^\m$)
\beq
[V, D_\a V]=\tb\tb (\s^{\n\m}\t)_\a [v_\m,v_\n]
+ i \t\t\tb\tb \s^\m_{\a\db} [v_\m, \lb^\db]
\eeq
and then, using again $\Db\Db\tb\tb=-4$
\beq\label{4Wexptwo}
{1\over 8} \Db\Db [V, D_\a V]
=\ha  (\s^{\m\n}\t)_\a [v_\m,v_\n] 
-{i\over 2} \t\t \s^\m_{\a\db} [v_\m, \lb^\db] \ .
\eeq
Adding this to (\ref{3Walphacomp}) simply turns ordinary derivatives 
of the fields into gauge covariant derivatives and we finally obtain
\beq\label{4Walphacomp}
W_\a=-i \l_\a(y) + \t_\a D(y) + i (\s^{\m\n}\t)_\a F_{\m\n}(y) 
+ \t\t (\s^\m D_\m \lb(y))_\a
\eeq
where now
\beq\label{3nonabelianfmn}
F_{\m\n}=\d_\m v_\n - \d_\n v_\m - {i\over 2} [v_\m, v_\n]
\eeq
and
\beq\label{4covderiv}
D_\m \lb = \d_\m \lb  - {i\over 2}  [v_\m, \lb] \ .
\eeq
The reader should not confuse the gauge covariant derivative $D_\m$ neither 
with the super covariant derivatives $D_\a$ and $\Db_\da$, nor with the auxiliary field $D$.

The gauge group generators $T^a$ satisfy
\beq\label{4structure}
[T^a,T^b]=i f^{abc} T^c
\eeq
with real structure constants $f^{abc}$. The field strength then is 
$F^a_{\m\n}=\d_\m v^a_\n -\d_n v^a_\m +\ha f^{abc} v^b_\m v^c_\n$.
We introduce the gauge coupling constant $g$ by scaling the superfield $V$ and hence all of its component fields as
\beq\label{4rescale}
V\to 2g\,  V \quad \Leftrightarrow
v_\m\to 2g\, v_\m\ , \ \l\to 2g\,  \l \ , \ D\to 2g\,  D
\eeq
so that then we have the rescaled definitions of gauge covariant derivative and field strength
\beq\label{4newfmn}
\begin{array}{rcl}
D_\m\l = \d_\m \l -ig[v_\m,\l] &\Rightarrow& 
(D_\m\l)^a=\d_\m\l^a + g f^{abc}v^b_\m\l^c\crbig
F_{\m\n}=\d_\m v_\n - \d_\n v_\m - i g [v_\m, v_\n]
&\Rightarrow& 
F^a_{\m\n}=\d_\m v^a_\n - \d_\n v^a_\m + g f^{abc}v^b_\m v^c_\n \ .
\crbig
\end{array}
\eeq
(We have implicitly assumed that the gauge group is simple so 
that there is a single coupling constant $g$. The generalisation to 
$G=G_1\times G_2\times \ldots$ and several $g_1,\ g_2, \ldots$ is 
straightforward.)
Then the component expansion (\ref{4Walphacomp}) of $W_\a$ remains 
unchanged, except for two things: there is on overall factor $2g$ 
multiplying the r.h.s. and $F_{\m\n}$ and $D_\m\l$ are now given by 
(\ref{4newfmn}). It follows that (\ref{3gaugekinetic}) also remains 
unchanged except for the replacements $f_{\m\n}\to F_{\m\n}$ and 
$\d_\m\lb\to D_\m\lb$ and an overall factor $4g^2$. One then introduces 
the complex coupling constant
\beq\label{4tau}
\tau={\Theta\over 2\pi} +{4\pi i\over g^2}
\eeq
where $\Theta$ stands for the $\Theta$-angle. (We use a capital $\Theta$ 
to avoid confusion with the superspace coordinates $\t$.) Then
\beq\label{4gaugelagr}
\begin{array}{rcl}
{\cal L}_{\rm gauge} &=&
{1\over 32\pi} {\rm Im}\, \left( \tau \Fint {\rm Tr}\, W^\a W_\a \right)
\crbig
&=&{\rm Tr} \left( -{1\over 4} F_{\m\n}F^{\m\n} 
- i\l\s^\m D_\m \lb + \ha D^2\right)
+{\Theta\over 32\pi^2} g^2\, {\rm Tr} F_{\m\n} \widetilde F^{\m\n} \crbig
\end{array}
\eeq
where
\beq\label{4fdual}
\widetilde F^{\m\n} = \ha \e^{\m\n\r\s} F_{\r\s}
\eeq
is the dual field strength. The single term $ {\rm Tr} W^\a W_\a $ has 
produced both, the conventionally normalised gauge kinetic term 
$-{1\over 4} {\rm Tr}F_{\m\n}F^{\m\n} $ and the instanton density 
${g^2\over 32\pi^2} {\rm Tr} F_{\m\n} \widetilde F^{\m\n}$ which multiplies 
the $\Theta$-angle!

\section{$N=1$ gauge theory with matter}

We now add chiral (matter) multiplets $\f^i$ transforming in some 
representation $R$ of the gauge group where the generators are 
represented by matrices $(T^a_R)^i_{\ j}$. Then
\beq\label{4mattertrans}
\f^i\to \left(e^{i\L}\right)^i_{\ j} \f^j \quad  , \quad
\f_i^\dag \to \f_j^\dag \left( e^{-i\L^\dag}\right)^j_{\ i}
\eeq
or simply $\f\to e^{i\L}\f,\ \f^\dag\to \f^\dag e^{-i\L^\dag}$ where 
$\L=\L^a T^a_R $ is understood. Then 
\beq\label{4ginvgen}
\f^\dag e^V\f\equiv \f^\dag e^{V^a T^a_R}\f \equiv 
\f_i^\dag \left( e^V\right)^i_{\ j} \f^j
\eeq
is the gauge invariant 
generalisation of the kinetic term and
\beq\label{4matterlagr}
{\cal L}_{\rm matter}=\Dint \f^\dag e^V\f +\Fint W(\f) + \Fbarint [W(\f)]^\dag
\ .
\eeq
Note that we have not yet scaled 
$V$ by $2g$, or equivalently we set $2g=1$ for the time being to 
simplify the formula. We want to compute the $\t\t\tb\tb$ component 
($D$-term) of $\f^\dag e^V\f=\f^\dag \f + \f^\dag V \f+ \ha \f^\dag V^2\f$.
The first term is given by (\ref{3kinDterm}). The second term is
\beq\label{4fvf}
\begin{array}{rcl}
\f^\dag V\f \Big\vert_{\t\t\tb\tb}
&=& {i\over 2} z^\dag v^\m \d_\m z -{i\over 2} \d_\m z^\dag v^\m z 
-\ha \pb \sb^\m v_\m \p \crbig
&+& {i\over \rd} z^\dag \l\p -{i\over \rd} \pb\lb z +\ha z^\dag D z \crbig
\end{array}
\eeq
and the third term is
\beq\label{4fvvf}
\f^\dag V^2\f \Big\vert_{\t\t\tb\tb} = {1\over 4} z^\dag v^\m v_\m z \ .
\eeq
Combining all three terms gives
\beq\label{4kinDterm}
\begin{array}{rcl}
\f^\dag e^V \f\Big\vert_{\t\t\tb\tb}&=& (D_\m z)^\dag D^\m z 
- i \pb\sb^\m D_\m \p + f^\dag f \crbig
&+&{i\over \rd} z^\dag \l\p -{i\over \rd} \pb\lb z +\ha z^\dag D z
+ {\rm total \ derivative}\ . \crbig
\end{array}
\eeq
with $D_\m z=\d_\m z-{i\over 2}v^a_\m T^a_R z$ and 
$D_\m\p=\d_\m\p-{i\over 2}v^a_\m T^a_R\p$.
We now rescale $V\to 2g V$ and use the first identity (\ref{1spinident}) 
to rewrite
$\pb\sb^\m D_\m \p= \p\s^\m D_\m \pb + {\rm total \ derivative}$. Then
this is replaced by
\beq\label{4kinDtermrescaled}
\begin{array}{rcl}
\f^\dag e^{2g V} \f\Big\vert_{\t\t\tb\tb}&=& (D_\m z)^\dag D^\m z 
- i \p\s^\m D_\m \pb + f^\dag f \crbig
&+&i \rd  g z^\dag \l\p - i\rd  g \pb\lb z + g z^\dag D z
+ {\rm total \ derivative}\ . \crbig
\end{array}
\eeq
now with $D_\m z=\d_\m z-i g v^a_\m T^a_R z$ and 
$D_\m\p=\d_\m\p- i g v^a_\m T^a_R\p$.
This part of the Lagrangian contains the kinetic terms for the scalar fields 
$z^i$ and the matter fermions $\p^i$, as well as specific interactions 
between the  $z^i$, the $\p^i$ and the gauginos 
$\l^a$. One has e.g. $z^\dag\l\p\equiv z_i^\dag (T^a_R)^i_{\ j} \l^a \p^j$.

What happens to the superpotential $W(\f)$? This must be a chiral superfield 
and hence must be constructed from the $\f^i$ alone. It must also be gauge 
invariant which imposes severe constraints on the superpotential. A term of 
the form $a_{i_1, \ldots i_n} \f^{i_1}\ldots \f^{i_n}$ will only be allowed 
if the $n$-fold product of the representation $R$ contains the trivial 
representation and then $a_{i_1, \ldots i_n}$ must be an invariant tensor 
of the gauge group. An example is $G={\rm SU}(3)$ with $R={\bf 3}$. Then 
${\bf 3}\times {\bf 3}\times {\bf 3}={\bf 1} + \ldots$ and the correponding 
${\rm SU}(3)$ invariant tensor is $\e_{ijk}$. In this example, however,
bilinears would not be gauge invariant. On the other hand, the representation 
$R$ need not be irreducible. Taking again the example of $G={\rm SU}(3)$ one 
may have $R={\bf 3}\oplus \ov{\bf 3}$ corresponding to a chiral superfield 
$\f^i$ transforming as ${\bf 3}$ (``quark") and a chiral superfield $\wt\f_i$
transforming as $\ov{\bf 3}$ (``antiquark"). Then one can form the gauge 
invariant chiral superfield $\wt\f_i \f^i$ which corresponds to a ``quark"
mass term.

There is a last type of term that may appear in case the gauge group 
simply is ${\rm U}(1)$ or contains ${\rm U}(1)$ factors.footnote{
If there is at least an extra ${\rm U}(1)$ factor the gauge group certainly is not simple and we have several coupling constants:
 These are the 
Fayet-Iliopoulos terms. Let $V^A$ denote the vector superfield in the 
abelian case, or the component corresponding to an abelian factor. Then 
under an abelian gauge transformation, $V^A\to V^A -i\L + i \L^\dag$, with 
$\L$ a chiral superfield. From the component expansion of such a chiral 
or anti chiral superfield (\ref{3chiralcompx}) or (\ref{3antichiralcompx}) 
one sees that the $D$-term (the term $\sim\t\t\tb\tb$) transforms as 
$D^A\to D^A + \d_\m\d^\m (\ldots)$, i.e. as a total derivative. Being a 
$D$-term, it also transforms as a total derivative under susy. It follows 
that
\beq\label{4FIterm}
{\cal L}_{\rm FI}=\sum_{A\in {\rm abelian\ factors}} \xi^A \Dint V^A = 
\ha \sum_{A\in {\rm abelian\ factors}} \xi^A D^A
\eeq
is a susy and gauge invariant Lagrangian (i.e. up to total derivatives).

We can finally write the full $N=1$ Lagrangian, being the sum of 
(\ref{4gaugelagr}), (\ref{4kinDtermrescaled}) and (\ref{4FIterm}):
\beq\label{4fullgaugelagr}
\begin{array}{rcl}
{\cal L}&=&  {\cal L}_{\rm gauge} + {\cal L}_{\rm matter}+{\cal L}_{\rm FI}
\crbig
&=&
{1\over 32\pi} {\rm Im}\, \left( \tau \Fint {\rm Tr}\, W^\a W_\a \right)
+2 g \sum_{A} \xi^A \Dint V^A  \crbig
&+&\Dint \f^\dag e^{2g V} \f
+\Fint W(\f) +\Fbarint [W(\f)]^\dag \crbig
&=&{\rm Tr} \left( -{1\over 4} F_{\m\n}F^{\m\n} 
- i\l\s^\m D_\m \lb + \ha D^2\right) 
+{\Theta\over 32\pi^2}g^2\, {\rm Tr} F_{\m\n} \widetilde F^{\m\n} 
+g \sum_{A} \xi^A D^A \crbig
&+& (D_\m z)^\dag D^\m z 
- i \p\s^\m D_\m \pb + f^\dag f 
+i \rd  g z^\dag \l\p - i\rd  g \pb\lb z + g z^\dag D z \crbig
&-& {\d W\over \d z^i} f^i + h.c. 
- \ha {\d^2 W\over \d z^i \d z^j} \p^i \p^j + h.c. 
+ {\rm total \ derivative}\ . \crbig
\end{array}
\eeq
The auxiliary field equations of motion are
\beq\label{4fequ}
f_i^\dag= {\d W\over \d z^i}
\eeq
and
\beq\label{4dequ}
D^a= - g z^\dag T^a z - g \xi^a
\eeq
where it is understood that $\xi^a=0$ if $a$ does not take values in an 
abelian factor of the gauge group. Substituting this back into the 
Lagrangian one finds
\beq\label{4fullgaugelagrtwo}
\begin{array}{rcl}
{\cal L}&=& 
{\rm Tr} \left( -{1\over 4} F_{\m\n}F^{\m\n} 
- i\l\s^\m D_\m \lb\right) 
+{\Theta\over 32\pi^2}g^2\, {\rm Tr} F_{\m\n} \widetilde F^{\m\n} \crbig
&+& (D_\m z)^\dag D^\m z 
- i \p\s^\m D_\m \pb  \crbig
&+&  i \rd  g z^\dag \l\p - i\rd  g \pb\lb z
- \ha {\d^2 W\over \d z^i \d z^j} \p^i \p^j 
- \ha \left({\d^2 W\over \d z^i \d z^j}\right)^\dag \pb^i \pb^j 
 \crbig
&-& V(z^\dag, z) + {\rm total \ derivative}\ , \crbig
\end{array}
\eeq
where the scalar potential $V(z^\dag,z)$ is given by
\beq\label{4scalarpot}
V(z^\dag, z)=f^\dag f + \ha D^2
=\sum_i \left| {\d W\over \d z^i}\right|^2
+{g^2\over 2} \sum_a \left|z^\dag T^a z + \xi^a \right|^2 \ .
\eeq

\section{Supersymmetric QCD}

At this point we have all the ingredients to write the action for 
supersymmetric QCD. The gauge group is ${\rm SU}(3)$. (More generally, 
one could consider ${\rm SU}(N)$.) There are then gauge bosons $v^a_\m$, 
$a=1, \ldots 8$ called the gluons, as well as their supersymmetric partners, 
the 8 gauginos or gluinos $\l^a$. If one considers pure $N=1$ ``glue", this 
is all there is. To describe $N=1$ QCD however, one also has to add quarks 
transforming in the ${\bf 3}$ of ${\rm SU}(3)$ as well as antiquarks in the 
$\ov{\bf 3}$. They are associated with chiral superfields. More precisely 
there are chiral superfields $Q^i_L=q^i_L +\rd \t\p^i_L -\t\t f^i_L$, 
$i=1,2,3$, and $L=1, \ldots N_f$ labels 
the flavours. These fields transform in the ${\bf 3}$ representation of the 
gauge group and correspond to left-handed quarks (or right-handed 
antiquarks). There are also chiral superfields 
$\wt Q_{i,L}=\wt q_{iL} +\rd \t\wt \p_{iL} -\t\t\wt f_{iL}$, $i=1,2,3$ and 
$L=1, \ldots N_f$. They transform in the $\ov{\bf 3}$ representation of the 
gauge group and correspond to left-handed antiquarks (or right-handed 
quarks). 

Note that the gauge group does not contain any ${\rm U}(1)$ factor, and 
hence no Fayet-Iliopoulos term can appear. The component
Lagrangian for massless susy QCD then is
given by (\ref{4fullgaugelagrtwo}) with $z=(q,\wt q)$ , $\xi^a=0$ 
and vanishing superpotential. 
Since all terms in the Lagrangian are diagonal in the flavour indices of 
the quarks and separately in the flavour indices of the antiquarks, there is an 
${\rm SU}(N_f)_L\times {\rm SU}(N_f)_R$ global symmetry. In addition there is 
a ${\rm U}(1)_V$ acting as $Q\to e^{iv} Q,\ \wt Q\to e^{-iv} \wt Q$, as well 
as an ${\rm U}(1)_A$ acting as $Q\to e^{ia} Q,\ \wt Q\to e^{ia} \wt Q$. We
also have the global ${\rm U}(1)_{\cal R}$ symmetry acting as
$Q(x,\t)\to e^{-iq} Q(x, e^{iq}\t)$,
$\wt Q(x,\t)\to e^{-iq} \wt Q(x, e^{iq}\t)$ and
$V(x,\t,\tb)\to V(x,e^{iq}\t, e^{-iq}\tb)$. Thus the global symmetry 
group in the massless case is
${\rm U}(N_f)_L\times {\rm U}(N_f)_R \times {\rm U}(1)_{\cal R}$

Due to the presence of both representations ${\bf 3}$ and $\ov{\bf 3}$ 
of the gauge group, one may add a gauge invariant superpotential
\beq\label{4qcdmass}
W(Q,\wt Q) =  m_{L,M} Q^i_L \wt Q_{i,M}\ .
\eeq
This is a quark mass term and $m_{L,M}$ is the $N_f\times N_f$ mass matrix.
Using the global symmetry of the other terms in the Lagrangian (which is just the massless Lagrangian) one can diagonalise the superpotential so that it reads
\beq\label{4qcdmassdiag}
W(Q,\wt Q) =  \sum_{L} m_L Q^i_L \wt Q_{i,L}\ .
\eeq
For the gauge group ${\rm SU}(3)$ one could also add the gauge invariant 
terms\break\hfill 
$a_{LMN} \e_{ijk} Q^i_L Q^j_M Q^k_N$ and 
$\wt a_{LMN} \e^{ijk} \wt Q_{iL} \wt Q_{jM} \wt Q_{kN}$. 
However, they explicitly
violate baryon number conservation and will not be considered.
Then finally one arrives at the following Lagrangian (where we suppress 
as much as possible all gauge and flavour indices):
\beq\label{4qcdlagr}
\begin{array}{rcl}
{\cal L}&=& 
{\rm Tr} \left( -{1\over 4} F_{\m\n}F^{\m\n} 
- i\l\s^\m D_\m \lb\right) 
+{\Theta\over 32\pi^2}g^2\, {\rm Tr} F_{\m\n} \widetilde F^{\m\n} \crbig
&+& (D_\m q)^\dag D^\m q + (D_\m \wt q)^\dag D^\m \wt q 
- i \p\s^\m D_\m \pb  - i {\wt\p} \s^\m D_\m \ov{\wt\p} \crbig
&+&  i \rd  g q^\dag \l\p +  i \rd  g \wt q^\dag \l\wt\p 
- i\rd  g \pb\lb q - i\rd  g \ov{\wt\p}\lb \wt q \crbig
&-& \ha \sum_L m_L \left(\p_L \p_L  + \ov{\wt\p}_L \ov{\wt\p}_L \right)
-V(q,\wt q, q^\dag, \wt q^\dag) + {\rm total \ derivative}\ , \crbig
\end{array}
\eeq
where the scalar potential $V$ is given by
\beq\label{4qcdscalarpot}
V(q,\wt q, q^\dag, \wt q^\dag)
=\sum_{L=1}^{N_f} m_L^2 \left( q_L^\dag q_L+\wt q_L^\dag \wt q_L\right) 
+{g^2\over 2} \sum_{a=1}^8 
\left\vert q^\dag T^a q + \wt q^\dag T^a \wt q \right\vert^2 \ .
\eeq

%% file: ch7.tex
%

\section{Vacua in susy theories}

Perturbation theory should be performed around a stable configuration. 
If quantum field theory is formulated using a euclidean functional 
integral, stable configurations correspond to minima of the euclidean 
action. A vacuum is a Lorentz invariant stable configuration. Lorentz 
invariance implies that all space-time derivatives and all fields that 
are nor scalars must vanish. Hence only scalar fields $z^i$ can have a 
non-vanishing value in a vacuum configuration, i.e. a non-vanishing vacuum 
expectation value (vev), denoted by $\la z^i \ra$. Minimality of the euclidean 
action (or else minimality of the energy functional) then is equivalent 
to the scalar potential $V$ having a minimum. Thus we have for a vacuum
\beq\label{7vacuumcond}
\la v_\m^a\ra = \la \l^a \ra = \la \p^i \ra = \d_\m \la z^i \ra = 0
\quad , \quad V(\la z^i \ra ,  \la z^\dag_i \ra) = {\rm minimum} \ .
\eeq
The minimum may be  the global minimum of $V$ in which case one has the 
true vacuum, or it may be a local minimum in which case one has a false 
vacuum that will eventually decay by quantum tunneling into the true 
vacuum (although the life-time may be extremely long). For a false or 
true vacuum one certainly has
\beq\label{7minvac}
{\d V\over \d z^i }(\la z^j \ra ,  \la z^\dag_j \ra)
= {\d V\over \d z^\dag_i }(\la z^j \ra ,  \la z^\dag_j \ra) = 0 \ .
\eeq
This shows again that a vacuum is indeed a solution of the equations 
of motion.

Now in a supersymmetric theory the scalar potential is given by
(\ref{4scalarpot}), namely 
\beq\label{7scalarpot}
V(z,z^\dag)=f^\dag_i f^i +\ha D^a D^a
\eeq
where
\beq\label{7fi}
f_i^\dag= {\d W(z)\over \d z^i}
\eeq
and
\beq\label{7da}
D^a=-g^a \left( z^\dag_i (T^a)^i_{\ j} z^j + \xi^a \right)
\eeq
where we allowd for Fayet-Iliopoulos terms $\sim \xi^a$ associated with 
possible ${\rm U}(1)$ factors and couplings $g^a$. Of course within each 
simple factor of the gauge group $G$ the $g^a$ are the same.\footnote{
If $G= G_1 \times \ldots \times G_k\times {\rm U}(1) \times \ldots \times
{\rm U}(1)$ with simple factors $G_l$ of dimension $d_l$ it is understood 
that $g^1 = \ldots =g^{d_1}$, $g^{d_1+1} = \ldots =g^{d_1+d_2}$, etc and 
$\xi^1=\ldots \xi^{d_1 + \ldots d_k}=0$.}
The potential (\ref{7scalarpot}) is non-negative and it will certainly be 
at its global minimum, namely $V=0$, if
\beq\label{7fdzero}
f^i(\la z^\dag\ra ) = D^a (\la z \ra ,  \la z^\dag \ra) = 0 \ .
\eeq
However, this system of equations does not necessarily have a solution as 
a simple counting argument shows: there are as many equations $f^i=0$ as 
unknown $\la z^\dag_i \ra$ (and as many complex conjugate equations 
$f^\dag_i=0$ as complex conjugate $\la z^i \ra $). On top of these there are 
${\rm dim} G$ equations $D^a=0$ to be satisfied. We now have two cases.

a) If the equations (\ref{7fdzero}) have a solution, then this solution is a 
global minimum of $V$ (since $V=0$) and hence a stable true vacuum. There 
can be many such solutions and then we have many degenerate vacuua. In 
addition to this true vacuum there can be false vacua satisfying 
(\ref{7minvac}) but not (\ref{7fdzero}).

b) If the equations (\ref{7fdzero}) have no solutions, the scalar
potential $V$ can never 
vanish and its minimum is strictly positive: $V\ge V_0 > 0$.  Now a vacuum 
with strictly positive energy necessarily breaks supersymmetry. This means 
that the vacuum cannot be invariant under all susy generators. The proof is 
very simple:   as in 
(\ref{2posenergy}) we have for any state $\ket \Omega$ 
\beq\label{7vacenergy}
\bra\Omega P_0 \ket \Omega = {1\over 4} || Q_\a \ket \Omega ||^2
+ {1\over 4} || Q^\dag_\a \ket \Omega ||^2 = 0 \ .
\eeq
Now assume that $\ket \Omega$ is invariant under all 
susy generators, i.e. $Q_\a \ket \Omega=0$. Then necessarily 
$\bra\Omega P_0 \ket \Omega  = 0$, and conversely if 
$\bra\Omega P_0 \ket \Omega > 0$ not all $Q_\a$ and $Q^\dag_\a$ 
can annihilate the state $\ket \Omega$. It is not surprising that 
an excited state, e.g. a one-particle state is not invariant under 
susy: indeed this is how susy transforms the different particles of 
a supermultiplet into each other. Non-invariance of the vacuum state  
has a different meaning: it implies that susy is really broken in the 
perturbation theory based on this vacuum. As usual, this is called 
spontaneous breaking of the (super)symmetry. 

There is also another way to see that susy is broken if either 
$f^i(\la z^\dag\ra )\ne 0$ or  $D^a (\la z \ra ,  \la z^\dag \ra) \ne 0$.
Looking at the susy transformations of the fields one has from 
(\ref{3susycompfieldstwo})
\beq\label{7susytransvac}
\begin{array}{rcl}
\delta \la z^i \ra &=& \rd \e \la \p^i\ra \crbig
\delta \la \p^i \ra &=& \rd  i \d_\m \la z^i\ra \s^\m \ov \e 
- \rd \la f^i\ra \e \crbig
\delta \la f^i \ra &=& \rd i \d_\m \la \p^i\ra \s^\m \ov \e \crbig
\end{array}
\eeq
which upon taking into account (\ref{7minvac}) reduces to
\beq\label{7susytransvactwo}
\begin{array}{rcl}
\delta \la z^i \ra &=& 0 \crbig
0=\delta \la \p^i \ra &=& - \rd \la f^i\ra \e \crbig
\delta \la f^i \ra &=& 0 \crbig
\end{array}
\eeq
which can be consistent only if $\la f^i \ra  \equiv f^i(\la z^\dag\ra) =0$. 
The argument similarly shows that $\delta \la \l^a\ra =0$ is only possible 
if $\la D^a\ra \equiv  D^a (\la z \ra ,  \la z^\dag \ra) = 0$. More generally, 
a necessary condition for unbroken susy is that the susy variations of the 
fermions vanish in the vacuum.

\section{The Goldstone theorem for susy}

Goldstone's theorem states that whenever a continuous global symmetry is 
spontaneously broken, i.e. the vacuum is not invariant, there is a massless 
mode in the spectrum, i.e. a massless particle. The quantum numbers carried 
by the Goldstone particle are related to the broken symmetry. Similarly, we 
will show that if supersymmetry is spontaneously broken there is a massless 
spin one-half particle, i.e. a massless spinorial mode, sometimes called the 
Goldstino.

As we have seen, a vacuum that breaks susy is such that 
${\d V\over \d z^i }(\la z^i \ra ,  \la z^\dag_i \ra) = 0$ (it is a vacuum)
and\footnote{
As before, $\la f^i\ra$ is shorthand for $f^i(\la z^\dag\ra)$ and $\la D^a\ra$
shorthand for $D^a(\la z\ra, \la z^\dag \ra)$.}
$\la f^i\ra \ne 0$ or $\la D^a \ra \ne 0$. Now from (\ref{7scalarpot})-(\ref{7da})
we have
\beq\label{7vderiv}
{\d V\over \d z^i} = f^j {\d^2 W\over \d z^i\d z^j} 
- g^a D^a z^\dag_j (T^a)^j_{\ i}
\eeq
and this must vanish for any vacuum. We will combine this with the statement 
of gauge invariance of the superpotential $W$ which reads
\beq\label{7Winvariance}
0=\delta^{(a)}_{\rm gauge} W = {\d W\over \d z^i} \delta^{(a)}_{\rm gauge} z^i
= f^\dag_i (T^a)^i_{\ j} z^j  \ .
\eeq
We can now combine the vanishing of (\ref{7vderiv}) in the vacuum with the vev 
of the complex conjugate equation (\ref{7Winvariance}) into the matrix equation
\beq\label{7eigenvalue}
M=
\pmatrix{
\la {\d^2 W\over \d z^i\d z^j}\ra &
-g^a \la z^\dag_l\ra (T^a)^l_{\ i} \cr
-g^b \la z^\dag_l\ra (T^b)^l_{\ j} &
0 \cr }
\quad , \quad
M \pmatrix{ \la f^j\ra \cr \la D^a \ra \cr }
=0
\eeq
stating that the matrix appearing here has a zero eigenvalue. But this matrix 
exactly is the fermion mass matrix. Indeed, the non-derivative fermion 
bilinears in the Lagrangian (\ref{4fullgaugelagr}) give rise in the vacuum to 
the following mass terms
\beq\label{7massterms}
\begin{array}{rcl}
& &\left( i\rd g^a \la z^\dag_j\ra (T^a)^j_{\ i} \l^a \p^i 
- \ha \la {\d^2 W\over \d z^i\d z^j}\ra \p^i \p^j \right) + h.c.
\crbig
&=& - \ha \left( \p^i, \rd i \l^b\right)
M \pmatrix{ \p^j \cr \rd i  \l^a \cr } + h.c.
\crbig
\end{array}
\eeq
with the same matrix $M$ as defined in (\ref{7eigenvalue}). This matrix has a 
zero eigenvalue, and this means that there is a zero mass fermion: the 
Goldstone fermion or Goldstino.

\section{Mechanisms for susy breaking}

We have seen that a minimum of $V$ with $\la f^i\ra\ne 0$ or $\la D^a\ra\ne 0$ 
is a vacuum that breaks susy. This can be a true or false vacuum. If there is 
{\it no} vacuum with $\la f^i\ra=\la D^a\ra=0$, i.e. no solution $\la z^i\ra$ 
to these equations, supersymmetry is necessarily broken by any vacuum. Whether 
or not there are solutions depends on the choice of superpotential $W$ and 
whether the Fayet-Iliopoulos parameters $\xi^a$ vanish or not.

\subsection{O'Raifeartaigh mechanism}

Assume first that no ${\rm U}(1)$ factors are present or else that the $\xi^a$
vanish. Susy will be broken if ${\d W\over \d z^i}=0$ and 
$z^\dag_j (T^a)^j_{\ l}\, z^l=0$ have no solution. If the superpotential 
$W$ has no linear term, $\la z^i\ra=0$ will always be a solution. So let's 
assume that there is a linear term $W=a_i z^i+\ldots$. But this can be gauge 
invariant only if the representation $R$ carried by the $z^i$ contains at 
least one singlet, say $z^1=Y$. As a simple example take
\beq\label{7oraifpot}
W=Y (a-X^2) + b Z X + w(X, z^i)
\eeq
with $X, Y, Z$ all singlets. Then $f^\dag_Y={\d W\over \d Y}= a-X^2$ 
and $f^\dag_Z={\d W\over \d Z} = b X$ cannot both vanish so that there 
is no susy preserving vacuum solution.

\subsection{Fayet-Iliopoulos mechanism}

Let there now be at least one  ${\rm U}(1)$ and non-vanishing $\xi$. The relevant part of $D=0$ is
\beq\label{7FI}
0=\sum_i q_i |z^i|^2 + \xi
\eeq
where the $q_i$ are the ${\rm U}(1)$ charges of $z^i$. If all charges $q_i$ 
had the same sign, taking a $\xi$ of the same sign as the $q_i$ would forbid 
the existence of solutions and break susy. However, absence of chiral anomalies 
for the ${\rm U}(1)$ imposes $\sum_i q_i^3=0$ so that charges of both signs 
must be present and there is always a solution to (\ref{7FI}). One needs further 
constraints from $f^i=0$ to break susy. To see how this works consider again a 
simple model. Take two chiral multiplets $\f^1$ and $\f^2$ with charges 
$q_1=-q_2=1$ so that (\ref{7FI}) reads $|z^1|^2-|z^2|^2+\xi=0$ and take a 
superpotential $W=m\f^1\f^2$. Then $f^1=m z^\dag_2$ and $f^2=m z^\dag_1$ and 
clearly, if $m\ne 0$ and $\xi\ne 0$, we cannot simultaneously have $f^1=f^2=D=0$ 
so that susy will be broken.

\section{Mass formula}

If supersymmetry is unbroken all particles within a supermultiplet have 
the same mass. Although this will no longer be true if supersymmetry is 
(spontaneously)
broken, but one can still relate the differences of the squared masses to the 
susy breaking parameters $\la f^i\ra$ and $\la D^a\ra$.

Let us derive the masses of the different particles: vectors, fermions and 
scalars. We begin with the vector fields. In the presence of non-vanishing 
vevs of the scalars, some or all of the vector gauge fields will become 
massive by the Higgs mechanism. Indeed the term $(D_\m z^i)^\dag (D^\m z^i)$ 
present in the Lagrangian (\ref{4fullgaugelagr}) gives rise to a mass term 
$g^2 \la z^\dag T^a T^b z\ra v^a_\m v^{a\m}$, while the gauge kinetic term 
is normalised in the standard way. Thus the mass matrix  for the spin-one 
fields is 
\beq\label{7vect}
\left( {\cal M}_1^2\right)^{ab}= 2 g^2  \la z^\dag T^a T^b z\ra  \ .
\eeq
It will be useful to introduce the notations
\beq\label{7di}
D_i^a={\d D^a\over \d z^i} = -g (z^\dag T^a)_i \quad , \quad
D^{ia}={\d D^a\over \d z^\dag_i} = -g (T^a z)^i
\eeq
as well as $D^{ai}_j=-g T^{ai}_j$, and similarly
\beq\label{7fip}
f^{ij}={\d f^i\over \d z^\dag_j}={\d^2 \ov W\over \d z^\dag_j \d z^\dag_i}
\quad , \quad
f_{ij}={\d f^\dag_i\over \d z^j}={\d^2  W\over \d z^j \d z^i}
\eeq
etc. Then eq. (\ref{7vect}) can be written as
\beq\label{7vectd}
\left( {\cal M}_1^2\right)^{ab}= 2 \la D^a_i D^{bi}\ra 
= 2 \la D^a_i\ra \la D^{bi}\ra \ .
\eeq

Next, for the spin-one-half fermions the mass matrix can be read from 
(\ref{7eigenvalue}) and (\ref{7massterms}) or again directly from 
(\ref{4fullgaugelagr}). The mass terms are
\beq\label{7ferm}
-\ha (\p^i\ \  \l^a) {\cal M}_\ha \pmatrix{\p^j\cr \l^b\cr} + {\rm h.c.} 
\quad , \quad
{\cal M}_\ha = \pmatrix{ \la f_{ij}\ra & \rd i \la D^b_i\ra \cr
\rd i \la D^a_j\ra & 0 \cr}
\eeq
with the squared masses of the fermions being given by the eigenvalues of 
the hermitian matrix
\beq\label{7fermsq}
\left( {\cal M}_\ha {\cal M}_\ha^\dag\right) =
\pmatrix{ \la f_{il}\ra \la f^{jl}\ra + 2 \la D^c_i\ra\la D^{cj}\ra &
-\rd i  \la f_{il}\ra \la D^{bl}\ra \cr
\rd i  \la D^a_l\ra \la f^{jl}\ra &
2\la D^a_l\ra \la D^{bl}\ra \cr } \ .
\eeq

Finally for the scalars the mass terms are\footnote{
The way the $z$ and $z^\dag$ are grouped as well as the $\ha$ may seem 
peculiar at the first sight, but they are easily explained by the example 
of a single complex scalar field for which the mass term is $m^2 z z^\dag$. 
Then simply ${\cal M}_0^2 = \pmatrix{m^2&0\cr 0&m^2\cr}$ and (\ref{7scal}) 
yields $-\ha z m^2 z^\dag -\ha z^\dag m^2 z=-m^2 z^\dag z$.
}
\beq\label{7scal}
-\ha (z^i\ \ z^\dag_j) {\cal M}_0^2 \pmatrix{z^\dag_k\cr z^l\cr}
\eeq
with
\beq\label{7scalsq}
{\cal M}_0^2=\pmatrix{
\la {\d^2 V\over \d z^i\d z^\dag_k}\ra &
\la {\d^2 V\over \d z^i\d z^l}\ra \cr
\la {\d^2 V\over \d z^\dag_j\d z^\dag_k}\ra &
\la {\d^2 V\over \d z^\dag_j\d z^l}\ra \cr } \ .
\eeq
Using (\ref{7scalarpot}) one finds that this matrix equals
\beq\label{7scalexpl}
\pmatrix{ 
\la f_{ip}\ra \la f^{kp}\ra + \la D^{ak}\ra\la D^a_i\ra + \la D^a\ra D^{ak}_i &
\la f^p\ra \la f_{ilp}\ra + \la D^a_i\ra \la D^a_l\ra \cr
\la f^\dag_p\ra \la f^{jkp}\ra + \la D^{aj}\ra \la D^{ak}\ra &
\la f_{lp}\ra \la f^{jp}\ra + \la D^{aj}\ra\la D^a_l\ra + \la D^a\ra D^{aj}_l
\cr }
\eeq

It is now straightforward to give the traces which yield the sums of the masses 
squared of the vectors, fermions and scalars, respectively.
\beq\label{7traces}
\begin{array}{rcl}
\tr {\cal M}_1^2 
&=& 2 \la D^a_i\ra \la D^{ai}\ra \crbig
\tr  {\cal M}_\ha {\cal M}_\ha^\dag 
&=& \la f_{il}\ra\la f^{il}\ra + 4 \la D^a_i\ra \la D^{ai}\ra \crbig
\tr {\cal M}_0^2 
&=& 2 \la f_{ip}\ra\la f^{ip}\ra + 2 \la D^a_i\ra \la D^{ai}\ra 
-2 g \la D^a\ra \tr T^a \crbig
\end{array}
\eeq
and
\beq\label{7suptr}
{\rm Str} {\cal M}^2 \equiv 3 \tr {\cal M}_1^2  
- 2 \tr  {\cal M}_\ha {\cal M}_\ha^\dag + \tr {\cal M}_0^2 
= -2 g \la D^a\ra \tr T^a \ .
\eeq
In this supertrace we have counted two degrees of freedom for spinors and
three for vectors as appropriate in the massive case (the massless states 
do not contribute anyhow). We see that if $\la D^a\ra=0$ or $\tr T^a=0$ (no 
${\rm U}(1)$ factor) this supertrace vanishes, stating that the sum of the 
squared masses of all bosonic degrees of freedom equals the sum for all 
fermionic ones. Without susy breaking this is a triviality. In the presence 
of susy breaking this supertrace formula is still a strong constraint on the 
mass spectrum. In particular, if susy is broken only by non-vanishing 
$\la f^i\ra$ (and $\la f^\dag_i\ra$), or if all gauge group generators are 
traceless, one must still have ${\rm Str} {\cal M}^2=0$.

Consider e.g. susy QCD. The gauge group is ${\rm SU}(3)$ and $\tr T^a=0$, 
while the gauge group must remain unbroken. Then ${\cal M}_1^2=0$ so that  
$\la D^a_i\ra = \la D^{ai}\ra =0$. Note from (\ref{7ferm}) that it is then 
obvious that also the gauginos (gluinos) remain massless, while supertrace 
formula tells us that the sum of 
the masses squared of the scalar quarks must equal (twice) the sum for the 
quarks. This means that the scalar quarks cannot all be heavier than the 
heaviest quark, and some must be substantially lighter. Since no massless 
gluinos and relatively light scalar quarks have been found experimentally, 
this scenario seems to be ruled out by experiment. However, it would be too 
quick to conclude that one cannot embed QCD into a susy theory. Indeed, 
there are two ways out. First, the mass formula derived here only give the 
tree-level masses and are corrected by loop effects. Typically, one introduces 
one or several additional chiral multiplets which trigger the susy breaking. 
Through loop diagrams this susy breaking then propagates to the gauge theory
we are interested in and, in principle, one can achieve heavy gauginos and 
heavy scalar quarks this way. leaving massless gauge fields and light 
fermions. Second, the susy theory may be part of a supergravity theory 
which is spontaneously broken, and in this case one rather naturally 
obtains experimentally reasonable mass relations.

Let's discuss a bit more the mass matrices derived above. As in the example 
of susy QCD just discussed, if the gauge symmetry is unbroken, ${\cal M}_1^2=0$
implying $\la D^a_i\ra=0$, so that the fermion mass matrix reduces to
$\left( {\cal M}_\ha {\cal M}_\ha^\dag\right) =
\pmatrix{ \la f_{il}\ra \la f^{jl}\ra &0\cr 0&0\cr}$,
showing again that the 
gauginos are massless, too. If we now suppose that there are no 
Fayet-Iliopoulos parameters, $\la D^a_i\ra=0$ implies that also 
$\la D^a\ra=0$ as is easily seen\footnote{
One has $D^a_j D^{bj}= z^\dag T^a T^b z$ and  
$D^{[a}_j D^{b]j}={i\over 2} f^{abc} z^\dag T^c z 
= -{i\over 2g} f^{abc} D^c$ (if there are no FI parameters). Thus if 
$\la D^a_i\ra =0$, also $\la D^{bj}\ra =0$ and this then implies 
$\la D^c\ra =0$.
}
so that the scalar mass matrix now is
\beq\label{7mo}
{\cal M}_0^2 
=\pmatrix{ 
\la f_{ip}\ra \la f^{kp}\ra  &
\la f^p\ra \la f_{ilp}\ra  \cr
\la f^\dag_p\ra \la f^{jkp}\ra  &
\la f_{lp}\ra \la f^{jp}\ra \cr } \ .
\eeq
The block diagonal terms are the same as for the  fermions $\p^i$, but the
block off-diagonal terms give an additional  contribution
\beq\label{7masssplit}
-\ha \la f^p\ra \la f_{ilp}\ra z^i z^l \ + \ {\rm h.c.} \ .
\eeq
The effect of this term typically is to lift the mass degeneracy between the 
real and the imaginary parts of the scalar fields, splitting the masses in a 
symmetric way with respect to the corresponding fermion masses. This is of 
course in agreement with ${\rm Str} {\cal M}^2 =0$ in this case.

%% file: ch5.tex
%

As long as one wants to formulate a fundamental, i.e. microscopic theory, 
one is guided by the principle of renomalisability. For the theory of 
chiral superfields $\f$ only this implies at most cubic superpotentials 
(leading to at most quartic scalar potentials) and  kinetic terms 
$K^i_{\ j} \f_i^\dag \f^j$ with some constant hermitian matrix $K$. 
After diagonalisation and rescaling of the fields this then reduces 
to the canonical kinetic term $\f_i^\dag \f^i$. Thus we are back to the 
Wess-Zumino model studied above.

In many cases, however, the theory on considers is an {\it effective} 
theory, valid at low energies only. Then renormalisability no longer is 
a criterion. The only restriction for such a low-energy effective theory  
is to contain no more than two (space-time) derivatives. Higher derivative 
terms are irrelevant at low energies. Thus we are led to study the 
supersymmetric non-linear sigma model. Another motivation comes from 
supergravity which is not renormalisable anyway. We will first consider 
the model for chiral multiplets only, and then extend the resulting theory 
to a gauge invariant one.

\section{Chiral multiplets only}

We start with the action
\beq\label{5sigmamod}
S=\xint \left( \Dint K(\f^i,\f_i^\dag) 
+\Fint w(\f^i) + \Fbarint w^\dag(\f_i^\dag)\right) \ .
\eeq
We have denoted the superpotential by $w$ rather than $W$. The function 
$K(\f^i,\f_i^\dag)$ must be real superfield, which will be the case if
$\ov K(z^i,z_j^\dag)= K(z_i^\dag,z^j)$. Derivatives with respect to its 
arguments will be denoted as
\beq\label{5Kderiv}
K_i={\d\over \d z^i} K(z,z^\dag) \quad , \quad
K^j={\d\over \d z_j^\dag} K(z,z^\dag) \quad , \quad
K_i^j={\d^2\over \d z^i \d z_j^\dag} K(z,z^\dag)
\eeq
etc. (Note that one does not need to distinguish indices like $K^i_{\ j}$ or 
$K^{\ i}_j$ since the partial derivatives commute.)
Similarly we have
\beq\label{5wderiv}
w_i={\d\over \d z^i} w(z) \quad , \quad
w_{ij}={\d^2\over \d z^i \d z^j} w(z)
\eeq
etc. We also use $w^i=[w_i]^\dag$, $w^{ij}=[w_{ij}]^\dag$.

The expansion of the $F$-terms in components was already given in 
(\ref{3superpot}). We may rewrite this as
\beq\label{5wexp}
w(\f) = w(z) + w_i \D^i + \ha w_{ij} \D^i\D^j
\eeq
with arguments $y^\m$ understood and
\beq\label{5deltai}
\D^i(y)=\f^i-z^i(y)=\rd \t\p^i(y) -\t\t f^i(y) \ .
\eeq    
Then extracting the $\t\t$-components of $\D^i$ and $\D^i\D^j$ yields
(\ref{3superpot}) again, i.e.
\beq\label{5Dterms}
\Fint w(\f^i) + h.c. = \left( -w_i f^i -\ha w_{ij} \p^i \p^j \right) 
+ h.c.
\ . \eeq

The component expansion of the $D$-term is more involved, since now 
$\D^i(x)=\f^i-z^i(x)$ and $\DD_i(x)=\f^\dag_i-z^\dag_i(x)$ appear. 
We have from (\ref{3chiralcompx})
\beq\label{5deltaix}
\begin{array}{rcl}
\D^j &=& \rd\t\p^j + i\t\s^\m\tb \d_\m z^j - \t\t f^j
-{i\over\rd} \t\t\d_\m\p^j \s^\m\tb - {1\over 4}\t\t\tb\tb \d^2 z^j \crbig 
\DD_j&=&\rd\tb\pb_j - i\t\s^\m\tb \d_\m z^\dag_j 
- \tb\tb  f^\dag_j 
+{i\over\rd} \tb\tb \t\s^\m\d_\m\pb_j - {1\over 4}\t\t\tb\tb \d^2 z^\dag_j
\crbig
\end{array}
\eeq
with all fields having $x^\m$ as argument. Note that 
$\D^i\D^j\D^k=\DD_i \DD_j \DD_k=0$ so that at most two $\D$ and two $\DD$ 
can appear in the expansion. One has the Taylor expansion of 
$K(\f^i,\f_i^\dag)$
\beq\label{5KTaylor}
\begin{array}{rcl}
K(\f^i,\f_i^\dag)&=& K(z^i,z_i^\dag) + K_i\D^i+K^i\DD_i
+\ha K_{ij}\D^i\D^j+\ha K^{ij}\DD_i\DD_j+K_i^j\D^i\DD_j\crbig
&+&\ha K_{ij}^k\D^i\D^j\DD_k +\ha K^{ij}_k \DD_i\DD_j\D^k
+{1\over 4}K_{ij}^{kl} \D^i\D^j\DD_k\DD_l \ , \crbig
\end{array}
\eeq
where
\beq\label{5deltaixpowers}
\begin{array}{rcl}
\D^i \D^j &=&-\t\t\p^i\p^j 
-{i\over \rd} \left( \p^i\s^\m\tb \d_\m z^j + \p^j\s^\m\tb \d_\m z^i\right)
- \ha\t\t\tb\tb \d_\m z^i \d^\m z^j \crbig 
\D^i \DD_j&=&\t\s^\m\tb \ \p^i\s_\m\pb_j \crbig
& &\ 
-\rd\t\t \left( \tb\pb_j f^i -{i\over 2} \p^i\s^\m\tb \d_\m z_j^\dag \right)
-\rd \tb\tb\left( \t\p^i f_j^\dag + {i\over 2} \t\s^\m\pb_j \d_\m z^i \right)
\crbig
& &\ 
+\t\t\tb\tb \left( f^i f_j^\dag +\ha \d_\m z^i \d^\m z_j^\dag 
-{i\over 2} \p^i\s^\m\d_\m \pb_j +{i\over 2}\d_\m\p^i\s^\m\pb_j \right)
\crbig
\D^i \D^j \DD_k&=&-\rd\t\t\, \tb\pb_k\, \p^i\p^j 
\crbig
& &\ +{i\over 2} \t\t\tb\tb \left( \p^i\s^\m\pb_k \d_\m z^j 
+ \p^j\s^\m \pb_k \d_\m z^i - 2i \p^i\p^j f_k^\dag \right)
\crbig
\D^i \D^j \DD_k  \DD_l &=&\t\t\tb\tb \p^i\p^j \, \pb_k\pb_l \ .
\crbig
\end{array}
\eeq
It is then easy to extract the $D$-term, i.e. the coefficient of $\t\t\tb\tb$
\beq\label{5KDterm}
\begin{array}{rcl}
\Dint K(\f^i,\f_i^\dag)&=& 
-{1\over 4} K_i \d^2 z^i -{1\over 4} K^i \d^2 z_i^\dag
-{1\over 4} K_{ij}\d_\m z^i \d^\m z^j + h.c. \crbig
&+&K_i^j \left( f^i f_j^\dag +\ha \d_\m z^i \d^\m z_j^\dag 
-{i\over 2} \p^i\s^\m\d_\m \pb_j +{i\over 2}\d_\m\p^i\s^\m\pb_j \right)
\crbig
&+&{i\over 4} K_{ij}^k \left( \p^i\s^\m\pb_k \d_\m z^j 
+ \p^j\s^\m \pb_k \d_\m z^i - 2i \p^i\p^j f_k^\dag \right) + h.c. 
\crbig
&+&{1\over 4} K_{ij}^{kl}  \p^i\p^j \, \pb_k\pb_l \ . 
\crbig
\end{array}
\eeq
Next note that 
\beq\label{5ddK}
\begin{array}{rcl}
\d_\m\d^\m K(z^i,z_j^\dag)&=&
K_i\d^2z^i + K^i \d^2 z_i^\dag +2 K_i^j\d_\m z_j^\dag \d^\m z^i \crbig
&+& K_{ij}\d_\m z^i \d^\m z^j + K^{ij} \d_\m z_i^\dag \d^\m z_j^\dag \crbig
\end{array}
\eeq
so that we can rewrite (\ref{5KDterm}) as
\beq\label{5KDtermtwo}
\begin{array}{rcl}
\Dint K(\f^i,\f_i^\dag)&=& 
K_i^j \left( f^i f_j^\dag + \d_\m z^i \d^\m z_j^\dag 
-{i\over 2} \p^i\s^\m\d_\m \pb_j +{i\over 2}\d_\m\p^i\s^\m\pb_j \right)
\crbig
&+&{i\over 4} K_{ij}^k \left( \p^i\s^\m\pb_k \d_\m z^j 
+ \p^j\s^\m \pb_k \d_\m z^i - 2i \p^i\p^j f_k^\dag \right) + h.c. 
\crbig
&+&{1\over 4} K_{ij}^{kl}  \p^i\p^j \, \pb_k\pb_l 
-{1\over 4} \d_\m\d^\m K(z^i,z_j^\dag)\ . 
\crbig
\end{array}
\eeq
where the last term is a total derivative and hence can be dropped 
from the Lagrangian.

Note that after discarding this total derivative, (\ref{5KDtermtwo}) no longer contains
the ``purely holomorphic" terms $\sim K_{ij}$ or the ``purely antiholomorphic" terms
$\sim K^{ij}$. Only the mixed terms with at least one upper and one lower index
remain. This shows that the transformation
\beq\label{5kahlertransf}
K(z,z^\dag) \to K(z, z^\dag) + g(z) + \ov g(z^\dag)
\eeq
does not affect the Lagrangian. Moreover, the metric of the kinetic terms for the
complex scalars is
\beq\label{5kahlermetric}
K_i^j={\d^2\over \d z^i \d z^\dag_j} K(z, z^\dag)\ .
\eeq
A metric like this obtained from a complex scalar function is called a K\"ahler
metric, and the scalar function $K(z, z^\dag)$ the K\"ahler potential. The metric is
invariant under K\"ahler transformations (\ref{5kahlertransf}) of this potential. Thus
one is led to interpret the complex scalars $z^i$ as (local) complex coordinates on a
K\"ahler manifold, i.e. the target manifold of the sigma-model is K\"ahler. The
K\"ahler invariance (\ref{5kahlertransf}) actually generalises to the superfield level
since
\beq\label{5kahlertransfsuper}
K(\f, \f^\dag)\to K(\f, \f^\dag) + g(\f) + \ov g(\f^\dag) 
\eeq
does not affect the resulting action because $g(\f)$ is again a chiral superfield
and its $\t\t\tb\tb$ component is a total derivative, see (\ref{3chiralcompx}), hence
$\Dint g(\f)=\Dint \ov g(\f^\dag)=0$.

Once $K_i^j$ is interpreted as a metric it is straightforward to compute the affine
connection and curvature tensor. However, in Riemannian geometry, indices are
lowered and rised by the metric and its inverse, while here we used upper and lower
indices to denote derivatives w.r.t. $z^i$ or $z_j^\dag$. To avoid confusion, we
temporarily switch conventions, replacing $z_j^\dag \to z^\jb$. Then
$K_i^j \to K_{i\jb}$ so that
\beq\label{5kmetric}
K_{i\jb}=K_{\jb i} \quad , \quad K_{ij}=K_{\ib\jb}=0 \ 
\eeq
and the inverse metric is  $K^{i\jb}=K^{\jb i}$, $K^{ij}=K^{\ib\jb}=0$. The affine
connection is given as usual by 
$\G^c_{ab} = \ha G^{cd}\left( \d_a G_{bd}+\d_b G_{ad}-\d_d
G_{ab}\right)$ which for the K\"ahler metric simplifies since ${\d\over \d z^i}
K_{j\mb}= {\d\over \d z^j} K_{i\mb}$, etc. One finds
\beq\label{5affineconn}
\G^l_{ij} = K^{l\mb} K_{ij\mb} \quad , \quad 
\G^{\ov l}_{\ib\jb} = K^{\ov l  m} K_{\ib\jb m} \
,
\eeq
all others with mixed indices like $\G^l_{i\jb}$ or $\G^{\ov l}_{ij}$ 
vanish.  The curvature 
tensor is given
in general by
\beq\label{5curvature}
(R_{ab})^c_{\ d} = \d_a\G^c_{bd} - \d_b \G^c_{ad} 
+\G^c_{af}\G^f_{bd} - \G^c_{bf}\G^f_{ad} \ .
\eeq           
It is easy to see that in the K\"ahler case the only nonvanishing components are
\beq\label{5curvaturecomp}
(R_{\kb i})^l_{\ j}=\d_\kb \G_{ij}^l = K^{l\ov p}
\left( K_{ij\ov p \kb} - K_{ij\mb}K^{\mb n} K_{n\ov p \kb}\right)
\eeq
and $(R_{i\kb})^l_{\ j} = - (R_{\kb i})^l_{\ j}$, and similarly
\beq\label{5curvaturecompbar}
(R_{i\kb})^{\ov l}_{\ \jb}=-(R_{\kb i})^{\ov l}_{\ \jb}
=\d_i \G^{\ov l}_{\kb\jb} = K^{\ov l p}
\left( K_{ip\kb\jb} - K_{ip\mb}K^{\mb n} K_{n\kb\jb}\right) \ .
\eeq
Reverting to our previous notation, we write
\beq\label{5kgr}
K_{i\jb}\to K_i^j \ , \quad
\G^l_{ij}\to \G_{ij}^l \ , \quad
\G_{\ib\jb}^{\ov l} \to \G_l^{ij} \ , \quad
(R_{\kb i})_{\ov l j} \to R_{ij}^{kl}  ,
\eeq
i.e.
\beq\label{5gr}
\begin{array}{rcl}
\G_{ij}^l&=& (K^{-1})^l_k K^k_{ij} \quad , \quad 
\G^{ij}_l= (K^{-1})_l^k K_k^{ij} \ ,\crbig
R_{ij}^{kl}&=& K_{ij}^{kl} - K_{ij}^m (K^{-1})_m^n K_n^{kl} \ . \crbig
\end{array}
\eeq
This allows us to rewrite various terms in the Lagrangian in a simpler and more
geometric form.

Define ``K\"ahler covariant" derivatives of the fermions as
\beq\label{5fermcov}
\begin{array}{rcl}
D_\m \p^i &=& \d_\m \p^i + \G^i_{jk} \d_\m z^j\, \p^k 
=\d_\m \p^i + (K^{-1})^i_l\,  K^l_{jk} \d_\m z^j\, \p^k 
\crbig
D_\m \pb_j&=& \d_\m \pb_j + \G^{ki}_j \d_\m z^\dag_k\,  \pb_i
=\d_\m \pb_j + (K^{-1})_j^l\,  K^{ki}_l \d_\m z^\dag_k\,  \pb_i \ .
\crbig
\end{array}
\eeq
The fermion bilinears in (\ref{5KDtermtwo}) then precisely are ${i\over 2}
K_i^j D_\m \p^i \s^\m \pb_j + h.c.$. The four fermion term is $K_{ij}^{kl}
\p^i\p^j\pb_k\pb_l$. The full curvature tensor will appear after we
eliminate the auxiliary fields $f^i$. To do this, we add the two pieces
(\ref{5KDtermtwo}) and  (\ref{5Dterms}) of the Lagrangian to see that the
auxiliary field equations of motion are \beq\label{5auxequs}
f^i=(K^{-1})^i_j w^j - \ha \G_{jk}^i \p^j\p^k \ .
\eeq
Substituting back into the sum of  (\ref{5KDtermtwo}) and  (\ref{5Dterms})  
we finally get the
Lagrangian
\beq\label{5fulllagr}
\begin{array}{rcl}
& & \int {\rm d}^4 x \left[ \Dint K(\f, \f^\dag) 
+ \Fint w(\f) + \Fbarint [w(\f)]^\dag\right]
\crbig
& & = \int {\rm d}^4 x \Big[ 
K_i^j \left( \d_\m z^i \d^\m z^\dag_j + {i\over 2} D_\m \p^i \s^\m \pb_j 
- {i\over 2} \p^i \s^\m D_\m \pb_j \right) 
- (K^{-1})^i_j w_i w^j 
\crbig
& & \quad - \ha \left( w_{ij}- \G_{ij}^k w_k \right) \p^i\p^j
- \ha \left( w^{ij} - \G_k^{ij} w^k \right) \pb_i\pb_j 
+{1\over 4} R_{ij}^{kl} \p^i\p^j\pb_k\pb_l \Big] \ .
\end{array}
\eeq

\section{Including gauge fields}

The inclusion of gauge fields changes two things. First, the kinetic term
$K(\f,\f^\dag)$ has to be modified so that, among others, all derivatives
$\d_\m$ are turned into gauge covariant derivatives as we did in section 4 when
we replaced $\f^\dag \f$ by $\f^\dag e^{2g V} \f$. Second, one has to add kinetic
terms for the gauge multiplet $V$. In the spirit of the $\s$-model, one will allow
a susy Lagrangian leading to terms of the form $f_{ab}(z) F^a_{\m\n} F^{b\m\n}$ etc.

Let's discuss the matter Lagrangian first. Since
\beq\label{5gtran}
\f\to e^{i\L}\f \ , \quad
\f^\dag\to\f^\dag e^{-i\L^\dag} \ , \quad
e^{2gV}\to e^{i\L^\dag} e^{2gV} e^{-i\L}
\eeq
one sees that
\beq\label{5fegvtrans}
\f^\dag e^{2gV}\to \f^\dag e^{2gV} e^{-i\L} \ .
\eeq
Then the combination $\left(\f^\dag e^{2gV}\right)_i \f^i$ is gauge invariant and
the same is true for any real (globally) $G$-invariant function $K(\f^i,
\f_i^\dag)$ if the argument $\f^\dag_i$ is replaced by 
$\left(\f^\dag e^{2gV}\right)_i$. We conclude that if $w(\f^i)$ is a $G$-invariant
function of the $\f^i$, i.e. if
\beq\label{5wtrans}
w_i (T^a)^i_{\ j} \f^j =0 \quad , \quad a=1, \ldots {\rm dim}G
\eeq
then
\beq\label{5matterlagr}
{\cal L}_{\rm matter}=\Dint K\left( \f^i, \left(\f^\dag e^{2gV}\right)_i\right)
+\Fint w(\f^i) +\Fbarint [w(\f^i)]^\dag
\eeq
is supersymmetric and gauge invariant.

To discuss the generalisation of the gauge kinetic Lagrangian 
(\ref{4gaugelagr}),
reall that $W_\a$ is defined by (\ref{4Wnonab}) with $V\to 2g V$ and in WZ 
gauge it
reduces to (\ref{4Walphacomp}) times $2g$. 
Note that any power of $W$ never contains more than
two derivatives, so we could consider a susy Lagrangian of the form 
$\Fint H(\f^i, W_\a)$ with an arbitrary $G$-invariant function $H$. 
We will be slightly less general and take
\beq\label{5gaugelagr}
{\cal L}_{\rm gauge} = {1\over 16 g^2} \Fint f_{ab}(\f^i) W^{a\a}W^b_\a
+ h.c.
\eeq
with $f_{ab}=f_{ba}$ transforming under $G$ as the symmetric product of the adjoint
representation with itself. To get back the standard Lagrangian (\ref{4gaugelagr}) one
only needs to take ${1\over g^2} f_{ab} ={\tau\over 4\pi i} {\rm Tr}\, T^a T^b$.
Expanding (\ref{5gaugelagr}) in components is straightforward and yields
\beq\label{5gaugelagrcomp}
\begin{array}{rcl}
{\cal L}_{\rm gauge}&=& {\rm Re} f_{ab}(z) 
\left( -{1\over 4} F^a_{\m\n}F^{b\m\n}-i\l^a\s^\m D_\m \lb^b +\ha D^a D^b\right)
-{1\over 4} {\rm Im} f_{ab}(z) F^a_{\m\n} \wt F^{b\m\n} 
\crbig
&+&{1\over 4} f_{ab,i}(z) \left( \rd i \p^i\l^a D^b -\rd\l^a\s^{\m\n}\p^i F^b_{\m\n}
+\l^a\l^b f^i\right) +h.c.
\crbig
&+&{1\over 8} f_{ab,ij}(z) \l^a\l^b\p^i\p^j + h.c. 
\crbig
\end{array}
\eeq
where $F_{\m\n}$ and $D_\m\l$ were defined in (\ref{4newfmn}) 
and $f_{ab,i}={\d\over \d z^i} f_{ab}(z)$ etc.

To obtain the component expansion of the matter Lagrangian (\ref{5matterlagr}) is a bit
lengthy. The computation parallels the one leading to (\ref{5KDtermtwo}) but paying
attention to the gauge field terms. The result can be read from (\ref{5KDtermtwo})
by gauge covariantising all derivatives and adding (\ref{5Dterms}). 
Furthermore, it is clear that one also obtains the Yukawa interactions that
already appeared in (\ref{4fullgaugelagr}) with the K\"ahler metric appropriately
inserted. Note also that the term $g z^\dag D z$ now is replaced by $g
z^\dag_i D K^i$. Taking all this into account it is easy to see that one obtains
\beq\label{5matterlagrcomp}
\begin{array}{rcl}
{\cal L}_{\rm matter}&=& K_i^j
\left[ f^i f_j^\dag + (D_\m z)^i (D^\m z)_j^\dag 
-{i\over 2} \p^i\s^\m \wt D_\m \pb_j +{i\over 2} \wt D_\m\p^i\s^\m\pb_j \right]
\crbig
&+&\ha K_{ij}^k\,  \p^i\p^j f_k^\dag + h.c. 
+{1\over 4} K_{ij}^{kl}\,  \p^i\p^j \, \pb_k\pb_l 
\crbig
&-&\left( w_i f^i +\ha w_{ij}\p^i\p^j\right) + h.c.
\crbig
&+& i\rd g K_j^i\, z_i^\dag \l \p^j - i\rd g K^i_j\, \pb_i \lb z^j 
+ g z_i^\dag D K^i \ ,
\end{array}
\eeq
where as before  gauge indices have been suppressed, e.g. 
$\pb_i \lb z^j\equiv\pb_i T^a_R z^j \lb^a
\equiv (\pb_i)_M (T^a_R)^M_{\ N} (z^j)^N \lb^a$ where $(T^a_R)^M_{\ N}$ are 
the matrices of the
representation carried by the matter fields $(z^j)^N$ and $(\p^i)^N$. 
The derivatives $\wt D_\m$ acting
on the fermions are gauge and K\"ahler covariant, i.e.
\beq\label{5fgaugekahlercov}
\begin{array}{rcl}
\wt D_\m \p^i &=& \d_\m \p^i -i g v^a_\m  T^a_R \p^i + \G^i_{jk} \d_\m z^j\, \p^k  
\crbig
\wt D_\m \pb_j&=& \d_\m \pb_j -i g v^a_\m T^a_R \pb_j + \G^{ki}_j \d_\m z^\dag_k\, 
\pb_i  \ .
\crbig
\end{array}
\eeq

The full Lagrangian is given by ${\cal L}={\cal L}_{\rm gauge}+{\cal L}_{\rm matter}$.
The auxiliary field equations of motion are
\beq\label{5auxeoms}
\begin{array}{rcl}
f^i&=& (K^{-1})^i_j \left( w^j -\ha K^j_{kl}\, \p^k\p^l 
-{1\over 4} (f_{ab,j})^\dag \lb^a\lb^b \right) \crbig
D^a&=& - ({\rm Re} f)^{-1}_{ab} \left( g z^\dag_i T^b K^i 
+{i\over 2\rd} f_{bc,i}\p^i\l^c -{i\over 2\rd} (f_{bc,i})^\dag \pb_i \lb^c \right) \ .
\crbig
\end{array}
\eeq
It is straightforward to substitute this into the Lagrangian ${\cal L}$ and we will
not write the result explicitly. Let us only note that the scalar potential is given by
\beq\label{5sclarpot}
V(z,z^\dag)=(K^{-1})^i_j w_i w^j 
+{g^2\over 2}  ({\rm Re} f)^{-1}_{ab}  (z^\dag_i T^a K^i)  (z^\dag_j T^b K^j) \ .
\eeq

%% file: ch6.tex
%

The $N=2$ multiplets with helicities not exceeding one are the massless 
$N=2$ vector multiplet and the hypermultiplet. The former contains an 
$N=1$ vector multiplet and an $N=1$ chiral multiplet, alltogether a 
gauge boson, two Weyl fermions and a complex scalar, while the 
hypermultiplet contains two $N=1$ chiral multiplets. The $N=2$ vector
multiplet is necessarily massless while the hypermultiplet can be 
massless or be a short (BPS) massive multiplet. Here we will concentrate 
on the $N=2$ vector multiplet.

\section{$N=2$ super Yang-Mills}

Given the decomposition of the $N=2$ vector multiplet into $N=1$ multiplets,
we start with a Lagrangian being the sum of the $N=1$ gauge and matter 
Lagrangians (\ref{4gaugelagr}) and (\ref{4kinDtermrescaled}). At present, 
however, all fields are in the same $N=2$ multiplet and hence must be in 
the same representation of the gauge group, namely the adjoint 
representation. The $N=1$ matter Lagrangian (\ref{4kinDtermrescaled})
then becomes, after rescaling $V\to 2g V$,
\beq\label{6matterlagr}
\begin{array}{rcl}
{\cal L}_{\rm matter}^{N=1}= \Dint \Tr \f^\dag e^{2gV} \f
&=& \Tr \Big[
(D_\m z)^\dag D^\m z 
- i \p\s^\m D_\m \pb + f^\dag f \crbig
&+&i \rd  g z^\dag \{\l,\p\} - i\rd  g \{\pb,\lb\} z + g D [z,z^\dag]
\Big]\, \crbig
\end{array}
\eeq
where now
\beq\label{6adjoint}
z=z^a T^a\ , \quad \p=\p^a T^a \ , \quad f= f^a T^a \ , 
\quad a=1, \ldots {\rm dim}G
\eeq
in addition to $\l=\l^a T^a,\ D=D^a T^a,\ v_\m = v_\m^a T^a$. The commutators
or anticommutators arise since the generators in the adjoint representation 
are given by
\beq\label{6adgen}
\left( T^a_{\rm ad}\right)_{bc} = -i f_{abc}
\eeq
and we normalise the generators by
\beq\label{6gennorm}
\Tr T^a T^b = \delta^{ab}
\eeq
so that
\beq\label{6trilin}
\begin{array}{rcl}
z^\dag \l\p &\to & z^\dag_b \l^a \left( T^a_{\rm ad}\right)_{bc}  \p^c
= -i  z^\dag_b \l^a f_{abc}\p^c =  i  z^\dag_bf_{bac}  \l^a \p^c \crbig
&=& z^\dag_b \l^a \p^c \Tr T^b [ T^a, T^c] = \Tr z^\dag \{ \l,\p\}
\end{array}
\eeq
and
\beq\label{6Dcomm}
z^\dag D z \to  z^\dag_b D^a \left( T^a_{\rm ad}\right)_{bc}  z^c
= -i  f_{abc} z^\dag_b D^a z^c
= - \Tr D[z^\dag, z] = \Tr D[z, z^\dag] \ .
\eeq
We now add (\ref{6matterlagr}) to the $N=1$ gauge lagrangian 
${\cal L}_{\rm gauge}^{N=1}$ (\ref{4gaugelagr})
and obtain
\beq\label{6N=2lagr}
\begin{array}{rcl}
{\cal L}_{\rm YM}^{N=2} 
&=&{1\over 32\pi} {\rm Im}\, \left( \tau \Fint {\rm Tr}\, W^\a W_\a \right)
+ \Dint \Tr \f^\dag e^{2gV} \f \crbig
&=& \Tr  \Big( -{1\over 4} F_{\m\n}F^{\m\n} 
- i\l\s^\m D_\m \lb - i \p\s^\m D_\m \pb + (D_\m z)^\dag D^\m z \crbig
&&+
{\Theta\over 32\pi^2} g^2\, {\rm Tr} F_{\m\n} \widetilde F^{\m\n}
+ \ha D^2 + f^\dag f \crbig
&&+
i \rd  g z^\dag \{\l,\p\} - i\rd  g \{\pb,\lb\} z + g D [z, z^\dag]\Big) \ .
\crbig
\end{array}
\eeq
A necessary and sufficient condition for $N=2$ susy is the existence of an 
${\rm SU}(2)_R$ symmetry that rotates the two supersymmetry generators 
$Q^1_\a$ and $Q^2_\a$ 
into each other. As follows from the construction of the supermultiplet in 
section 2, the same symmetry must act between the two fermionic fields $\l$ 
and $\p$. Now the relative coefficients of ${\cal L}_{\rm gauge}^{N=1}$ and 
${\cal L}_{\rm matter}^{N=1}$ in (\ref{6N=2lagr}) have been chosen precisely 
in such a way to have this ${\rm SU}(2)_R$ symmetry: the $\l$ and $\p$ 
kinetic terms have the same coefficient, and the Yukawa couplings 
$z^\dag \{\l,\p\}$ and $\{\pb,\lb\} z$ also exhibit this symmetry. The 
Lagrangian (\ref{6N=2lagr}) is indeed $N=2$ supersymmetric.

Note that we have not added a superpotential. Such a term (unless 
linear in $\f$) would  break the ${\rm SU}(2)_R$ invariance and not lead to 
an $N=2$ theory.

The auxiliary field equations of motion are simply
\beq\label{6auxeoms}
\begin{array}{rcl}
f^a&=&0\crbig
D^a&=& - g\, [z, z^\dag]^a
\end{array}
\eeq
leading to a scalar potential
\beq\label{6N=2scalarpot}
V(z,z^\dag)= \ha g^2\,  \Tr \left( [z,z^\dag]\right)^2 \ .
\eeq
This scalar potential is fixed and a consequence solely of the auxiliary 
$D$-field of the $N=1$ gauge multiplet.

\section{Effective $N=2$ gauge theories}

As for the non-linear $\s$-model, if one considers effective theories, 
disregarding renormalisability, one may allow more general gauge and matter 
kinetic terms and start with an appropriate sum of (\ref{5matterlagr}) 
(with $w(\f^i)=0$) and (\ref{5gaugelagr}). It is clear however that the 
functions $f_{ab}$ cannot be independent from the K\"ahler potential $K$. 
Indeed, the ${\rm SU}(2)_R$ symmetry equates ${\rm Re} f_{ab}$ with the 
K\"ahler metric $K_a^b$. It turns out that this requires the following 
identification
\beq\label{6prepot}
\begin{array}{rcl}
{16\pi\over (2g)^2}\, f_{ab}(z)
&=&-i {\d^2\over \d z^a \d z^b} \F(z) \equiv -i \F_{ab}(z) \crbig
{16\pi\over (2g)^2}\, K(z, z^\dag)
&=& - {i\over 2}\, z^\dag_a {\d\over \d z^a} \F(z) + h.c. \equiv 
- {i\over 2}\, z^\dag_a \F_a(z) + {i\over 2}\, \left[ \F_a(z)\right]^\dag z_a 
\crbig
\end{array}
\eeq
where the holomorphic function $\F(z)$ is called the $N=2$ prepotential. We 
have pulled out a factor ${16\pi\over (2g)^2} $ for later convenience. Also, 
we again absorb the factor $2g$ into the normalisation of the field. This 
makes sense since ${\rm Im} \F_{ab}$ will play the role of an effective 
generalised coupling. Hence we set 
\beq\label{62g=1}
2g=1.
\eeq
Then the full general $N=2$ Lagrangian is
\beq\label{6effN=2lagr}
\begin{array}{rcl}
{\cal L}_{\rm eff}^{N=2} 
&=& \left[ {1\over 64 \pi i} \Fint \F_{ab}(\f) W^{a\a} W^b_\a
+{1\over 32 \pi i}  \Dint \left( \f^\dag e^V \right)^a \F_a(\f) \right] + h.c. 
\crbig
&=& {1\over 16\pi} {\rm Im} \left[ \ha  \Fint \F_{ab}(\f) W^{a\a} W^b_\a + 
\Dint \left( \f^\dag e^V \right)^a \F_a(\f) \right] \ . \crbig
\end{array}\eeq
Note that with the K\"ahler potential $K$ given by (\ref{6prepot}), the K\"ahler metric is proportional to ${\rm Im} \F_{ab}$ as required by ${\rm SU}(2)_R$ :
\beq\label{6kahlermetric}
K_a^b={1\over 16\pi} {\rm Im} \F_{ab} 
={1\over 32\pi i} \left( \F_{ab}-\F_{ab}^\dag \right) \ .
\eeq
The component expansion follows from the results of the previous section on 
the non-linear $\s$-model, using the identifications (\ref{6prepot}) and 
(\ref{6kahlermetric}), and taking vanishing superpotential $w(\f)$. In 
particular, the scalar potential is given by (cf (\ref{5sclarpot}))
\beq\label{6effscalarpot}
V(z,z^\dag)= -{1\over 2\pi} ({\rm Im} \F)^{-1}_{ab} \, 
[z^\dag, \F_c(z)T^c]^a\, [z^\dag, \F_d(z) T^d]^b \ .
\eeq

Let us insist that the full effective $N=2$ action written in (\ref{6effN=2lagr}) 
is determined by a single holomorphic function $\F(z)$. Holomorphicity will turn 
out to be a very strong requirement. Finally note that $\F(z)= \ha \tau\, \Tr z^2$ 
gives back the  standard Yang-Mills Lagrangian (\ref{6N=2lagr}).

%% file: ch11.tex
%
\def\a{\alpha}
\def\adot{{\dot\alpha}}
\def\b{\beta}
\def\m{\mu}
\def\s{\sigma}
\def\n{\nu}
\def\r{\rho}
\def\l{\lambda}
\def\lb{{\bar\lambda}}
\def\L{\Lambda}
\def\g{\gamma}
\def\G{\Gamma}
\def\t{\theta}
\def\tb{{\bar\theta}}
\def\sm{\sigma^\m}
\def\smb{\bar\sigma^\m}
\def\d{\partial}
\def\rmd{{\rm d}}
\def\vf{\varphi}
\def\f{\phi}
\def\F{\phi}
\def\fd{\phi_D}
\def\Fd{\phi_D}
\def\ix{\int {\rm d}^4 x \, }
\def\dt{{\rm d}^2 \t \, }
\def\dtb{{\rm d}^2 \tb \, }
\def\dtt{{\rm d}^2 {\tilde\t} \, }
\def\P{\Psi}
\def\cf{{\cal F}}
\def\cfd{{\cal F}_D}
\def\ad{a_D}
\def\e{\epsilon}
\def\rmd{{\rm d}}
\def\rd{\sqrt{2}}
\def\la{\langle}
\def\ra{\rangle}
\def\vac{\vert 0\rangle}
\def\P{\Psi}
\def\p{\psi}
\def\pb{{\bar \psi}}
\def\dd #1 #2{{\delta #1\over \delta #2}}
\def\tr{{\rm tr}\ }
\def\til{\widetilde}
\def\nm{\nabla_\m}
\def\fmn{F_{\m\n}}
\def\fmnt{{\tilde F}_{\m\n}}
\def\im{{\rm Im}\, }
\def\M{{\cal M}}

In this section, I will discuss how electric-magnetic duality is realised in 
an effective low-energy $N=2$ gauge theory. This was pioneered by Seiberg and 
Witten in 1994 \cite{SW} who considered the
simplest case of pure $N=2$ supersymmetric ${\rm SU}(2)$ Yang-Mills theory. 
This work
was then generalized to other gauge groups and to theories including extra
matter fields (susy QCD). In the mean time, it became increasingly clear that
dualities in string theories play an even more fascinating role (as
is discussed by others at this school). Here I focus on the simplest 
${\rm SU}(2)$ case which most clearly
examplifies the beauty of duality. This section is based on an earlier
introduction into the subject by the present author \cite{AB} where further 
references can be found.

The idea of duality probably goes back to Dirac who observed that the
source-free Maxwell equations are symmetric under the exchange of the
electric and magnetic fields. More precisely, the symmetry is $E\to B,\ B\to
-E$, or $\fmn\to \fmnt=\half
\e_{\m\n}^{\phantom{\m\n}\rho\sigma}F_{\rho\sigma}$. To maintain
this symmetry in the presence of sources, Dirac introduced, somewhat ad hoc,
magnetic monopoles with magnetic charges $q_m$ in addition to the electric
charges $q_e$, and showed that consistency of the quantum theory requires a
charge quantization condition $q_mq_e=2\pi n$ with integer $n$. Hence the
minimal charges obey $q_m={2\pi\over q_e}$. Duality exchanges $q_e$ and
$q_m$, i.e. $q_e$ and ${2\pi\over q_e}$. Now recall that the electric charge
$q_e$ also is the coupling constant. So duality exchanges the coupling
constant with its inverse (up to the factor of $2\pi$), hence exchanging
strong and weak coupling. This is the reason why we are so much interested in
duality: the hope is to learn about strong-coupling physics from the
weak-coupling physics of a dual formulation of the theory. Of course, in
classical Maxwell theory we know all we may want to know, but this is no
longer true in quantum electrodynamics.

Actually, quantum electrodynamics is not a good candidate for exhibiting a
duality symmetry since there are no magnetic monopoles, but the latter
naturally appear in spontaneously broken non-abelian gauge theories. 
Unfortunately, electric-magnetic duality in its simplest form cannot
be a symmetry of the quantum theory due to the running of the coupling
constant (among other reasons). Indeed, if duality exchanges $\a(\L)
\leftrightarrow {1\over\a(\L)}$ (where $\a(\L)={e^2(\L)\over 4\pi}$) at some
scale $\L$, in general this won't be true at another scale. This argument is avoided if
the coupling does not run, i.e. if the $\b$-function vanishes as is the case
in certain ($N=4$) supersymmetric extensions of the Yang-Mills theory. This
and other reasons led Montonen and Olive  to conjecture that duality
might be an exact symmetry of $N=4$ susy Yang-Mills theory. 
The Seiber-Witten duality concerns a different type of theory: it deals with
an $N=2$ susy low-energy {\it effective} action and duality exchanges the 
effective coupling $\a_{\rm eff}(\L)$ with a dual coupling 
$\a_{\rm eff}^D(\L_D) \sim {1\over \a_{\rm eff}(\L)}$ at a dual scale $\L_D$. 
The dependence of this dual scale $\L_D$ on the original scale $\L$ precisely
takes into account the running of the coupling.
Let me insist that the Seiberg-Witten duality is an {\it exact}  
symmetry of the abelian low-energy {\it effective} theory, not of the 
microscopic $SU(2)$ theory. This is different from the Montonen-Olive 
conjecture about an exact duality symmetry of a microscopic gauge theory.

A somewhat similar duality symmetry appears in the two-dimensional
Ising model where it exchanges the temperature with a dual temperature, thereby
exchanging high and low temperature analogous to strong and weak coupling.
For the Ising model, the sole existence of the duality symmetry led to the 
exact determination of the critical temperature as the self-dual point, well prior to
the exact solution by Onsager. One may
view the existence of this self-dual point as the requirement that the dual
high and low temperature regimes can be consistently ``glued" together.
Similarly, in the Seiberg-Witten theory, as will be explained below, duality
allows us to obtain the full effective action for the light fields at any
coupling (the analogue of the Ising free energy at any temperature) from
knowledge of its weak-coupling limit and the behaviour at certain
strong-coupling ``singularities", together with a holomorphicity 
requirement that tells
us how to patch together the different limiting regimes.

\section{Low-energy effective action of $N=2$  $SU(2)$ YM theory}

Following Seiberg and Witten we want to study and determine the low-energy
effective action of the $N=2$ susy Yang-Mills theory with gauge group ${\rm SU}(2)$. The
latter theory is the microscopic theory which controls the high-energy behaviour. 
It was discussed in section 6 and its Lagrangian is given by (\ref{6N=2lagr}).
This theory
is renormalisable and well-known to be asymptotically free. The low-energy effective
action will turn out to be quite different.

\subsection{Low-energy effective actions}

There are two types of effective actions. One is the standard generating functional
$\Gamma[\vf]$ of one-particle irreducible Feynman diagrams (vertex functions). It is
obtained from the standard renormalised generating functional $W[\vf]$ of connected
diagrams by a Legendre transformation. Momentum integrations in loop-diagrams are from
zero up to a UV-cutoff which is taken to infinity after renormalisation.
$\Gamma[\vf]\equiv \Gamma[\m,\vf]$ also depends on
the scale $\m$ used to define the renormalized vertex functions.

A quite different object is the Wilsonian effective action $S_{\rm W}[\m,\vf]$. It is
defined as $\Gamma[\m,\vf]$, except that all loop-momenta are only integrated down 
to $\m$
which serves as an infra-red cutoff.  In theories with
massive particles only, there is no big difference between $S_{\rm W}[\m,\vf]$ and
$\Gamma[\m,\vf]$ (as long as $\m$ is less than the smallest mass). When massless 
particles are present, as is the case for gauge
theories, the situation is  different. In particular, in supersymmetric gauge
theories there is the so-called Konishi anomaly which can be viewed as an IR-effect.
Although $S_{\rm W}[\m,\vf]$ depends holomorphically on $\m$, this is not the case for
$\Gamma[\m,\vf]$ due to this anomaly.

\subsection{The ${\rm SU}(2)$ case, moduli space}

Following Seiberg and Witten, we want to
determine the Wilsonian effective action in the case where the microscopic theory 
is the ${\rm SU}(2)$, $N=2$ super Yang-Mills theory. As
explained above, classically this theory has a scalar potential $V(z)=\half g^2 \tr
([z^\dag,z])^2$ as given in (\ref{6N=2scalarpot}). 
Unbroken susy requires that $V(z)=0$ in the vacuum, but this still
leaves the possibilities of non-vanishing $z$ with $[z^\dag,z]=0$.
We are interested in determining the gauge inequivalent vacua. 
A general $z$ is of the form $z(x)=\half\sum_{j=1}^3  \left( a_j(x)+i b_j(x)\right)
\sigma_j$ with real fields $a_j(x)$ and $b_j(x)$ (where I assume that not all three $a_j$
vanish, otherwise exchange the roles of the $a_j$'s and $b_j$'s in the sequel). By a
${\rm SU}(2)$ gauge transformation one can always arrange $a_1(x)=a_2(x)=0$.
Then $[z, z^\dag]=0$ implies
$b_1(x)=b_2(x)=0$ and hence, with $a= a_3 + i b_3$, one has $z=\half a \sigma_3$.
Obviously, in the vacuum $a$ must be a constant.
Gauge transformation from the Weyl group (i.e. rotations by
$\pi$ around the 1- or 2-axis of ${\rm SU}(2)$) can still change $a\to -a$, so $a$ and $-a$ are gauge
equivalent, too. The gauge invariant quantity describing inequivalent vacua is $\half
a^2$, or $\tr z^2$, which
 is the same, semiclassically.
When quantum fluctuations are important this is no longer so.
In the sequel, we will use the following definitions for $a$ and $u$:
\beq\label{swmoduli}
u=\la \tr z^2\ra\quad , \quad \la z \ra = \half a\sigma_3\ .
\eeq
The complex parameter $u$ labels gauge inequivalent vacua. The manifold of gauge
inequivalent vacua is called the moduli space $\M$ of the theory. Hence $u$ is a
coordinate on $\M$, and $\M$ is essentially the complex $u$-plane. We will see in the
sequel that $\M$ has certain singularities, and the knowledge of the behaviour of the
theory near the singularities will eventually allow the determination of the effective
action $S_{\rm W}$.

Clearly, for non-vanishing $\la z \ra$, the ${\rm SU}(2)$ gauge 
symmetry is broken by the
Higgs mechanism, since the $z$-kinetic term 
$\vert D_\m z\vert^2$ generates masses for
the gauge fields. With the above conventions, $v_\m^b,\ b=1,2$ become massive with
masses given by $\half m^2 = g^2 \vert a\vert^2$, i.e $m=\rd  g \vert a\vert$. 
Similarly
due to the $\f,\l,\p$ interaction terms, $\p^b, \l^b,\ b=1,2$ become massive with the
same mass as the $v_\m^b$, as required by supersymmetry. Obviously, $v_\m^3,\p^3$ and
$\l^3$, as well as the mode of $\f$ describing the flucuation of $\f$ in the
$\sigma_3$-direction, remain massless. These massless modes are described by a
Wilsonian low-energy effective action which has to be $N=2$ supersymmetry invariant,
since, although the gauge symmetry is broken, ${\rm SU}(2)\to {\rm U}(1)$, the $N=2$ susy remains
unbroken. Thus it must be of the general form (\ref{6effN=2lagr})
where the indices $a,b$
now take only a single value ($a,b=3$) and will be suppressed since the gauge
group is ${\rm U}(1)$. Also,  in an
abelian theory there is no self-coupling of the gauge boson and the same arguments
extend to all members of the $N=2$ susy multiplet: they do not carry electric charge.
Thus for a ${\rm U}(1)$-gauge theory, from (\ref{6effN=2lagr}) we simply get 
\beq\label{swu1action}
{1\over 16\pi} \im\ix \left[ \ha \int\dt \cf''(\f)W^\a W_\a
+\int\dt\dtb \f^\dag\cf'(\f) \right] \ . 
\eeq

\subsection{Metric on moduli space}

As shown in (\ref{6kahlermetric}), the K\"ahler metric of the present 
$\s$-model is given by $K_{z \ov z} = {1\over 16\pi} \im
\cf''(z)$.
By the same token this defines the metric in the space of
(inequivalent) vacuum
configurations, i.e. the metric on moduli space as ($\bar a$ denotes the
complex conjugate of $a$)
\beq\label{swmodspacemetric}
\rmd s^2=\im \cf''(a) \rmd a \rmd \bar a = \im \tau(a) \rmd a \rmd \bar a
\eeq
where $\tau(a)=\cf''(a)$ is the effective (complexified) coupling constant 
according to the remark after eq. (\ref{5gaugelagr}). The $\s$-model 
metric $K_{z \ov z}$ has been replaced on
the moduli space $\M$ by ($16\pi$ times) its expectation value in the 
vacuum corresponding to the
given point on $\M$, i.e. by $\im \cf''(a)$.

The question now is whether the description of the effective action in terms of the
fields $\f, W$ and the function $\cf$ is appropriate for all vacua, i.e. for all value
of $u$, i.e. on all of moduli space. In particular the kinetic terms or
what is the same, the metric  on moduli space should be positive definite,
translating into $\im \tau(a) >0$. However, a simple argument shows that this cannot be
the case: since $\cf(a)$ is holomorphic, $\im\tau(a)=\im {\d^2\cf(a)\over \d a^2}$ is a
harmonic function and as such it cannot have a minimum, and hence (on the compactified
complex plane)  it cannot obey
$\im\tau(a)>0$ everywhere (unless it is a constant as in the classical case). The way
out is to allow for different local descriptions: the coordinates $a, \bar a$ and the
function $\cf(a)$ are appropriate  only in a certain region of $\M$. When a singular
point with $\im\tau(a)\to 0$ is approached one has to use a different set of
coordinates $\hat a$ in which $\im\hat\tau (\hat a)$ is non-singular (and
non-vanishing). This is possible provided the singularity of the metric is only a
coordinate singularity, i.e. the kinetic terms of the effective action are not
intrinsically singular, which will be the case.

\subsection{Asymptotic freedom and the one-loop formula}

Classically the function $\cf(z)$ is given by $\half\tau_{\rm class} z^2$. The one-loop
contribution has been determined by Seiberg. The combined tree-level and one-loop result
is
\beq\label{tvi}
\cf_{\rm pert}(z)={i\over 2\pi} z^2\ln {z^2\over \L^2} \ .
\eeq
Here $\L^2$ is some combination of $\m^2$ and numerical factors chosen so as to fix the
normalisation of $\cf_{\rm pert}$. Note that due to non-renormalisation theorems for
$N=2$ susy there are no corrections from two or more loops to the Wilsonian effective
action $S_{\rm W}$ and (\ref{tvi}) is the full perturbative result. There are however
non-perturbative corrections that will be determined below.

For very large $a$ the dominant contribution when computing $S_{\rm W}$ from the
microscopic ${\rm SU}(2)$
 gauge theory comes from regions of large momenta ($p\sim a$) where the microscopic
theory is asymptotically free. Thus, as $a\to\infty$ the effective coupling constant
goes to zero, and the perturbative expression (\ref{tvi}) for $\cf$ becomes an excellent
approximation. Also $u\sim \half a^2$ in this limit.\footnote{
One can check from the explicit solution in section 6 that one indeed has 
$\half a^2 - u = {\cal O}(1/u)$ as $u\to\infty$.
}
Thus
\beq\label{tvii}
\begin{array}{rcl}
\cf(a  )&\sim& {i\over 2\pi} a^2\ln {a^2\over \L^2} \crbig
\tau(a) &\sim& {i\over \pi} \left( \ln {a^2\over \L^2} +3\right)  \quad {\rm as}\
u\to\infty \ .\crbig
\end{array}
\eeq
Note that due to the logarithm appearing at one-loop, $\tau(a)$ is a multi-valued
function of $a^2\sim 2u$. Its imaginary part, however, $\im\tau(a)\sim {1\over
\pi}\ln{\vert a\vert^2\over \L^2}$ is single-valued and positive (for $a^2\to \infty$).

\section{Duality}

As already noted, $a$ and $\bar a$ do provide local coordinates on the moduli space
$\M$ for the region of large $u$. This means that in this region $\f$ and
$W^\a$ are appropriate fields to describe the low-energy effective action. As also
noted, this description cannot be valid globally, since $\im\cf''(a)$, being a harmonic
function, must vanish somewhere, unless it is a constant - which it is not. Duality will
provide a different set of (dual) fields $\f_{\rm D}$ and $W^\a_{\rm D}$ that provide an
appropriate description for a different region of the moduli space.

\subsection{Duality transformation}

Define a dual field $\Fd$ and a dual function  $\cfd(\Fd)$ by
\beq\label{qi}
\Fd=\cf'(\F) \quad , \quad
\cfd'(\Fd)=-\F \ .
\eeq
These duality
transformations simply constitute a Legendre transformation \break\hfill
$\cfd(\Fd)$ $=\cf(\F)-\F\Fd$. Using these
relations, the second term in the $\f$ kinetic term of the action 
can be written as
\beq\label{qiii}
\begin{array}{rcl}
\im\ix\dt\dtb\F^+\cf'(\F)&=&\im\ix\dt\dtb \left(-\cfd'(\Fd)\right)^+ \Fd \crbig
&=&
\im\ix\dt\dtb  \Fd^+ \cfd'(\Fd)\ .\crbig
\end{array}
\eeq
We see that this second term in the effective action is invariant under the
duality transformation.


Next, consider the $\cf''(\F)W^\a W_\a$-term in the effective action 
(\ref{swu1action}). 
While the
duality transformation on $\F$ is local, this will not be the case for the
transformation of $W^\a$. Recall that $W$ contains the ${\rm U}(1)$ field strength $\fmn$.
This $\fmn$ is not arbitrary but of the form $\d_\m v_\n-\d_\n v_\m$ for
some $v_\m$. This can be translated into the Bianchi identity $\half\e^{\m\n\rho\sigma}
\d_\n F_{\rho\sigma}\equiv \d_\nu \tilde F^{\m\n}=0$. The corresponding constraint in
superspace is $\im (D_\a W^\a)=0$.  In the functional integral 
one has the choice of integrating
over $V$ only, or over $W^\a$ and imposing the constraint $\im (D_\a W^\a)=0$ by a real
Lagrange multiplier superfield which we call $V_D$:
\beq\label{qiv}
\begin{array}{rcl}
&&\int{\cal D}V \exp\left[ {i\over 32\pi}\im\ix\dt\cf''(\F)W^\a W_\a\right]\crbig
&&\simeq \int{\cal D}W {\cal D}V_D
 \exp\Big[ {i\over 32\pi}\im\ix\Big(\int\dt \cf''(\F)W^\a W_\a\crbig
&&\phantom{\simeq \int{\cal D}W {\cal D}V_D
 \exp\Big[ {i\over 32\pi}\im\ix\Big(}
+\ha \int\dt\dtb V_D D_\a W^\a\Big) \Big] \crbig
\end{array}
\eeq
Observe that 
\beq\label{qiva}
\begin{array}{rcl}
\int\dt\dtb V_D D_\a W^\a&=& - \int\dt\dtb D_\a V_D W^\a
=+\int\dt \bar D^2 (D_\a V_D W^\a)\crbig
&=&\int\dt (\bar D^2 D_\a V_D) W^\a
=-4\int\dt (W_D)_\a W^\a\crbig
\end{array}
\eeq
where we used $\bar D_{\dot\b} W^\a=0$ and where the dual $W_D$ is defined from $V_D$
by $(W_D)_\a=-{1\over 4} \bar D^2 D_\a V_D$, as appropriate in the abelian case. 
Then one can do the
functional integral over $W$ and one obtains
\beq\label{qv}
\int{\cal D}V_D \exp\left[ {i\over 32\pi}\im\ix\dt
\left( -{1\over\cf''(\F)} W_D^\a W_{D\a}\right)\right] \ .
\eeq

This reexpresses the ($N=1$) supersymmetrized  Yang-Mills action in terms of a dual
Yang-Mills action with the effective coupling $\tau(a)=\cf''(a)$ replaced by $-{1\over
\tau(a)}$. Recall that $\tau(a)={\theta(a)\over 2\pi}+{4\pi i \over g^2(a)}$, so that
$\tau\to -{1\over \tau}$ generalizes the inversion of the coupling constant discussed
in the introduction. Also, it can be shown that the replacement $W\to W_D$ 
corresponds to replacing
$\fmn\to \tilde F_{\m\n}$, the electromagnetic dual,  so that the manipulations
leading to (\ref{qv}) constitute a  duality transformation that 
generalizes the old electromagnetic duality of Montonen and Olive. 
Expressing the $-{1\over \cf''(\F)}$ in terms of $\Fd$ one sees from
(\ref{qi}) that $\cfd''(\Fd)=-{\rmd \F\over\rmd\Fd}=-{1\over\cf''(\F)}$ so that
\beq\label{qvi}
-{1\over \tau(a)}=\tau_D(\ad) \ .
\eeq
The whole action can then equivalently be written as
\beq\label{qvii}
{1\over 16\pi} \im\ix \left[ \ha \int\dt \cfd''(\Fd)W_D^\a W_{D\a}
+\int\dt\dtb \Fd^+\cfd'(\Fd) \right] \ . 
\eeq

\subsection{The duality group}

To discuss the full group of duality transformations of the action it is most
convenient to write it as
\beq\label{qviia}
{1\over 16\pi} \im \ix\dt {\rmd\Fd\over \rmd \F}W^\a W_\a +
{1\over 32 i \pi} \ix\dt\dtb \left( \F^+\Fd-\Fd^+\F\right) \ .
\eeq
While we have shown in the previous subsection that there is a duality symmetry
\beq\label{qviii}
\pmatrix{\Fd\cr \F\cr} \to \pmatrix{0&1\cr -1&0\cr} \pmatrix{\Fd\cr \F\cr} \ ,
\eeq
the form (\ref{qviia}) shows that there  also is a symmetry
\beq\label{qix}
\pmatrix{\Fd\cr \F\cr} \to \pmatrix{1&b\cr 0&1\cr} \pmatrix{\Fd\cr \F\cr} \quad ,
\quad b\in {\bf Z} \ .
\eeq
Indeed,  in (\ref{qviia}) the second term remains invariant since
$b$ is real, while the first term 
gets shifted by
\beq\label{qx}
{b\over 16\pi} \im\ix\dt W^\a W_\a = - {b\over 16\pi}  \ix \fmn\tilde F^{\m\n} =-2\pi
b \n
\eeq
where $\n\in {\bf Z}$ is the instanton number. Since the action appears as $e^{iS}$ in
the functional integral, two actions differing only by $2\pi  {\bf Z}$ are equivalent,
and we conclude that(\ref{qix}) with integer $b$ is a symmetry of the effective action. The
transformations (\ref{qviii}) and (\ref{qix}) together generate the group $Sl(2,{\bf Z})$. This is
the group of duality symmetries.

Note that the metric (\ref{swmodspacemetric}) on moduli space can be written as
\beq\label{qxa}
\rmd s^2 =\im(\rmd \ad\rmd \bar a) = {i\over 2} (\rmd a\rmd\bar \ad - \rmd \ad \rmd
\bar a)
\eeq
where $\la z_{\rm D}\ra = \half \ad\sigma_3$ and $\ad=\d\cf(a)/\d a$, and that this metric
obviously also is invariant under the duality group  $Sl(2,{\bf Z})$

\subsection{Monopoles, dyons and the BPS mass spectrum}

At this point, I will have to add a couple of ingredients without much further
justification and refer the reader to the literature for more details.

In a spontaneously broken gauge theory as the one we are considering, typically there
are solitons (static, finite-energy solutions of the equations of motion) that carry
magnetic charge and behave like non-singular magnetic monopoles (for a
pedagogical treatment, see Coleman's lectures). The duality transformation 
(\ref{qviii}) constructed above
exchanges electric and magnetic degrees of freedom, hence electrically charged states,
as would be described by hypermultiplets of our $N=2$ supersymmetric version, with
magnetic monopoles.

As for any theory with extended supersymmetry, there are long and short (BPS)
multiplets in the present $N=2$ theory.
small (or short) multiplets have 4 
helicity states and large (or long) ones have 16 helicity states. 
As discussed earlier, massless states must be 
in short
multiplets, while massive states are in short ones if they satisfy the BPS condition
$m^2=2\vert Z\vert^2$, or in long ones if
$m^2>2\vert Z\vert^2$. Here $Z$ is the central charge of the $N=2$ susy algebra 
rescaled by a factor of $\rd$ with respect to our earlier conventions of section 2 
(in order to conform with the normalisation used by Seiberg and Witten).
The states that become massive by the Higgs mechanism must
be in short multiplets since they were before the symmetry breaking 
and the Higgs mechanism cannot generate
the missing $16-4=12$ helicity states. The heavy gauge bosons\footnote{
Again, to conform with the Seiberg-Witten normalisation, we have absorbed a factor 
of $g$ into $a$ and $a_D$, so that the masses of the heavy gauge bosons now are $m=\rd |a|$  rather than $\rd g |a|$.}
have masses $m=\rd |a|= \rd |Z|$ and hence $Z=a$. This generalises to all purely electrically charged states as
$Z=a n_e$ where $n_e$ is the (integer) electric charge. Duality then implies that a
purely magnetically charged state has $Z=\ad n_m$ where $n_m$ is the (integer)
magnetic charge. A state with both types of charge, called a dyon,
 has $Z=a n_e + \ad n_m$ since the
central charge is additive. All this applies to states in short multiplets, so-called
BPS-states. The mass formula  for these states then is
\beq\label{qxi}
m^2=2\vert Z\vert^2\quad , \quad Z= (n_m,n_e)\pmatrix{\ad\cr a\cr} \ .
\eeq
It is clear that under a $Sl(2,{\bf Z})$ transformation
$M=\pmatrix{\a&\b\cr\g&\delta\cr}
\in Sl(2,{\bf Z})$ acting on $\pmatrix{\ad\cr a\cr}$, the charge vector gets
transformed to $ (n_m,n_e) M =  (n'_m,n'_e)$ which are again integer charges. In
particular, one sees again at the level of the charges that
 the transformation (\ref{qviii}) exchanges purely electrically charged states with purely
magnetically charged ones. It can be shown that precisely those BPS
states are stable for which $n_m$ and $n_e$ are  relatively prime, i.e. for stable
states $(n_m,n_e) \ne (qm,qn)$ for integer $m,n$ and $q\ne \pm 1$.

\section{Singularities and Monodromy}

In this section we will study the behaviour of $a(u)$ and $\ad(u)$ as $u$ varies on the
moduli space $\M$. Particularly useful information will be obtained from their
behaviour as $u$ is taken around a closed contour. If the contour does not encircle
certain singular points to be determined below, $a(u)$ and $\ad(u)$ will return to
their initial values once $u$ has completed its contour. However, if the $u$-contour
goes around these singular points, $a(u)$ and $\ad(u)$ do not return to their initial
values but rather to certain linear combinations thereof: one has a non-trivial
monodromy for the multi-valued functions $a(u)$ and $\ad(u)$.

\subsection{The monodromy at infinity}

This is immediately clear from the  behaviour near $u=\infty$. As already
explained in section 3.4, as $u\to \infty$, due to asymptotic freedom,
the perturbative expression for $\cf(a)$ is valid and one has from (\ref{tvi})
for $\ad=\d\cf(a)/\d a$
\beq\label{ci}
\ad(u)={i\over \pi} a \left( \ln{a^2\over \L^2}+1\right)\quad , \quad
u\to\infty \ .
\eeq
Now take $u$ around a counterclockwise contour of very large radius in
the complex $u$-plane, often simply written as $u\to e^{2\pi i}u$. This
is equivalent to having $u$ encircle the point at $\infty$ on the Riemann
sphere in a {\it clockwise} sense. In any case, since $u=\half a^2$ (for
$u\to\infty$) one has $a\to -a$ and
\beq\label{cii}
\ad\to {i\over\pi} (-a) \left( \ln{e^{2\pi i}a^2\over \L^2}+1\right)
=-\ad+2a
\eeq
or
\beq\label{ciii}
\pmatrix{\ad(u)\cr a(u)\cr}\to M_\infty  \pmatrix{\ad(u)\cr a(u)\cr}
\quad , \quad M_\infty=\pmatrix{-1&2\cr 0&-1\cr} \ .
\eeq
Clearly, $u=\infty$ is a branch point of $\ad(u)\sim {i\over
\pi}\sqrt{2u} \left(\ln{u\over\L^2}+1\right)$. This is why this point is 
referred to as a
singularity of the moduli space.

\subsection{How many singularities?}

Can $u=\infty$ be the only singular point? Since a branch cut has to
start and end somewhere, there must be at least one other singular point.
Following Seiberg and Witten, I will argue that one actually needs three
singular points at least. To see why two cannot work, let's suppose for a
moment that there are only two singularities and show that this leads to
a contradiction.

Before doing so, let me note that there is an important 
so-called ${\rm U}(1)_R$-symmetry in the classical theory
that takes $z\to e^{2i\a}z$, $\f \to e^{2i\a}\f$,
$W \to e^{i\a}W$, $\t\to e^{i\a}\t$,
$\tb\to e^{i\a}\tb$, thus $\dt \to e^{-2i\a} \dt$,
$\dtb \to e^{-2i\a}\dtb$. Then the
classical action is invariant under this global
symmetry. More generallly, the action  will be invariant if
$\cf(z) \to e^{4i\a} \cf(z)$. This symmetry is broken by the one-loop
correction and also by instanton contributions. The latter give
corrections to $\cf$ of the form $z^2\sum_{k=1}^\infty c_k \left( \L^2/
z^2 \right)^{2k}$, and hence are invariant only for 
$\left(e^{4i\a}\right)^{2k} =1$, i.e. $\a={2\pi n\over 8},\ n\in{\bf Z}$.
Hence instantons break the ${\rm U}(1)_R$-symmetry  to a dicrete ${\bf Z}_8$.
The one-loop corrections behave as 
${i\over 2\pi}z^2\ln {z^2\over\L^2}\to e^{4i\a}\left(
{i\over 2\pi}z^2\ln {z^2\over\L^2} - {2\a\over \pi}z^2\right)$. As 
before one shows that this only changes the
action by $2\pi\n \left({4\a\over \pi}\right)$ where $\n$ is integer, so
that again this change is irrelevant as long as ${4\a\over \pi}=n$ or
$\a={2\pi n\over 8}$. Under this ${\bf Z}_8$-symmetry, $z\to e^{i\pi
n/2}z$, i.e. for odd $n$ one has $z^2\to -z^2$. The non-vanishing
expectation value $u=\la\tr z^2\ra$ breaks this ${\bf Z}_8$ further to
${\bf Z}_4$. Hence for a given vacuum, i.e. a given point on moduli space
there is only a ${\bf Z}_4$-symmetry left from the ${\rm U}(1)_R$-symmetry.
However, on the manifold of all possible vacua, i.e. on $\M$, one has
still the full ${\bf Z}_8$-symmetry, taking $u$ to $-u$.

Due to this global symmetry $u\to -u$, singularities of $\M$ should come
in pairs: for each singularity at $u=u_0$ there is another one at
$u=-u_0$. The only fixed points of $u\to -u$ are $u=\infty$ and $u=0$. We
have already seen that $u=\infty$ is a singular point of $\M$. So if
there are only two singularities the other must be the fixed point $u=0$.

If there are only two singularities, at $u=\infty$ and $u=0$, then by contour
deformation (``pulling the contour over the back of the sphere")\footnote{
It is well-known from complex analysis that monodromies are associated with contours
around branch points. The precise from of the contour does not matter, and it can be
deformed as long as it does not meet another branch point. Our singularities precisely
are the branch points of $a(u)$ or $\ad(u)$.}
 one sees that the
monodromy around 0 (in a counterclockwise sense) is the same as the above monodromy
around $\infty$: $M_0=M_\infty$. But then $a^2$ is not affected by any monodromy and
hence is a good global coordinate, so
one can take $u=\half a^2$ on all of $\M$, and furthermore one must have
\beq\label{civ}
\begin{array}{rcl}
\ad&=&{i\over \pi} a \left( \ln {a^2\over \L^2}+1\right) + g(a)\crbig
a&=&\sqrt{2u}\crbig
\end{array}
\eeq
where $g(a)$ is some entire function of $a^2$. This implies that
\beq\label{cv}
\tau={\rmd\ad\over\rmd a}={i\over \pi} \left( \ln {a^2\over \L^2}+3\right) +
{\rmd g\over\rmd a} \ .
\eeq
The function $g$ being entire, $\im {\rmd g\over\rmd a}$ cannot have a minimum (unless
constant) and it is clear that $\im\tau$ cannot be positive everywhere. As already
emphasized, this means that $a$ (or rather $a^2$) cannot be a good global coordinate and
(\ref{civ}) cannot hold globally. Hence, two singularities only cannot work.

The next simplest choice is to try 3 singularities. Due to the $u\to -u$ symmetry,
these 3 singularities are at $\infty, u_0$ and $-u_0$ for some $u_0\ne 0$. 
In particular, $u=0$ is no
longer a singularity of the quantum moduli space. To get a singularity also at $u=0$
one would need at least four singularities at $\infty, u_0, -u_0$ and $0$. As discussed
later, this is not possible, and more generally, exactly 3 singularities seems to be
the only consistent possibility.

So there is no singularity at $u=0$ in the quantum moduli space $\M$.
Classically, however, one precisely expects that $u=0$ should be a singular point,
since classically $u=\half a^2$, hence $a=0$ at this point, and
then there is no Higgs mechanism any more. Thus all (elementary) massive states, i.e.
the gauge
bosons $v_\m^1, v_\m^2$ and their susy partners $\p^1, \p^2, \l^1, \l^2$
become massless. Thus the description of the lights fields in terms of
the previous Wilsonian effective action should break down, inducing a singularity on
the moduli space. As already stressed, this is the clasical picture. While $a\to
\infty$ leads to asymptotic freedom  and the microscopic ${\rm SU}(2)$ theory is weakly
coupled, as $a\to 0$ one goes to a strong coupling regime where the classical
reasoning has no validity any more, and $u\ne \half a^2$. By the BPS mass formula 
(\ref{qxi})
massless gauge bosons still are possible at $a=0$, but this does no longer correspond to
$u=0$.

So where has the singularity due to massless gauge bosons at $a=0$ 
moved to? One might be tempted to think that
$a=0$ now corresponds to the singularities at $u=\pm u_0$, but this is not the case
as I will show in a moment. The answer is that the point $a=0$ no
longer belongs to the quantum moduli space (at least not to the component connected to
$u=\infty$ which is the only thing one considers). This can be seen explicitly from
the form of the solution for $a(u)$ given in the next section.

\subsection{The strong coupling singularities}

Let's now concentrate on the case of three singularities at $u=\infty, u_0$ and $-u_0$.
What is the interpretation of the (strong-coupling) singularities at finite $u=\pm
u_0$? One might first try to consider that they are still due to the gauge bosons
becoming massless. However, as Seiberg and Witten point out, massless gauge bosons
would imply an asymptotically conformally invariant theory in the infrared limit and
conformal invariance implies $u=\la\tr z^2\ra=0$ unless $\tr z^2$ has dimension zero
and hence would be the unity operator - which it is not. So the singularities at $u=\pm
u_0\ (\ne 0)$ do not correspond to massless gauge bosons.

There are no other elementary $N=2$ multiplets in our theory. The next thing to try
is to consider collective excitations - solitons, like the magnetic monopoles or dyons.
Let's first study what happens if a magnetic monopole  of unit magnetic charge becomes
massless. From the BPS mass formula (\ref{qxi}), the mass of the magnetic monopole is
\beq\label{cvi}m^2=2\vert \ad\vert^2
\eeq
and hence vanishes at $\ad=0$. We will see that this produces one of the two
stron-coupling singularities. So call $u_0$ the value of $u$ at whiche $\ad$ vanishes.
Magnetic monopoles are described by hypermultiplets $H$ of $N=2$ susy that couple
locally to the dual fields $\Fd$ and $W_D$, just as electrically charged ``electrons"
would be described by hypermultiplets that couple locally to $\F$ and $W$. So in the
dual description we have $\Fd, W_D$ and $H$, and, near $u_0$, $\ad\sim \la \Fd\ra$ is
small. This theory is exactly $N=2$ susy QED with very light electrons (and a 
subscript
$D$ on every quantity). The latter theory is not asymptotically free, but has a
$\b$-function given by
\beq\label{cvii}
\m{\rmd\over \rmd\m} g_D={g_D^3\over 8\pi^2}
\eeq
where $g_D$ is the coupling constant. But the scale $\m$ is proportional to $\ad$ and
${4\pi i\over g_D^2(\ad)}$ is $\tau_D$ for $\t_D=0$ (of course, super QED, unless
embedded into a larger gauge group,  does not
allow for a non-vanishing theta angle). One concludes that for $u\approx u_0$ or
$\ad\approx 0$
\beq\label{cviii}
\ad {\rmd\over \rmd \ad} \tau_D=-{i\over \pi} \ \Rightarrow \ \tau_D = -{i\over \pi}
\ln\ad \ .
\eeq
Since $\tau_D={\rmd (-a)\over \rmd\ad}$ this can be integrated to give
\beq\label{cix}
a\approx a_0+{i\over \pi} \ad\ln\ad \qquad (u\approx u_0)
\eeq
where we dropped a subleading term $-{i\over \pi}\ad$. Now, $\ad$ should be a good
coordinate in the vicinity of $u_0$, hence depend linearly\footnote{
One might want to try a more general dependence like $\ad\approx c_0 (u-u_0)^k$ with
$k>0$. This leads to a monodromy in $Sl(2,{\bf Z})$ only for integer $k$. 
The factorisation
condition below, together with the form of $M(n_m,n_e)$ also given below, 
then imply that
$k=1$ is the only possibility.
}
on $u$. One concludes
\beq\label{cx}
\ad\approx c_0 (u-u_0)\quad , \quad
a\approx  a_0+{i\over \pi} c_0 (u-u_0)\ln (u-u_0) \ .
\eeq
From these expressions one immediately reads the monodromy as $u$ turns around $u_0$
counterclockwise, $u-u_0\to e^{2\pi i} (u-u_0)$:
\beq\label{cxi}
\pmatrix{\ad\cr a\cr} \to \pmatrix{ \ad\cr a-2\ad \cr} = M_{u_0} 
\pmatrix{\ad\cr a\cr} \quad , \quad
 M_{u_0}=\pmatrix{1&0\cr -2&1\cr} \ . 
\eeq

To obtain the monodromy matrix at $u=-u_0$ it is enough to observe that the contour
around $u=\infty$ is equivalent to a counterclockwise contour of very large radius in
the complex plane. This contour can be deformed into a contour encircling $u_0$ and a
contour encircling $-u_0$, both counterclockwise. It follows the factorisation
condition on the monodromy matrices\footnote{
There is an ambiguity concerning the ordering of $M_{u_0}$ and $M_{-u_0}$
which will be resolved below.}
\beq\label{cxii}
M_\infty=M_{u_0} M_{-u_0}
\eeq
and hence
\beq\label{cxiii}
M_{-u_0} = \pmatrix{-1&2\cr -2&3\cr} \ .
\eeq

What is the interpretation of this singularity at $u=-u_0$? To discover this, consider
the behaviour under monodromy of the BPS mass formula $m^2=2\vert Z\vert^2$ with $Z$
given by (\ref{qxi}), i.e. $Z=(n_m,n_e)\pmatrix{\ad\cr a\cr}$. The monodromy transformation
$\pmatrix{\ad\cr a\cr}\to M \pmatrix{\ad\cr a\cr}$ can be interpreted as changing the
magnetic and electric quantum numbers as
\beq\label{cxiv}
(n_m,n_e)\to (n_m,n_e) M \ .
\eeq
The state of vanishing mass responsible for a singularity should be invariant under the
monodromy, and hence be a left eigenvector of $M$ with unit eigenvalue. This is clearly
so for the magnetic monopole: $(1,0)$ is a left eigenvector of $\pmatrix{1&0\cr
-2&1\cr}$ with unit eigenvalue. This simply reflects that $m^2=2\vert \ad\vert^2$ is
invariant under (\ref{cxi}). Similarly, the left eigenvector of (\ref{cxiii}) with unit eigenvalue
is $(n_m, n_e)=(1,-1)$ This is a dyon. Thus the sigularity at $-u_0$ is interpreted as
being due to a $(1,-1)$ dyon becoming massless.

More generally, $(n_m, n_e)$ is the left eigenvector with unit eigenvalue\footnote{
Of course, the same is true for any $(q n_m, q n_e)$ with $q\in {\bf Z}$, but according
to the discussion in section 4.3 on the stability of BPS states, states with $q\ne \pm
1$ are not stable.}
of
\beq\label{cxv}
M(n_m,n_e)=\pmatrix{ 1+2n_m n_e& 2 n_e^2\cr -2 n_m^2 & 1- 2 n_m n_e\cr}
\eeq
which is the monodromy matrix that should appear for any singularity due to a massless
dyon with charges $(n_m, n_e)$. Note that $M_\infty$ as given in (\ref{ciii}) is not of this
form, since it does not correspond to a hypermultiplet becoming massless.

One notices that the relation (\ref{cxii}) does not look invariant under $u\to -u$, i.e
$u_0\to -u_0$ since $M_{u_0}$ and $M_{-u_0}$ do not commute. The apparent contradiction
with the ${\bf Z}_2$-symmetry is resolved by the following remark. The precise
definition of the composition of two monodromies as in (\ref{cxii}) 
requires a choice of
base-point $u=P$ (just as in the definition of homotopy groups). Using a different
base-point, namely $u=-P$, leads to 
\beq\label{cxvi}
M_\infty =M_{-u_0}M_{u_0}
\eeq
instead. Then one would obtain $M_{-u_0}=\pmatrix{3&2\cr -2&-1}$, and comparing with
(\ref{cxv}), this would be interpreted as due to a $(1,1)$ dyon. Thus the ${\bf
Z}_2$-symmetry $u\to -u$ on the quantum moduli space also acts on the base-point $P$,
hence exchanging (\ref{cxii}) and (\ref{cxvi}). At the same time it exchanges the $(1,-1)$ dyon
with the $(1,1)$ dyon.

Does this mean that the $(1,1)$ and  $(1,-1)$ dyons play a privileged role? Actually
not. If one first turns $k$ times around $\infty$, then around $u_0$, and then $k$
times around $\infty$ in the opposite sense, the corresponding monodromy is\break\hfill
$M_\infty^{-k} M_{u_0} M_\infty^k = \pmatrix{1-4k&8k^2\cr -2& 1+4k\cr}=M(1,-2k)$ and
similarly
$M_\infty^{-k} M_{-u_0} M_\infty^k$
$=\pmatrix{-1-4k&2+8k+8k^2\cr -2& 3+4k\cr} =M(1,-1-2k)$. 
So one sees that these monodromies correspond to dyons with $n_m=1$ and any
$n_e\in {\bf Z}$ becoming massless. Similarly one has e.g.
$M_{u_0}^{k} M_{-u_0} M_{u_0}^{-k}$ $=M(1-2k,-1)$, etc.

Let's come back to the question of how many singularities there are. Suppose there are
$p$ strong coupling singularities at $u_1, u_2, \ldots u_p$ in addition to the one-loop
perturbative singularity at  $u=\infty$. Then one has a factorisation analogous to
(\ref{cxii}):
\beq\label{cxvii}
M_\infty =  M_{u_1}  M_{u_2} \ldots  M_{u_p}
\eeq
with $M_{u_i}=M(n_m^{(i)}, n_e^{(i)})$ of the form (\ref{cxv}). 
It thus becomes a problem of
number theory to find out whether, for given $p$, there exist solutions to  
(\ref{cxvii}) with integer 
$n_m^{(i)}$ and $n_e^{(i)}$. For several low values of $p>2$ it has been 
checked
that there are no such solutions, and it seems likely that the same is true for all
$p>2$.

\section{The solution}

Recall that our goal is to determine the exact non-perturbative 
low-energy effective action, i.e. determine the function $\cf(z)$ 
locally. This will be achieved, at least in principle, once we know 
the functions $a(u)$ and $a_D(u)$, since one then can invert the first 
to obtain $u(a)$, at least within a certain domain of the moduli 
space. Substituting this into $a_D(u)$ yields $a_D(a)$ which upon 
integration gives the desired $\cf(a)$.

So far we have seen that $\ad(u)$ and $a(u)$ are single-valued except for the
monodromies around $\infty, u_0$ and $-u_0$. As is well-known from complex analysis,
this means that  $\ad(u)$ and $a(u)$ are really multi-valued functions with branch
cuts, the branch points being  $\infty, u_0$ and $-u_0$. A typical example is 
$f(u)=\sqrt{u} F(a,b,c;u)$, where $F$ is the hypergeometric function. The latter has a
branch cut from $1$ to $\infty$. Similarly, $\sqrt{u}$ has a branch cut from $0$ to
$\infty$ (usually taken along the negative real axis), so that $f(u)$ has two branch
cuts joining the three singular points $0,1$ and $\infty$. When $u$ goes around any of
these singular points there is a non-trivial monodromy between $f(u)$ and one other
function $g(u)= u^d F(a',b',c';u)$. The three monodromy matrices are in (almost) one-to-one
correspondence with the pair of functions $f(u)$ and $g(u)$.

In the physical problem at hand one knows the monodromies, namely
\beq\label{si}
M_{\infty}=\pmatrix{-1&2\cr 0&-1\cr}\ , \quad 
M_{u_0}=\pmatrix{1&0\cr -2&1\cr}\ , \quad
M_{-u_0}=\pmatrix{-1&2\cr -2&3\cr}
\eeq
and one wants to determine the corresponding functions $\ad(u)$ and $a(u)$. As will be
explained, the monodromies fix $\ad(u)$ and $a(u)$ up to normalisation, which will be
determined from the known asymptotics (\ref{ci}) at infinity.

The precise location of $u_0$ depends on the renormalisation
conditions which can be chosen such that $u_0=1$. Assuming this choice in the
sequel will simplify somewhat the equations. If one wants to keep $u_0$, essentially
all one has to do is to replace $u\pm 1$ by ${u\pm u_0\over u_0}={u\over u_0}\pm 1$.

\subsection{The differential equation approach}

Monodromies typically arise from differential equations with periodic coefficients.
This is well-known in solid-state physics where one considers a Schr\"odinger equation
with a periodic potential\footnote{
The constant energy has been included into the potential, and the mass has been
normalised to $\half$.}
\beq\label{sia}
\left[ -{\rmd^2\over \rmd x^2} + V(x)\right] \p(x)=0 \quad , \quad V(x+2\pi)=V(x) \ .
\eeq
There are two independent solutions $\p_1(x)$ and $\p_2(x)$. One wants to compare
solutions at $x$ and at $x+2\pi$. Since, due to the periodicity of the potential $V$,
the differential equation at $x+2\pi$ is exactly the same as at $x$, the set of
solutions must be the same. In other words, $\p_1(x+2\pi)$ and $\p_2(x+2\pi)$ must be
linear combinations of $\p_1(x)$ and $\p_2(x)$:
\beq\label{sii}
\pmatrix{\p_1\cr \p_2\cr} (x+2\pi) = M \pmatrix{\p_1\cr \p_2\cr} (x) 
\eeq
where $M$ is a (constant) monodromy matrix.

The same situation arises for differential equations in the complex plane with
meromorphic coefficients. Consider again the Schr\"odinger-type equation
\beq\label{siii}
\left[ -{\rmd^2\over \rmd z^2} + V(z)\right] \p(z)=0
\eeq
with meromorphic $V(z)$, having poles at $z_1, \ldots z_p$ and (in general) also at
$\infty$. The periodicity of the previous example is now replaced by the
single-valuedness of $V(z)$ as $z$ goes around any of the poles of $V$ (with $z-z_i$
corresponding roughly to $e^{ix}$). So, as $z$  goes once around any one of the $z_i$, the
differential equation (\ref{siii}  does not change. So by the same argument as above, the two solutions
$\p_1(z)$ and $\p_2(z)$, when continued along the path surrounding $z_i$ must again be
linear combinations of $\p_1(z)$ and $\p_2(z)$:
\beq\label{siv}
\pmatrix{\p_1\cr \p_2\cr} \left(z+e^{2\pi i}(z-z_i)\right) 
= M_i \pmatrix{\p_1\cr \p_2\cr} (z) 
\eeq
with a constant $2\times 2$-monodromy matrix $M_i$ for each of the poles of $V$. Of
course, one again has the factorisation condition (\ref{cxvii}) for $M_\infty$. It is
well-known, that non-trivial constant monodromies correspond to poles of $V$ that are
at most of second order. In the language of differential equations, (\ref{siii}) then only
has {\it regular} singular points.

In our physical problem, the {\it two} multivalued functions $\ad(z)$ and $a(z)$ have 3
singularities with non-trivial monodromies at $-1, +1$ and $\infty$. Hence they must be
solutions of a second-order differential equation (\ref{siii}) with the potential $V$ having
(at most) second-order poles precisely at these points. 
The general form of this potential is\footnote{
Additional terms in $V$ that naively look like first-order poles 
($\sim {1\over z-1}$ or ${1\over z+1}$)
cannot appear since they correspond to third-order poles at $z=\infty$.
}
\beq\label{sv}
V(z)=-{1\over 4} \left[ {1-\l_1^2\over (z+1)^2} +  {1-\l_2^2\over (z-1)^2 }
-{1-\l_1^2-\l_2^2+\l_3^2\over (z+1)(z-1)} \right]
\eeq
with double poles at $-1, +1$ and $\infty$. The corresponding residues are 
$-{1\over 4}(1-\l_1^2)$, $-{1\over 4}(1-\l_2^2)$ and $-{1\over 4}(1-\l_3^2)$. Without
loss of generality, I assume $\l_i\ge 0$. The corresponding differential equation
(\ref{siii}) is well-known in the mathematical literature 
since it can be transformed into the hypergeometric differential equation. The
transformation to the standard hypergeometric equation is readily performed by setting
\beq\label{svi}
\p(z)=(z+1)^{\half (1-\l_1)} (z-1)^{\half (1-\l_2)}\,  f\left( {z+1\over 2}\right) \ .
\eeq
One then finds that $f$ satisfies the hypergeometric differential equation
\beq\label{sviii}
x(1-x) f''(x)+[c-(a+b+1)x]f'(x)-abf(x)=0
\eeq
with
\beq\label{six}
a=\half (1-\l_1-\l_2+\l_3)\ , \quad
b=\half (1-\l_1-\l_2-\l_3)\ , \quad
c=1-\l_1 \ . 
\eeq
The solutions of the hypergeometric equation (\ref{sviii}) can be written in many different
 ways due to the various identities between the hypergeometric function $F(a,b,c;x)$
and products with powers, e.g. $(1-x)^{c-a-b} F(c-a,c-b,c;x)$, etc. A convenient choice for the two
independent solutions is the following 
\beq\label{sx}
\begin{array}{rcl}
f_1(x)&=&(-x)^{-a}F(a,a+1-c,a+1-b;{1\over x})\crbig
f_2(x)&=&(1-x)^{c-a-b} F(c-a,c-b,c+1-a-b;1-x) \ . \crbig
\end{array}
\eeq
$f_1$ and $f_2$ correspond to Kummer's solutions denoted $u_3$ and $u_6$. The choice of
$f_1$ and $f_2$ is motivated by the fact that $f_1$ has simple monodromy properties
 around
$x=\infty$ (i.e. $z=\infty$) and $f_2$ has simple monodromy properties
around $x=1$ (i.e. $z=1$),
so they are good candidates to be identified with $a(z)$ and $\ad(z)$.

One can extract a great deal of information from the asymptotic forms of $\ad(z)$ and
$a(z)$. As $z\to\infty$ one has $V(z)\sim -{1\over 4} \, {1-\l_3^2\over z^2}$, so that
the two independent solutions behave asymptotically as $z^{\half (1\pm \l_3)}$ if $\l_3
\ne 0$, and as $\sqrt{z}$ and $\sqrt{z}\ln z$ if $\l_3=0$. Comparing with (\ref{civ}) 
(with $u\to z$) we see
that the latter case is realised. Similarly, with $\l_3=0$, as $z\to 1$, one has
$V(z)\sim -{1\over 4} \left(  {1-\l_2^2\over (z-1)^2}-{1-\l_1^2-\l_2^2\over 2(z-1)}
\right)$, where I have kept the subleading term. From the logarithmic asymptotics
(\ref{cx})
one then concludes $\l_2=1$ (and from the subleading term also $-{\l_1^2\over
8}={i\over \pi}{c_0\over a_0}$). The ${\bf Z}_2$-symmetry ($z\to -z$) on the moduli
space then implies that, as $z\to -1$, the potential  $V$ does not  have a double pole
either, so that also $\l_1=1$. Hence we conclude
\beq\label{sxi}
\l_1=\l_2=1\ ,\ \ \l_3=0 \ \Rightarrow \ V(z)= -{1\over 4}\, {1\over (z+1)(z-1)}
\eeq
and $a=b=-\half,\ c=0$. Thus from (\ref{svi}) one has $\p_{1,2}(z)=f_{1,2}\left({z+1\over
2}\right)$. One can then verify that the two solutions
\beq\label{sxiii}
\begin{array}{rcl}
\ad(u)&=&i \p_2(u)=i{u-1\over 2} F\left({1\over 2},{1\over 2},2;{1-u\over
2}\right)\crbig
a(u)&=&-2i \p_1(u)=\sqrt{2} (u+1)^{1\over 2} 
F\left(-{1\over 2},{1\over 2},1;{2\over u+1}\right)\crbig
\end{array}
\eeq
indeed have the required monodromies (\ref{si}), as well as the correct asymptotics.

It might look as if we have not used the monodromy properties to determine $\ad$ and
$a$ and that they have been determined only from the asymptotics. This is not entirely
true, of course. The very fact that there are non-trivial monodromies only at $\infty,
+1$ and $-1$ implied that $\ad$ and $a$ must satisfy the second-order differential
equation (\ref{siii}) with the potential (\ref{sv}). 
To determine the $\l_i$ we then used the
asymptotics of $\ad$ and $a$. But this is (almost) the same as using the monodromies
since the
latter were obtained from the asymptotics.

Using the integral representation of the hypergeometric function, the solution
(\ref{sxiii}) can be nicely rewritten as 
\beq\label{sxiv}
\ad(u)={\sqrt{2}\over \pi} \int_1^u {\rmd x\ \sqrt{x-u}\over \sqrt{x^2-1} } \ , \quad
a(u)={\sqrt{2}\over \pi} \int_{-1}^1 {\rmd x\ \sqrt{x-u}\over \sqrt{x^2-1} } \ . 
\eeq

One can invert the second equation (\ref{sxiii})  to obtain $u(a)$, within a 
certain domain, and 
insert the result into $\ad(u)$ to obtain $\ad(a)$. Integrating
with respect to $a$ yields $\cf(a)$ and hence the low-energy effective action. 
I should
stress that this expression for $\cf(a)$ is not globally
valid but only on a certain portion of the moduli space. Different analytic
continuations must be used on other portions.

\subsection{The approach using elliptic curves}

In their paper, Seiberg and Witten do not use the differential equation approach just
described, but rather introduce an auxiliary construction: a certain elliptic curve by
means of which two functions with the correct monodromy properties are constructed. I
will not go into details here, but simply sketch this approach.

To motivate their construction {\it a posteriori}, we notice the following: from the
integral representation (\ref{sxiv}) it is natural to consider the complex $x$-plane. More
precisely, the integrand has square-root branch cuts with branch points at $+1, -1, u$
and $\infty$. The two branch cuts can be taken to run from $-1$ to $+1$ and from $u$ to
$\infty$. The Riemann surface of the integrand is two-sheeted with the two sheets
connected through the cuts. If one adds the point at infinity to each of the two
sheets, the topology of the Riemann surface is that of two spheres connected by two
tubes (the cuts), i.e. a torus. So one sees that the Riemann
surface of the integrand in (\ref{sxiv}) has genus one. This is the elliptic curve considered
by Seiberg and Witten.

As is well-known, on a torus there are two independent non-trivial closed paths
(cycles). One cycle ($\g_2$) can be taken to go once around the cut $(-1,1)$, and the
other cycle ($\g_1$) to go from $1$ to $u$ on the first sheet and back from $u$ to $1$
on the second sheet. The solutions $\ad(u)$ and $a(u)$ in (\ref{sxiv}) are precisely the
integrals of some suitable differential $\l$ along the two cycles $\g_1$ and $\g_2$:
\beq\label{sxiva}
\ad=\oint_{\g_1} \l  \quad , \quad a=\oint_{\g_2} \l  \quad , \quad
\l={\sqrt{2}\over 2\pi} {\sqrt{x-u}\over \sqrt{x^2-1} } \rmd x \ .
\eeq
These integrals are called period integrals. They are known to satisfy a
second-order differential equation, the so-called Picard-Fuchs equation, that is
nothing else than our Schr\"odinger-type equation (\ref{siii}) with $V$ given by
(\ref{sxi}).

How do the monodromies appear in this formalism? As $u$ goes once around $+1,-1$ or
$\infty$, the cycles $\g_1, \g_2$ are changed into linear combinations of themselves
with integer coefficients:
\beq\label{sxv}
\pmatrix{\g_1\cr \g_2} \to M \pmatrix{\g_1\cr \g_2} \quad , \quad
M\in Sl(2, {\bf Z}) \ .
\eeq
This immediately implies
\beq\label{sxvi}
\pmatrix{\ad\cr a} \to M \pmatrix{\ad\cr a}
\eeq
with the same $M$ as in (\ref{sxv}). The advantage here is that one automatically gets
monodromies with {\it integer} coefficients. The other advantage is that
\beq\label{sxvii}
\tau(u)={\rmd \ad/\rmd u\over  \rmd a/\rmd u}
\eeq
can be easily seen to be the $\tau$-parameter describing the complex structure of the
torus, and as such is garanteed to satisfy
$\im\tau(u) >0$
which was the requirement for positivity of the metric on moduli space.

To motivate the appearance of the genus-one elliptic curve (i.e. the torus) {\it a
priori} - without knowing the solution (\ref{sxiv}) from the differential equation approach -
Seiberg and Witten remark that the three monodromies are all very special: they do not
generate all of $Sl(2, {\bf Z})$ but only a certain subgroup $\G(2)$ of matrices 
in
$Sl(2, {\bf Z})$ congruent to $1$ modulo $2$. Furthermore, they remark that the
$u$-plane with punctures at $1,-1,\infty$ can be thought of as the  quotient of the
upper half plane $H$ by $\G(2)$, and that $H/\G(2)$ naturally parametrizes (i.e. is the
moduli space of) elliptic curves described by
\beq\label{sxix}
y^2=(x^2-1)(x-u) \ .
\eeq
Equation (\ref{sxix}) corresponds to the genus-one Riemann surface discussed above, and it is
then natural to introduce the cycles $\g_1, \g_2$ and the differential $\l$ from
(\ref{sxiv}).
The rest of the argument then goes as I just exposed.

\section{Summary}

 We have seen realised a version of
electric-magnetic duality accompanied by a duality transformation on the expectation
value of the scalar (Higgs) field, $a\leftrightarrow \ad$. There is a manifold of
inequivalent vacua, the moduli space $\M$, corresponding to different Higgs expectation
values. The duality relates strong coupling regions in $\M$ to the perturbative region
of large $a$ where the effective low-energy action is known asymptotically
 in terms of $\cf$. Thus duality allows us to determine the latter also at strong
coupling. The holomorphicity condition from $N=2$ supersymmetry then puts such strong
constraints on $\cf(a)$, or equivalently on $\ad(u)$ and $a(u)$ that the full functions
can be determined solely from their asymptotic behaviour at the strong and weak
coupling singularities of $\M$.